\documentclass[reprint, aps, nofootinbib, prb]{revtex4-1}
\usepackage{amssymb}
\usepackage{appendix}
\usepackage{bm}
\usepackage{dcolumn}
\usepackage{floatrow}
\usepackage[dvipdfmx]{graphicx}
\usepackage[hidelinks]{hyperref}
\usepackage{mathtools}
\usepackage{tikz}
\usepackage{makecell}
\usepackage{epstopdf}
\usepackage{multirow}
\usepackage[section]{placeins}

\hypersetup{
  colorlinks,
  linkcolor={red!50!black},
  citecolor={blue!50!black},
  urlcolor={blue!80!black}
}

\graphicspath{{img/}}

\begin{document}
  
\newlength{\figurewidth} 
\setlength{\figurewidth}{\columnwidth}
\setlength\parindent{20pt}

\newcommand{\prtl}{\partial} \newcommand{\la}{\left\langle}
  \newcommand{\ra}{\right\rangle} \newcommand{\dla}{\la \! \! \!
  \la} \newcommand{\dra}{\ra \! \! \! \ra}
\newcommand{\we}{\widetilde}
\newcommand{\smfp}{{\mbox{\scriptsize mfp}}}
\newcommand{\smp}{{\mbox{\scriptsize mp}}}
\newcommand{\sph}{{\mbox{\scriptsize ph}}}
\newcommand{\sinhom}{{\mbox{\scriptsize inhom}}}
\newcommand{\sneigh}{{\mbox{\scriptsize neigh}}}
\newcommand{\srlxn}{{\mbox{\scriptsize rlxn}}}
\newcommand{\svibr}{{\mbox{\scriptsize vibr}}}
\newcommand{\smicro}{{\mbox{\scriptsize micro}}}
\newcommand{\scoll}{{\mbox{\scriptsize coll}}}
\newcommand{\sattr}{{\mbox{\scriptsize attr}}}
\newcommand{\sth}{{\mbox{\scriptsize th}}}
\newcommand{\sauto}{{\mbox{\scriptsize auto}}}
\newcommand{\seq}{{\mbox{\scriptsize eq}}}
\newcommand{\teq}{{\mbox{\tiny eq}}}
\newcommand{\sinn}{{\mbox{\scriptsize in}}}
\newcommand{\suni}{{\mbox{\scriptsize uni}}}
\newcommand{\tin}{{\mbox{\tiny in}}}
\newcommand{\scr}{{\mbox{\scriptsize cr}}}
\newcommand{\tstring}{{\mbox{\tiny string}}}
\newcommand{\sperc}{{\mbox{\scriptsize perc}}}
\newcommand{\tperc}{{\mbox{\tiny perc}}}
\newcommand{\sstring}{{\mbox{\scriptsize string}}}
\newcommand{\stheor}{{\mbox{\scriptsize theor}}}
\newcommand{\sGS}{{\mbox{\scriptsize GS}}}
\newcommand{\sBP}{{\mbox{\scriptsize BP}}}
\newcommand{\sNMT}{{\mbox{\scriptsize NMT}}}
\newcommand{\sbulk}{{\mbox{\scriptsize bulk}}}
\newcommand{\tbulk}{{\mbox{\tiny bulk}}}
\newcommand{\sXtal}{{\mbox{\scriptsize Xtal}}}
\newcommand{\sliq}{{\text{\tiny liq}}}

\newcommand{\smin}{\text{min}} \newcommand{\smax}{\text{max}}

\newcommand{\saX}{\text{\tiny aX}} \newcommand{\slaX}{\text{l,{\tiny
      aX}}}

\newcommand{\svap}{{\mbox{\scriptsize vap}}}
\newcommand{\sjam}{J} \newcommand{\Tm}{T_m}
\newcommand{\sTS}{{\mbox{\scriptsize TS}}}
\newcommand{\sDW}{{\mbox{\tiny DW}}}
\newcommand{\cN}{{\cal N}} \newcommand{\cB}{{\cal B}}
\newcommand{\br}{\bm r} \newcommand{\be}{\bm e}
\newcommand{\cH}{{\cal H}} \newcommand{\cHlt}{\cH_{\mbox{\scriptsize
      lat}}} \newcommand{\sthermo}{{\mbox{\scriptsize thermo}}}

\newcommand{\bu}{\bm u} \newcommand{\bk}{\bm k} \newcommand{\bX}{\bm
  X} \newcommand{\bY}{\bm Y} \newcommand{\bA}{\bm A}
\newcommand{\bb}{\bm b}

\newcommand{\lintf}{l_\text{intf}}

\newcommand{\DV}{\delta V_{12}}
\newcommand{\sout}{{\mbox{\scriptsize out}}} \newcommand{\dv}{\Delta
  v_{1 \infty}} \newcommand{\dvin}{\Delta v_{2 \infty}}

\newcommand*\xbar[1]{%
  \hbox{%
    \vbox{%
      \hrule height 0.5pt 
      \kern0.5ex
      \hbox{%
        \kern-0.1em
        \ensuremath{#1}%
        \kern-0.1em
      }%
    }%
  }%
}

\def\Xint#1{\mathchoice {\XXint\displaystyle\textstyle{#1}}%
  {\XXint\textstyle\scriptstyle{#1}}%
  {\XXint\scriptstyle\scriptscriptstyle{#1}}%
  {\XXint\scriptscriptstyle\scriptscriptstyle{#1}}%
  \!\int} \def\XXint#1#2#3{{\setbox0=\hbox{$#1{#2#3}{\int}$}
    \vcenter{\hbox{$#2#3$}}\kern-.5\wd0}} \def\ddashint{\Xint=}
\def\dashint{\Xint-}
\newcommand{\HRule}{\rule{\linewidth}{0.5mm}}

\title{Structural origin of the midgap electronic states and the
  Urbach tail in pnictogen-chalcogenide glasses}

\author{Alexey Lukyanov} \affiliation{Department of Chemistry,
  University of Houston, Houston, TX 77204-5003}

\author{Jon C. Golden} \affiliation{Department of Chemistry,
  University of Houston, Houston, TX 77204-5003}
\affiliation{Department of Physics, University of Houston, Houston, TX
  77204-5005}

\author{Vassiliy Lubchenko} \email{vas@uh.edu}
\affiliation{Department of Chemistry, University of Houston,
  Houston, TX 77204-5003} \affiliation{Department of Physics,
  University of Houston, Houston, TX 77204-5005}

\date{\today}

\begin{abstract}

  We determine the electronic density of states for
  computationally-generated bulk samples of amorphous chalcogenide
  alloys As$_{x}$Se$_{100-x}$.  The samples were generated using a
  structure-building algorithm reported recently by us ({\em
    J. Chem. Phys.} {\bf 147}, 114505).  Several key features of the
  calculated density of states are in good agreement with experiment:
  The trend of the mobility gap with arsenic content is reproduced.
  The sample-to-sample variation in the energies of states near the
  mobility gap is quantitatively consistent with the width of the
  Urbach tail in the optical edge observed in experiment.  Most
  importantly, our samples consistently exhibit very deep-lying midgap
  electronic states that are delocalized significantly more than what
  would be expected for a deep impurity or defect state; the
  delocalization is highly anisotropic. These properties are
  consistent with those of the topological midgap electronic states
  that have been proposed by Zhugayevych and Lubchenko as an
  explanation for several puzzling opto-electronic anomalies observed
  in the chalcogenides, including light-induced midgap absorption and
  ESR signal, and anomalous photoluminescence.  In a complement to the
  traditional view of the Urbach states as a generic consequence of
  disorder in atomic positions, the present results suggest these
  states can be also thought of as intimate pairs of topological
  midgap states that cannot recombine because of disorder. Finally,
  samples with an odd number of electrons exhibit neutral, spin $1/2$
  midgap states as well as polaron-like configurations that consist of
  a charge carrier bound to an intimate pair of midgap states; the
  polaron's identity---electron or hole---depends on the preparation
  protocol of the sample.

\end{abstract}

\maketitle

\section{Introduction}

\begin{figure}[t] 
  \centering
  \includegraphics[width=\columnwidth]{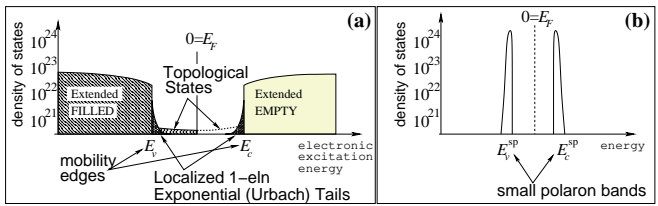}
  \caption{\label{DOS} \small {\bf (a)} Hypothetical electronic
    structure of amorphous chalcogenides, see also
    Refs.~\cite{PWA_negU2, Emin_rev, ZL_JCP} One-electron excitations
    occur between two bands, separated by a mobility gap. Localized
    states exist in the gap. The topological states, whose
    one-particle energy is distributed throughout the gap, absorb
    light at sub-gap frequencies when singly-occupied. {\bf (b)} The
    hypothetical density of states of itinerant electrons and holes
    that have relaxed to form small polarons, after
    Ref.~\cite{Emin_rev}}
\end{figure}

In contrast with their periodic counterparts, amorphous materials are
expressly non-Bloch solids. Despite this complication, many amorphous
semiconductors purvey electricity similarly to the corresponding
crystals. Indeed, from the viewpoint of the wave-packet representing a
charge carrier, the pertinent molecular orbitals are virtually
indistinguishable from true, infinitely-extended Bloch states as long
as the mean-free path of the carrier is less than the extent of the
orbitals. Thus one may still speak of {\em mobility} bands even in the
absence of strict periodicity.~\cite{AndersonLoc, Mott1982, CFO,
  Mott1990} In addition, many families of crystalline and amorphous
compounds alike are expected to exhibit conduction by strongly
localized, ``polaronic'' charge carriers, when the electron-lattice
interaction is sufficiently strong.~\cite{Emin_rev, Emin_revII} These
ideas are graphically summarized in Fig.~\ref{DOS}.

Because of the electron-lattice coupling, on the one hand, and the
constant thermal motion of atoms, on the other hand, the optical edge
in semiconductors is not sharp whether the solid is periodic or not:
This is because optical excitations are much faster than nuclear
motions implying that, effectively, electrons are always subject to an
aperiodic Born-Oppenheimer potential. The effective band edge turns
out to be nearly exponential and is often called the Urbach
tail.~\cite{PhysRev.92.1324, Toyozawa1961, PhysRevB.5.594,
  PhysRev.145.602, Kostadinov, 0022-3719-13-12-005,
  0022-3719-11-8-006, PhysRevLett.57.1777, PhysRev.148.741,
  PhysRevLett.25.520} Now in amorphous materials, there is no
underlying long-range order even with regard to vibrationally-averaged
atomic positions. Thus the disorder in the atomic locations is
partially frozen-in. Because such frozen glasses can be very far away
from equilibrium, the distribution of the energies of the localized
states is generally decoupled from the ambient temperature (and
pressure) and, furthermore, depends on the preparation protocol of the
sample.~\cite{LW_aging, LWjamming} Appropriately, exponential tails of
localized states generically emerge in models with quenched disorder,
within the venerable Anderson-Mott framework of electron localization
in disordered media.~\cite{AndersonLoc, Mott1982, CFO, Mott1990} Those
approaches assume generic forms for the random one-electron potential.
Non-withstanding the seemingly general character of the resulting
predictions, such generic approaches are not fully constructive in
that the random potential they postulate may not be consistent with
the actual molecular field in a stable structure. In a constructive
treatment, such a potential must arise self-consistently.

An additional, seemingly separate set of electronic excitations have
been observed in glassy chalcogenides. These excitations are
apparently activated by exposing the sample to macroscopic quantities
of photons at supra-gap frequencies. The excitations reveal themselves
as optical absorption at midgap frequencies and concomitant emergence
of ESR signal,~\cite{BiegelsenStreet, PhysRevB.38.11048,
  ShimakawaElliott, PhysRevB.15.2278, Mollot1980} and anomalous
photo-luminescence.~\cite{TadaNinomiya, TadaNinomiya2, TadaNinomiya3}
The concentration of these seemingly intrinsic defect-like states is
estimated at $\sim 10^{20}$~cm$^{-3}$,~\cite{BiegelsenStreet} i.e. one
per several hundred atoms. This is much greater, for instance, than
the typical amount of dopants in crystalline semiconductors, and leads
to a very efficient pinning of the Fermi energy. At the same time, the
glassy chalcogenides are {\em not} amenable to conventional
doping.~\cite{Kolomiets1981, Mott1993}

According to an early microscopic proposal, due to
Anderson~\cite{PWA_negU, PWA_negU2}, such behavior would be observed,
if electrons exhibited effective mutual attraction when occupying a
localized orbital so that per electron, the energy of a filled orbital
is lower than that of a singly-occupied orbital. (The subset of such
special orbitals would be relatively small.) The effective attraction
could stem, for instance, from lattice polarization and would amount
to an effective negative Hubbard $U$.  Shortly thereafter, Street and
Mott~\cite{PhysRevLett.35.1293} argued that given a certain, large
number of dangling bonds, nearby {\em pairs} of such bonds will be
unstable toward the formation of intimate pairs of malcoordinated
configurations, one negatively and one positively charged. (The
dangling bonds themselves are electrically neutral.) Kastner et
al.,~\cite{KAF} put forth specific microscopic proposals as to the
specific atomic motifs that could host such valence alternation pairs
(VAP). In addition, Vanderbilt and
Joannopoulos~\cite{PhysRevB.23.2596} proposed candidate malcoordinated
configurations that do not have to come in pairs but are
standalone. In these approaches, the defect states are viewed as
essentially defects in an otherwise perfect crystalline lattice. In a
more recent effort, Li and Drabold~\cite{PhysRevLett.85.2785} have
produced candidate defected configurations by modeling light-induced
formation of relatively malcoordinated motifs in computer generated
aperiodic samples.

In a distinct approach, Zhugayevych and Lubchenko~\cite{ZL_JCP,
  ZLMicro2} (ZL) have argued that chalcogenide glasses must host
special midgap electronic states that would be intrinsic to any glass
exhibiting spatially-inhomogeneous bond saturation. (In the case of
the chalcogenides, the bond strength varies between that of a
canonical single bond and a formally closed-shell,\cite{Pyykko}
secondary bond.)  These midgap states are tied to relatively strained
regions that are intimately related to transition-state configurations
for activated transport in an equilibrated glassy liquid, as well as
for aging in a frozen glass.~\cite{LW_aging, LWjamming} The strained
configurations can be thought of as domain walls separating distinct
aperiodic minima of the free energy; they must be present in
thermodynamic quantities. The equilibrium concentration of the domain
walls just above the glass transition has been estimated at $\sim
10^{20}$~cm$^{-3}$, using the random first order transition (RFOT)
theory;~\cite{XW, LW, LW_ARPC, L_AP} this figure matches well the
apparent quantity of the light-activated midgap states. Since the
structure of the liquid becomes largely arrested below the glass
transition---apart from some aging~\cite{LW_aging} and a minor
decrease in the vibrational amplitude---the concentration of the
strained regions remains steady upon cooling the glass.  When
expressed in terms of the size of a rigid molecular unit, this
concentration is nearly universal for two reasons: On the one hand, it
depends only logarithmically on the time scale of the glass
transition. On the other hand, the dependence of the concentration of
the midgap states on the material constants is only through the so
called Lindemann ratio. This ratio is defined as the relative
vibrational displacement near the onset of activated transport and
happens to be a nearly universal quantity.~\cite{L_Lindemann, RL_Tcr}
Thus doping the material may shift the glass transition temperature,
but will not significantly affect the concentration of the intrinsic
midgap states.  The robustness of the ZL midgap states has a
topological aspect, too: They can be thought of as stemming from an
extra or missing bond on an atom while the malcoordination cannot be
removed by elastically deforming the lattice. At most, lattice
relaxation results in the malcoordination being ``smeared'' over a
substantial region. This smearing is accompanied by further
delocalization of an already surprisingly extended wave-function of
the midgap electronic state.  The overall extent of the wavefunction
can be in excess of a dozen lattice spacings, which is much greater
than what one would expect for a very deep impurity state.

Analogous in many ways to the solitonic midgap states in
trans-polyacetylene,~\cite{RevModPhys.60.781} the midgap states can be
thought of as composed of an equal measure of the states from the
valence and conduction band. As a result, their energy is efficiently
pinned near the center of the mobility gap, if the states are
singly-occupied and, thus, neutral. Such neutral states absorb light
at sub-gap frequencies. However in a pristine sample, the midgap
states are occupied or vacant---corresponding with being negatively
and positively charged respectively---and thus stabilized owing to
lattice polarization.~\cite{ZL_JCP, ZLMicro2} Because of the lattice
distortion, charged midgap states absorb at supra-gap
frequencies.~\cite{PWA_negU, PWA_negU2, ZLMicro2} These notions
underlie the light-induced emergence of ESR signal and midgap
absorption:~\cite{ZLMicro2} Supra-gap irradiation excites electrons
from filled midgap states into the conduction band; in addition filled
midgap states can capture the oppositely charged free carriers
produced by the irradiation. As a result, the midgap states become
electrically neutral and begin to absorb light at sub-gap
frequencies. Thus in the ZL scenario, the specific atomic motifs
giving rise to the midgap states are not light generated defects;
instead they are intrinsically present in the structure. Light only
serves to make the defects ESR- and optically-active---at sub-gap
frequencies---by making them half-filled.

Some of these aspects of the topological midgap states are reminiscent
of the negative-$U$ model of Anderson~\cite{PWA_negU, PWA_negU2} and
subsequent ad hoc defect-based models mentioned above.\cite{KAF,
  PhysRevLett.35.1293, PhysRevB.22.2927} In fact, some of those defect
states can be viewed as an ultra-local limit of the ZL
theory.~\cite{ZLMicro2}
In those earlier developments, however, the concentration of the
defects is tied to the number of specific molecular motifs whose
quantity would seem to depend on the stoichiometry.  Nor is it clear
whether such defected configurations could combine to form a stable
lattice. In the ZL treatment, the concentration is predicted to be
{\em inherently} at roughly one defect per several hundred atoms,
irrespective of the precise stoichiometry.  The defect concentration
is determined by an interplay between the enthalpic cost of forming
the domain walls, on the one hand, and their entropic advantage, on
the other hand. This entropic advantage can be ultimately be traced to
the excess liquid entropy of the supercooled liquid relative to the
corresponding crystal.


ZL~\cite{ZLMicro2} have also proposed specific malcoordinated motifs
underlying the midgap states.  The motifs consist of relatively
extended $pp\sigma$-bonded chains connecting an odd number of
orbitals, at half filling.  The presence of such motifs is expected on
a systematic basis, insofar as one may regard the chalcogenides as
distorted versions of relatively symmetric, parent structures defined
on the simple cubic lattice.~\cite{ZLMicro1} Parent structures defined
on the simple cubic lattice have been indeed obtained for all known
lattice types found in stoichiometric crystalline compounds of the
type Pn$_{2}$Ch$_{3}$ (Pn $=$ P, As, Sb, Bi, Ch $=$ S, Se,
Te),~\cite{GoldenThesis} and are themselves periodic of course. ZL
anticipated that {\em aperiodic} parent structures, upon deformation,
would give rise to bona fide glassy structures. Thus the aperiodicity
and the sporadic malcoordination of the lattice would arise
self-consistently. ZL~\cite{ZLMicro2} have generated standalone,
molecular malcoordinated motifs, which were properly passivated to
emulate proper coordination appropriate for a 3D solid.  It was argued
that the interaction of defect-bearing chains with the surrounding
solid amounts to a renormalization of the on-site energies and
electron hopping elements along the chain but would not change the
physics qualitatively---a notion that will be revised in this
article. No attempt was made to generate a {\em bulk}
three-dimensional structure that would host such motifs while obeying
local chemistry throughout.

Here we report what we believe is the first realization of the
topological midgap states in bulk samples. The samples were generated
using a structure-building algorithm reported earlier by
us.~\cite{LL1} In this algorithm, one first generates so called {\em
  parent} aperiodic structures that exhibit octahedral coordination,
locally.  A subset of the lattice sites in the parent structure form a
{\em random}-close packed structure,~\cite{RevModPhys.82.789} thus
ensuring that within a certain range of wavelengths, the structures
are at what is believed to be the highest achievable density for
aperiodic solids.  These parent structures are subsequently optimized
using quantum-chemical force fields while coordination becomes
distorted-octahedral.~\cite{ZLMicro1} Thus the sample is {\em not}
generated by quenching a melt that had been equilibrated at some high
temperature. In principle, such quenched samples should approximate
real materials subject only to the accuracy of the effective
inter-atomic force fields and, possibly, finite-size effects. In
practice, however, the dynamical range of molecular dynamics
simulations is very limited even for relatively simple model liquids
let alone the chalcogenides, where the effective force fields between
the atoms must be determined using computationally expensive,
quantum-chemical approximations. As a result, one can equilibrate a
chalcogenide melt only at temperatures much exceeding the laboratory
glass transition. The resulting samples are thus hyper-quenched while
their structure exhibits exaggerated effects of mixing entropy in the
form of excess homo-nuclear contacts even in the stoichiometric
compound As$_{40}$Se$_{60}$, in contrast with
observation.~\cite{doi:10.1021/acs.jpcc.5b02423}

The parent structures generated in Ref.~\cite{LL1} do not contain
homonuclear contacts by construction, nor do such defects seem to
appear in significant quantities following the geometric optimization,
at least in the stoichiometric compound As$_{40}$Se$_{60}$. We have
argued~\cite{LL1} this circumstance explains why the resulting
amorphous samples consistently exhibit the first sharp diffraction
peak (FSDP) in the structure factor, including its trends with
pressure and arsenic content. (The FSDP is the hallmark of the poorly
understood medium-range-order~\cite{ElliottNature1991, Salmon_ZnCl2}
in inorganic glasses.) In contrast, samples generated using first
principles molecular dynamics often fail to exhibit the
FSDP.~\cite{fpmdMic}

Likewise, the electronic density of states (DOS) for the presently
generated samples exhibit properties expected of the amorphous
chalcogenides.  On the one hand, the gross part of the electronic
spectrum does not vary significantly from sample to sample; this part
of the DOS thus can be attributed to the mobility bands. On the other
hand, the states near the band edges are found to fluctuate in energy
substantially, as is expected for the localized Urbach-tail states. In
fact, the magnitude of the fluctuation matches well the width of the
Urbach tail observed in experiment.  Perhaps more interestingly,
samples generated according to the procedure from Ref.~\cite{LL1} do
indeed exhibit very deep-lying midgap states that are close to the gap
center, closer than could possibly take place for an Urbach-tail
state.

Both the presence of the midgap states in the computed spectra and the
value of the gap itself are found to depend rather sensitively on the
detailed quantum-chemical approximation, consistent with earlier
studies,~\cite{bandGap-problem1,bandGap-problem2} requiring the
relatively higher-end, computationally expensive hybrid DFT to solve
for the electronic spectrum. (Plain DFT seems to suffice for structure
optimization.)  Consistent with the general predictions by
ZL,~\cite{ZL_JCP} the wavefunctions of a subset of the midgap states
are elongated preferentially in one direction, in contrast with the
extended states comprising the mobility bands, which are isotropic.
In addition, here we argue that a somewhat distinct physical
possibility can be realized in which the wave function consists of a
few linear fragments emanating from the same spot in space; this
possibility was overlooked by ZL. Some of the samples we have
generated appear to exhibit this urchin-like shape.  At least in one
case, we were able to clearly identify a motif that can be thought of
as a passivated odd-numbered chain that must necessarily host a
topological midgap state. Such states correspond to ZL chains with
closed ends. Now, the aforementioned deep midgap states were unforced
in the sense that the structures contained an even number of electrons
and were geometrically optimized so that there are no intentionally
broken bonds in the sample. By way of contrast, we have also generated
samples containing an odd number of electrons, thus {\em forcing} the
system to have a dangling bond. Here we find the resulting ``defect''
is often very deep in the forbidden gap. This is in contrast with what
would be expected for a generic impurity state. In some cases, the
defect states are so deep inside the gap that they can be thought of
as electrically-neutral entities, where the electron charge is largely
compensated by lattice polarization. Yet in some cases, the
configurations are more reminiscent of the polarons in conjugated
polymers.\cite{RevModPhys.60.781, ISI:A1981MD41000002} Finally, we
have established that the Urbach states turn out to be
intermediate---both in terms of their shape and degree of
localization---between the extended band states and the topological
states thus suggesting the Urbach states and the topological states
are intimately related.

The article is organized as follows: Section~\ref{extended} discusses
the salient features of the presently obtained spectra, the forbidden
gap and the Urbach tail states.  In Section~\ref{midgap1} we first
briefly survey pertinent general properties of the topological midgap
states for isolated chains and then use model calculations to
demonstrate that such midgap states are robust even if the defected
chain are coupled to other chains. In Section~\ref{midgap2}, we
quantitatively characterize the very deep midgap states obtained in the
present study and argue that they are, in fact, the topological midgap
states predicted earlier by ZL.  Section~\ref{summary} provides a
brief summary.

\section{Salient features of the density of states: Mobility bands and
  the Urbach tail of localized states}
\label{extended}

\begin{figure}[t]
  \centering
  \includegraphics[width=1\figurewidth]{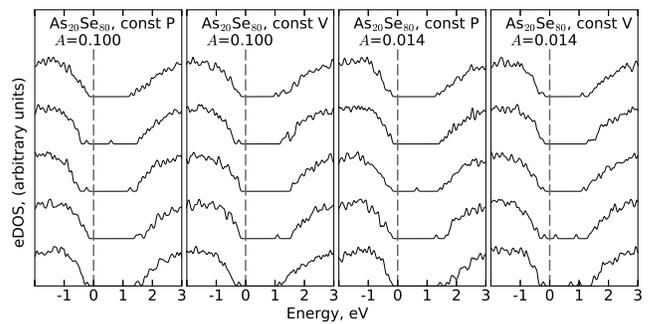}
  \caption{\label{band-gaps-As20Se80} Electronic density of states for
    five distinct amorphous samples of the compound As$_{20}$Se$_{80}$
    and two distinct values of the parameter $A$, optimized at
    constant pressure and volume, respectively. The vertical dashed
    line denotes the Fermi energy.  }
\end{figure}

We have generated disordered structures for the binary compound
arsenic selenide As$_{x}$Se$_{100-x}$ for three distinct
stoichiometries $x=20, 40$, and $50$, with the help of the algorithm
described in Ref.~\onlinecite{LL1}. In this algorithm, we first
generate a chemically-motivated aperiodic parent structure, in which a
subset of atoms are placed at the vertices of a random close-packed
(RCP) lattice, called the ``primary lattice.'' A complementary subset
of atoms is then placed at the vertices of the so called ``secondary
lattice,'' which is generated according to a detailed algorithm to
maximize octahedrality in local bonding while maintaining desired
stoichiometry. This detailed algorithm uses a threshold parameter $A$,
which prescribes the precise way in which we break up the interstitial
space of the primary lattice into cavities. For instance, for very
small values of $A$, the cavities will be all tetrahedral, the
tetrahedra determined using the Delaunay triangulation. For
sufficiently large values of $A$, on the other hand, some of these
tetrahedra are merged into higher order polyhedra.  These higher order
polyhedra host the vertices of the secondary lattice.  To minimize the
number of homonuclear bonds, the primary and secondary lattice are
populated by distinct species. Here we focus exclusively on the so
called C-type structures, for which the primary lattice is made of
$C$halcogens (selenium in this case).

By varying the value of the parameter $A$, one can effectively control
the amount of vacancies in the parent structures. This amount is equal
to the number of non-tetrahedral cavities in excess of the number of
pnictogens (for C-type structures).  For instance, $A=0.1$ is
convenient for the stoichiometric compound As$_{40}$Se$_{60}$ because
the number of vacancies that must be made is small yet not too small
so that one can generate sufficiently distinct structures by randomly
choosing the locations for the vacancies. In addition to $A=0.1$, we
have also generated samples with $A=0.014$ and $0.24$ for the
As$_{20}$Se$_{80}$ and As$_{50}$Se$_{50}$ compounds, respectively.
Five samples were generated for each value of $A$, for each compound,
except for As$_{40}$Se$_{60}$, where we have generated eleven.
Finally, the parent structures are geometrically optimized using
plane-wave DFT as implemented in the package
VASP~\cite{vasp1,vasp2,vasp3,vasp4}
in the Perdew-Wang~\cite{vasp-gga-pw91-1,vasp-gga-pw91-2} (PW91)
generalized gradient approximation (GGA) for the exchange-correlation
energy.  Optimization was performed both at constant pressure and
volume; the corresponding structures are labelled const-$P$ and
const-$V$, respectively. We note that optimization at constant
pressure, which thus allows for unit cell optimization, produces much
better results for the first sharp diffraction peak.  Throughout the
paper, the samples As$_{20}$Se$_{80}$, As$_{40}$Se$_{60}$, and
As$_{50}$Se$_{50}$ contain a total of 250, 330, and 304 atoms,
respectively.

While producing good structures---this we have tested for crystalline
As$_2$Se$_3$~\cite{LL1}---GGA-based methods are known to substantially
underestimate the value of the band gap in semiconductors and
insulators.~\cite{bandGap-problem1,bandGap-problem2} Thus to evaluate
the electronic density of states, we use here a more accurate, but
computationally costly hybrid functional B3LYP.  The latter
approximation does much better in reproducing experimental figures for
the band gap than plain DFT and some hybrid functionals; see the
Supplemental Information for more detail. In the rest of the article,
we report spectra that were obtained using the hybrid functional
B3LYP.  We set the one-electron energy reference at the Fermi energy
as reported by VASP.

\begin{figure}[t]
  \centering
  \includegraphics[width=1\figurewidth]{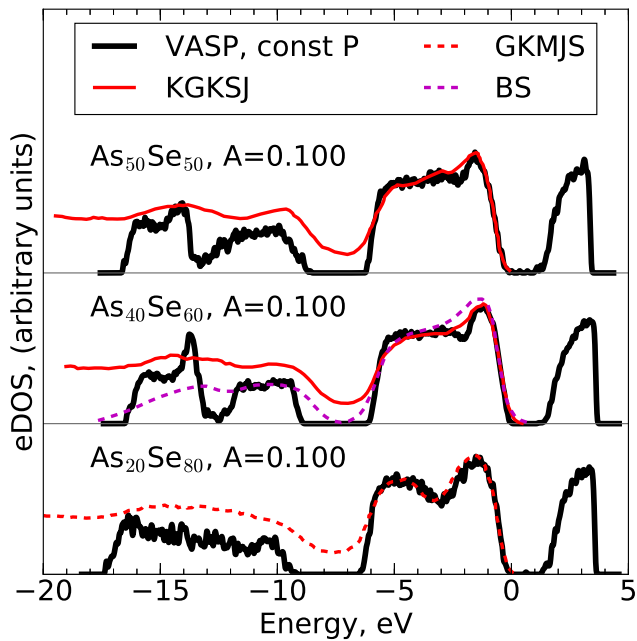}
  \caption{\label{edos-vasp} Total electronic density of states (eDOS)
    of As$_{x}$Se$_{100-x}$ glasses for compositions $x=\left\{0.2,
      0.4, 0.5\right\}$ (const-$P$) alongside experimental eDOS
    inferred from X-ray photo-emission spectroscopy studies due to
    Kozyukhin {\em at al.}  (KGKSJ)~\cite{kozyukhin-2011-edos},
    Golovchak {\em et al.}  (GKMJS)~\cite{golovchak-2007-edos}, and
    Bishop and Shevchik (BS)\cite{bishop-1975-edos}.  The presently
    predicted eDOS is averaged over 5 samples for As$_{20}$Se$_{80}$,
    10 for As$_{40}$Se$_{60}$, and 5 for As$_{50}$Se$_{50}$.  }
\end{figure}

The electronic spectra, as illustrated for several As$_{20}$Se$_{80}$
samples in Fig.~\ref{band-gaps-As20Se80}, are overall similar yet
clearly vary in detail from sample to sample, and especially so around
the forbidden gap. The corresponding figures for the compounds
As$_{40}$Se$_{60}$ and As$_{50}$Se$_{50}$ can be found in the
Supplemental Material. There, we also provide an animation that allows
one to efficiently survey how the shape and extent of the wave
function is correlated with the corresponding energy eigenvalue.  The
aforementioned sample-to-sample variation is indeed expected for
disordered samples. The variation is expected to be relatively small
for electronic states sufficiently away from the
gap,~\cite{RevModPhys.50.191} where the density of states is truly
continuous and self-averaging.  It is this averaged spectrum that is
pertinent to experimentally observed absorption spectra because
absorption experiments represent a bulk measurement.  The electronic
density of states averaged over several samples are shown in
Fig.~\ref{edos-vasp}, along with experimental data, for all three
stoichiometries. Clearly, the present results reproduce the gross
features found in actual spectra, particularly in the higher energy
part of the valence band.  The agreement is especially notable for the
20-80 mixture.

\begin{figure}[t]
  \centering
  \includegraphics[width=1\figurewidth]{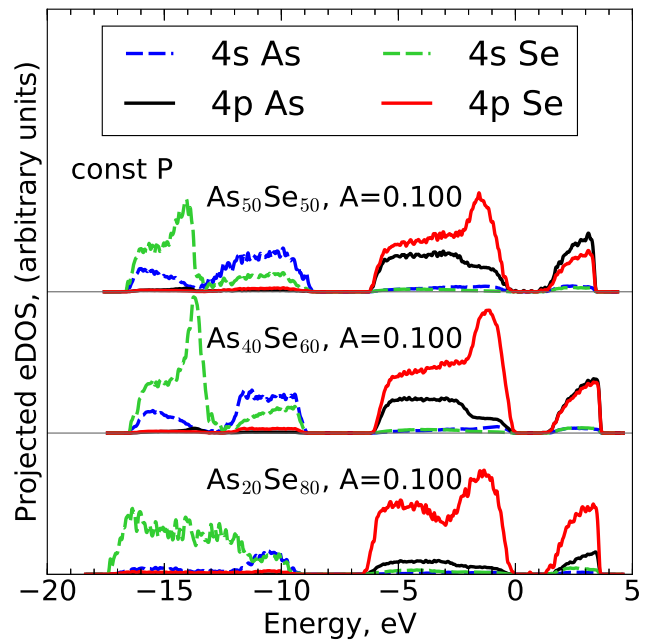}
  \caption{\label{pdos-vasp} Partial contributions of the atomic $s$
    and $p$ orbitals to the overall density of states. The setup is the
    same as in Fig.~\ref{edos-vasp}.  }
\end{figure}

To infer the atomic-orbital makeup of the electronic density of
states, we consider the $s$ and $p$ atomic-orbital contributions to
the total density of states as determined using VASP's built-in
projection technique, see Fig.~\ref{pdos-vasp}. According to the
latter Figure, the higher energy portion of the valence band is
largely due to the $p$-orbitals. We note that the present spectra for
As$_{40}$Se$_{60}$ and As$_{50}$Se$_{50}$ seem to overestimate the
``dip'' near -2~eV, which separates the bonding $p$-orbitals from the
the selenium-based lone pairs. According to Fig.~\ref{pdos-vasp}, this
may stem from a (modest) underestimation of the density of states
corresponding to the As-based $p$-orbitals. The overall good agreement
between theory and experiment in this part of the spectrum is
reassuring since the bonding is indeed expected to be largely due to
the $pp\sigma$-network.~\cite{ZLMicro1} In contrast with $p$-orbitals,
the $s$-orbitals are filled and thus affect the bonding and geometry
less directly, via $sp$-mixing.~\cite{SeoHoffmann1999} In the
Supplemental Material, we compare the present results with the
electronic spectra obtained in two earlier studies, due to Bauchy et
al.~\cite{fpmdMic} and Li et al.~\cite{Drabold-2002-edos-theory} All
methodologies seem to reproduce well the $p$-portion of the spectrum,
Li et al.'s results~\cite{Drabold-2002-edos-theory} for
As$_{50}$Se$_{50}$ standing out. There is significantly less agreement
with experiment in the $s$-orbital portion.

In contrast with the states comprising the mobility bands, states near
the gap are expected to be relatively localized; their quantity and
energy should strongly depend on the specific realization. Only upon
averaging over many realizations, these localized states should yield
a relatively smooth spectrum, which is expected to be
exponential.~\cite{PhysRev.92.1324, Toyozawa1961, PhysRevB.5.594,
  PhysRev.145.602, Kostadinov, 0022-3719-13-12-005,
  0022-3719-11-8-006, PhysRevLett.57.1777, PhysRev.148.741,
  PhysRevLett.25.520} This is indeed born out by the present data.  In
view of limited statistics and modest sample sizes, however, it is
difficult to determine, based on the spectra alone, which states
comprising the individual spectra in Fig.~\ref{band-gaps-As20Se80}
should be assigned to the mobility band and which to the Urbach tails
of localized states or, potentially, to the topological midgap states
predicted by ZL.~\cite{ZL_JCP, ZLMicro2} This complication makes
determination of the width of the forbidden gaps in
Fig.~\ref{band-gaps-As20Se80} ambiguous. In experiment, one
conventionally places the edges of the mobility bands at energies
corresponding to the onset of the Urbach tail, the latter located by
fitting, see Figs. 1 and 2 of
Ref.~\onlinecite{Slusher-gapAs40Se60}. The functional form used to fit
the Urbach-tail states is:
\begin{equation} \label{nEUrbach} n(\epsilon) \propto
  e^{-\epsilon/E_U},
\end{equation}
where $\epsilon$ stands for the depth of a localized state relative to
the edge of the corresponding mobility band, into the gap.

\begin{table}[t]
  \begin{tabular}{|c|c|c|c|m{3cm}|c|}
    \cline{2-6}
    \multicolumn{1}{c|}{}              &  const  & $E_g^{I}$ & $E_g^{II}$ & \parbox{3cm}{\centering $E_g^\text{expt}$} & $\sigma_{U}$\\ \hline
    \multirow{2}{*}{As$_{20}$Se$_{80}$}  &  {\it P}    &  2.00      & 1.91      & \multirow{2}{*}{\parbox{3cm}{
        1.84\cite{Dahshan-gapAs20Se80},
        1.84\cite{Fang-gapAs20-40-50}}} & 0.1\\ \cline{2-4}\cline{6-6}
    &  {\it V}    &  1.71      & 1.74      &                                 & 0.1\\ \hline
    \multirow{2}{*}{As$_{40}$Se$_{60}$}  &  {\it P}    &  1.94      & 1.82      & \multirow{2}{*}{\parbox{3cm}{
        1.78\cite{Slusher-gapAs40Se60},
        1.76\cite{Behera-gapAs40Se60},
        1.74\cite{FELTY1967555},
        1.64\cite{Fang-gapAs20-40-50}}} & 0.08\\ \cline{2-4}\cline{6-6}
    &  {\it V}   &  1.74      & 1.78      &                                 & 0.07\\ \hline
    \multirow{2}{*}{As$_{50}$Se$_{50}$}  &  {\it P}   &  1.94      & 1.60      & \multirow{2}{*}{\parbox{3cm}{
        1.74\cite{Fang-gapAs20-40-50},
        1.72\cite{Behera-gapAs50Se50},
        [1.64, 1.69, 1.85, 1.76
        ]\cite{Nemec-gapAs50-60}}} & 0.27\\ \cline{2-4}\cline{6-6}
    &  {\it V}   &  1.81      & 1.70      &                              & 0.17\\ \hline
    \end{tabular}
    \caption{\label{table-of-gaps} Band gap values $E_g$ estimated using 
      two different approaches  for samples optimized at constant 
      pressure (const-$P$) and volume (const-$V$) at three 
      distinct stoichiometries. The corresponding experimental values are
      also provided. The quantity $\sigma_U$ yields the steady state
      value of the standard deviation for the level position and is used
      to estimate the width $E_U$ of the Urbach tail, see text.}
\end{table}

\begin{figure}[t!]
  \centering
  \includegraphics[width=1\figurewidth]{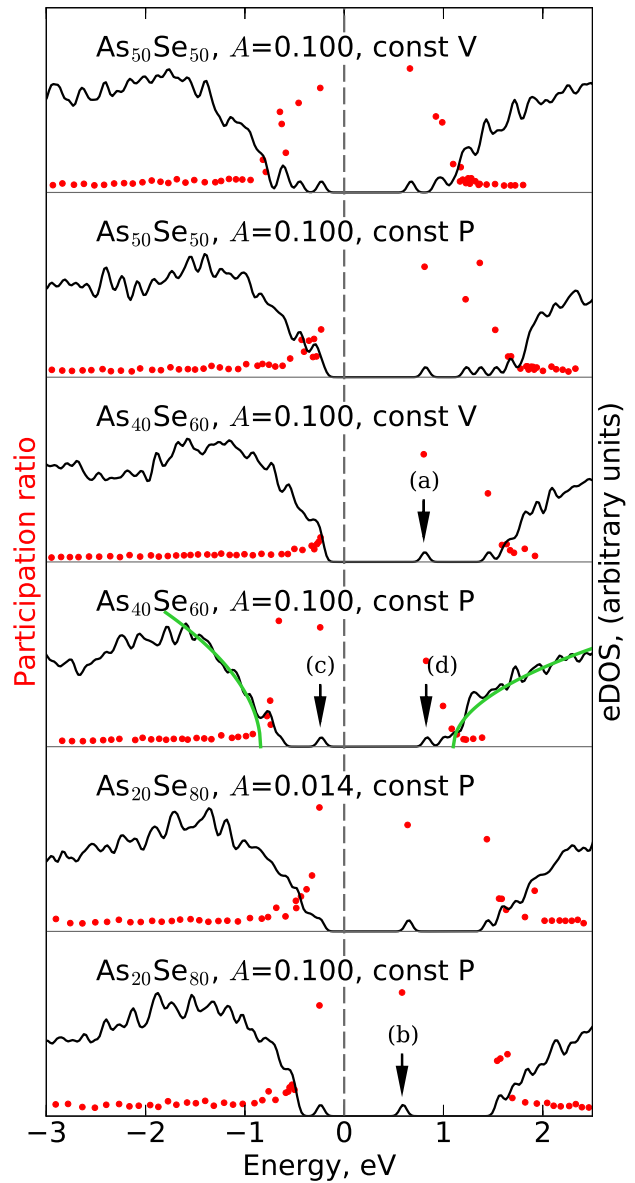}
  \caption{\label{participation-ratio} Solid black line: electronic
    DOS for specific realizations of samples for three distinct
    stoichiometries. Red dots: Inverse participation ratios for select
    orbitals, the sampling is nearly complete near band edges, see
    Section~\ref{midgap2}. Green solid line: The result of a fit using
    the idealized forms $(E_v - E)^{1/2}$ and $(E - E_c)^{1/2}$, as
    pertinent to the valence and conduction band, respectively. The
    midgap states labeled (a)-(d) refer to the respective panels in
    Fig.~\ref{defects}, where the corresponding wavefunctions-squared
    are visualized.  The vertical dashed line corresponds with the
    Fermi energy.  The stoichiometry for each sample is provided in
    its respective panel.}
\end{figure}

According to Ref.~\onlinecite{Slusher-gapAs40Se60}, the onset of the
exponential tail of the localized states corresponds, by convention,
to roughly 1\% of the absorption strength characteristic of
transitions between bona fide extended states.  The full range of
absorption strength available in experiment spans a range of at least
five orders of magnitude thus permitting one to identify the edge
states relatively unambiguously. In contrast, the present DOS barely
spans two orders of magnitude.  To work around this complication, we
employ two distinct methods to assess the gap.  In method one, we fit
the DOS, near but not exactly at the gap edge, using the functional
form $(E_v - E)^{1/2}$ and $(E - E_c)^{1/2}$ for the valence and
conduction band, respectively.~\cite{Tauc1974} The square root scaling
would be exact for a translationally invariant system in 3D, of
course.~\cite{Kittel} An example of the resulting fit is shown in one
of the panels in Fig.~\ref{participation-ratio}. (As a practical
matter, we obtain the fits by plotting the square of the DOS.)  The
resulting values of the gap, $E_g = E_c - E_v$, are given as $E_g^{I}$
in Table~\ref{table-of-gaps}. We note the apparent deviation of the
computed DOS from the simple square-root scaling is consistent with
the expectation that the presently generated disordered samples should
host localized states.

\begin{figure}[t]
  \centering
  \includegraphics[width=0.8 \figurewidth]{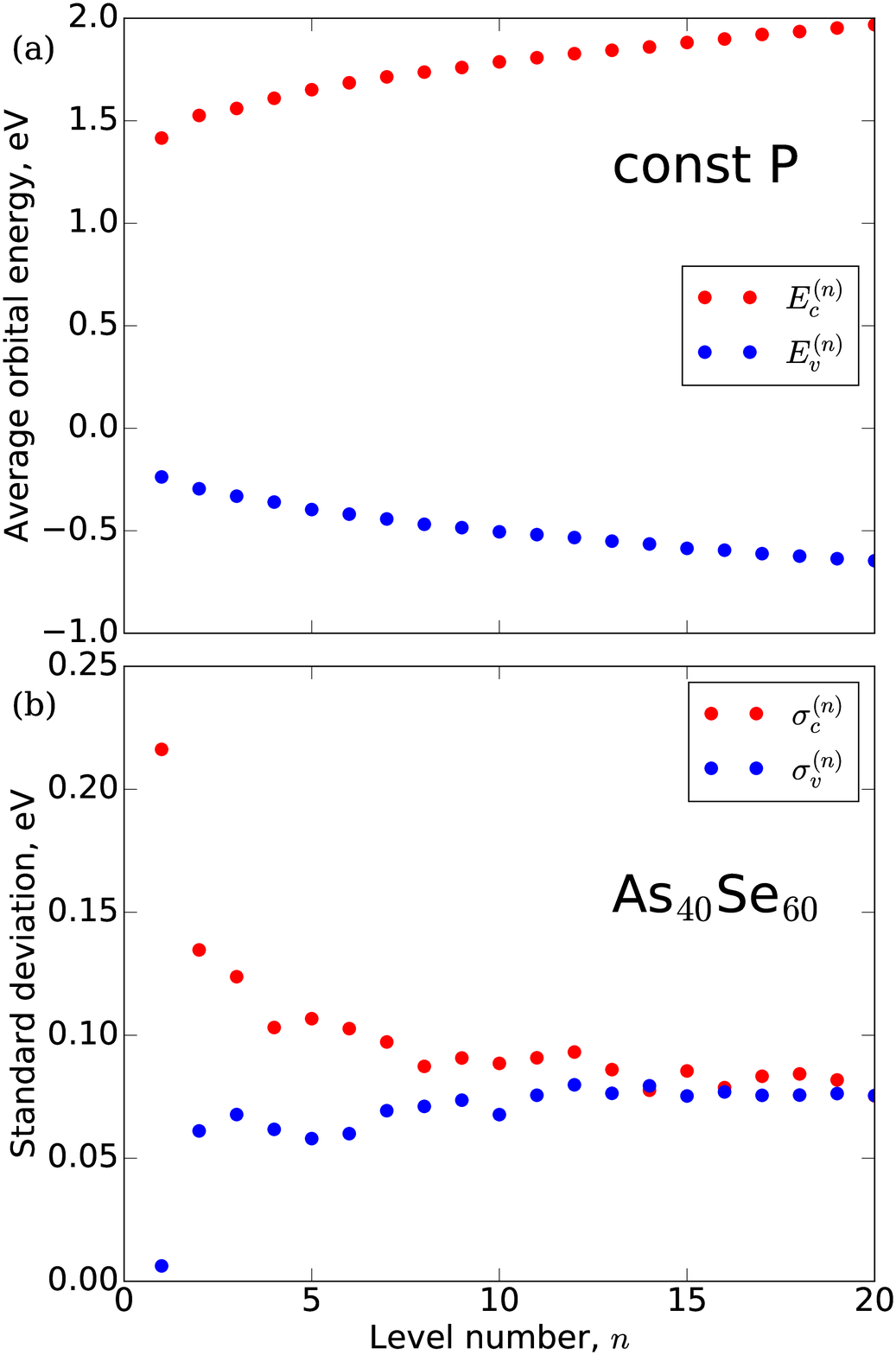}
  \caption{\label{EnSigma} (a) The average energies $E_v^{(n)}$ and
    $E_c^{(n)}$ of the $n$-th occupied and empty level,
    respectively. $E_v^{(1)}$ ($E_c^{(1)}$) corresponds with the
    energy of the HOMO (LUMO). (b) The corresponding standard
    deviation. As$_{40}$Se$_{60}$, const-$P$.}
\end{figure}

In an alternative strategy, we do not assume any specific functional
form for the spectrum either within mobility bands or for the
localized states. Instead, we first order the {\em occupied} states
according to their distance, energy-wise, from the HOMO for each
individual sample. Likewise we order, for each sample, the {\em
  vacant} levels according to their separation from the LUMO. Next, we
determine the average energy $E_v^{(n)}$ of the $n$-th occupied state
and the average energy $E_c^{(n)}$ of the $n$-th empty state.  Call
the corresponding standard deviations $\sigma^{(n)}_v$ and
$\sigma^{(n)}_c$, respectively.  We graph these average energies and
the corresponding standard deviations in Fig.~\ref{EnSigma} for levels
1 through 20.  Data for the other stoichiometries, both at constant
pressure and volume, can be found in the Supplemental
Material. Although the detailed trends of $\sigma_U$ vary somewhat
depending on the stoichiometry and the optimization
conditions---i.e. const-$P$ vs. const-$V$---the following two features
are securely reproduced in all cases: (a) The energies $E_v^{(n)}$
($E_c^{(n)}$) decrease (increase) with $n$. (b) The standard
deviations $\sigma^{(n)}_v$ and $\sigma^{(n)}_c$ saturate at large $n$
at some value $\sigma_U$.  We also note that $\sigma^{(1)}_v$ is
always relatively small, the likely reason being is that the Fermi
level is tied to the HOMO.

Already the energies of the next-to-HOMO and LUMO levels, $n=2$,
exhibit the amount of variation comparable to the steady value of
$\sigma_U$.  At the same time, even if it so happens that these $n=2$
levels belong to the mobility band, they would not be too deep into
the band.  In addition, there are good reasons to believe the HOMO and
LUMO themselves can be sufficiently often associated with the
topological midgap states, as we shall see in
Section~\ref{midgap2}. Based on these notions and for the sake of
concreteness, we settle on quantifying the band gap using the energy
difference between the $n=2$ states:
\begin{equation} E_g^{II} \equiv  E_c^{(2)} - E_v^{(2)},
\end{equation}
see Table~\ref{table-of-gaps}. Given the scatter in the reported
experimental values of the gap, the present estimates seem rather
satisfactory. Significantly, we observe that the samples optimized at
constant pressure reproduce the experimental trend that the gap
decreases modestly with arsenic content. (We note that some older data
suggest that the gap width, as a function of arsenic content,
experiences a shallow ``dip'' around
As$_{43}$Se$_{57}$.~\cite{BandGap-composition}) In contrast, the
const-$V$ samples do not follow this trend. Interestingly, samples
optimized at constant pressure also did much better~\cite{LL1} with
regard to the first sharp diffraction peak (FSDP), which is a {\em
  structural} feature. In contrast, const-$V$ samples often exhibited
a small shoulder instead of a well-defined peak. That the detailed
characteristics of the band edges would have a structural signature is
consistent with the apparent correlation between the strength of the
FSDP, pressure, and the phenomenon of photodarkening. (The term
photodarkening~\cite{Pfeiffer} refers to a light-induced narrowing of
the optical gap.)  This correlation was discussed in detail in
Ref.~\onlinecite{LL1}.

We next turn to the standard variation $\sigma_U$. First we note that
for the exponential distribution $p(x) = \sigma^{-1} e^{-x/\sigma}$,
the standard deviation is equal to $\sigma$. The number of samples we
have generated is too modest to conclusively make out the shape of the
distribution of the tail states, as already mentioned. Yet insofar the
sampling can be regarded as sufficient to infer the standard
deviation, we may associate the quantity $\sigma_U$ with the width
parameter $E_U$ of the Urbach tail from Eq.~(\ref{nEUrbach}).  The
thus estimated value for the width of the Urbach tail for the
stoichiometric compound As$_{40}$Se$_{60}$, viz. 0.07-0.08~eV, is
consistent with the experimentally reported value of
0.066~eV.~\cite{Harea-2003-urbach-energy} We were unable to find the
Urbach energy $E_U$ in the literature for the other two
stoichiometries.

We now turn our attention to the features deep in the mobility gap
that are sporadically found in the density of states for specific
realizations; these are exemplified in
Fig.~\ref{participation-ratio}. Note that such sporadic features
readily become part of the background upon averaging, as in
Fig.~\ref{edos-vasp}. While our statistics are clearly limited,
examination of the available spectra for individual realizations,
Fig.~\ref{band-gaps-As20Se80} and Figs.~\ref{band-gaps-As40Se60} and
\ref{band-gaps-As50Se50} in the Supplemental Material, suggest that
roughly only one in ten samples display such deep midgap states
already in the stoichiometric compound As$_{40}$Se$_{60}$, the number
of incidents seemingly increasing away from the exact 2:3
stoichiometry.  We immediately point out that on purely statistical
grounds, observing a midgap state like that in
Fig.~\ref{inertia-ratio} is not at all likely. Indeed, given that the
width of the Urbach tail is numerically close to $0.08$~eV, our
chances of observing that midgap state are roughly one in $e^{0.6
  \text{eV}/0.08 \text{eV}}$ in view of Eq.~(\ref{nEUrbach}). That is,
roughly one per thousand samples.

Note that the defect-like midgap states in
Fig.~\ref{band-gaps-As20Se80} and \ref{participation-ratio} are
unforced in that the samples contain an even number of electrons and
are fully geometrically optimized; no broken bonds are deliberately
put in the system.  Conversely we note that when occupied, such deep
states are very costly---at up to $(E_g/2)\times 2$ per state,
viz. ca. 2~eV, and thus would be presumably stabilized during
geometric optimization unless prevented from doing so for some special
reason. Zhugayevych and Lubchenko~\cite{ZL_JCP, ZLMicro2} (ZL) have
argued exactly such a reason exists in glassy chalcogenides.  The
reader is referred to those publications for detailed discussion. In
the following Section, we briefly reiterate some of those notions
and present new results that will be helpful in interpreting some of
the midgap-state features revealed in this study.

\section{The topological electronic midgap states: General results}
\label{midgap1}

\begin{figure}[t]
  \centering
  \includegraphics[width=.8 \figurewidth]{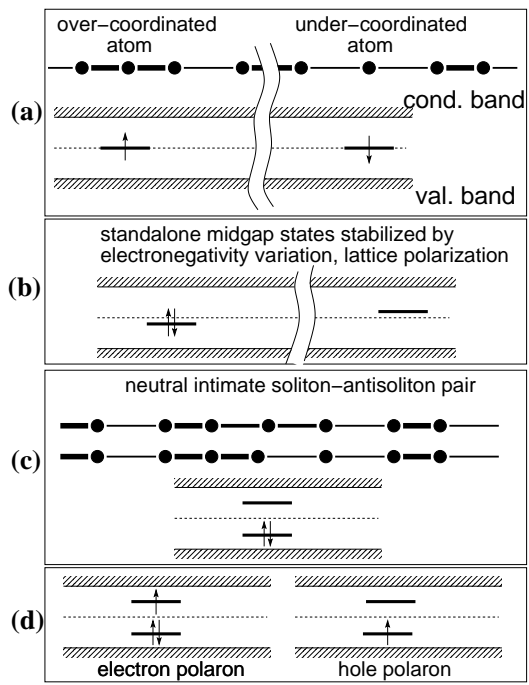}
  \caption{\label{intimate} Schematic of quasi-linear motifs---the
    surrounding solid matrix not shown---which host solitonic midgap
    states. Zigzag-like distortions of the motif, which would be
    present in actual materials, are not shown for clarity. The dots
    denote atoms, either chalcogens or pnictogens. The thickest line
    stands for a single covalent bond, the thinnest line for the
    secondary interaction, and the line intermediate in thickness for
    a bond intermediate in strength between those two interactions.
    (a) Malcoordination leads to the appearance of a midgap state. (b)
    In the presence of lattice polarization, such states will be
    stabilized by becoming filled with electrons (holes). (c) When
    close-by, but not recombined---two pertinent atomic configurations
    shown schematically---the midgap states form a resonance, the
    electrons occupying the lower term. Upon complete recombination,
    the lower and upper term merge with the valence and conduction
    band, respectively. (d) The configuration from (c) with an added
    electron or hole corresponds with a polaron. }
\end{figure}

ZL have argued~\cite{ZL_JCP} that at some places along the domain
walls separating regions occupied by distinct aperiodic free energy
minima in a glassy liquid, local coordination will differ from its
optimal value. The missing or extra bond should be perpendicular to
the domain wall.  Specifically in the chalcogenides,
quasi-one-dimensional chain-like motifs can be
identified~\cite{ZLMicro2} based on a structural
model,~\cite{ZLMicro1} in which these materials are regarded as
distorted, symmetry-broken version of more symmetric parent structures
locally defined on a simple cubic lattice.  In the simplest case, a
defect-free distorted chain exhibits an exact alternation pattern of a
covalent bond and a secondary interaction and can be thought of as a
chain of weakly interacting dimers.  At a malcoordination defect, one
atom or more will be have one too many or one too few bonds. As was
understood in the context of conjugated
polymers,\cite{RevModPhys.60.781, ISI:A1981MD41000002} such
malcoordination defects must host very special midgap electronic
states, see illustration in Fig.~\ref{intimate}. If singly occupied
and close to the middle of the gap, they formally correspond with a
neutral particle that has spin 1/2, see Fig.~\ref{intimate}(a).  (The
electron charge is exactly compensated by the polarization of the
lattice.) The deviation of the bond length from its value in a
perfectly dimerized chain shows a sigmodal dependence on the
coordinate. The strain has a solitonic profile, hence the midgap
states are often called ``solitonic.''

\begin{figure}[t]
  \centering
  \includegraphics[width =  \figurewidth]{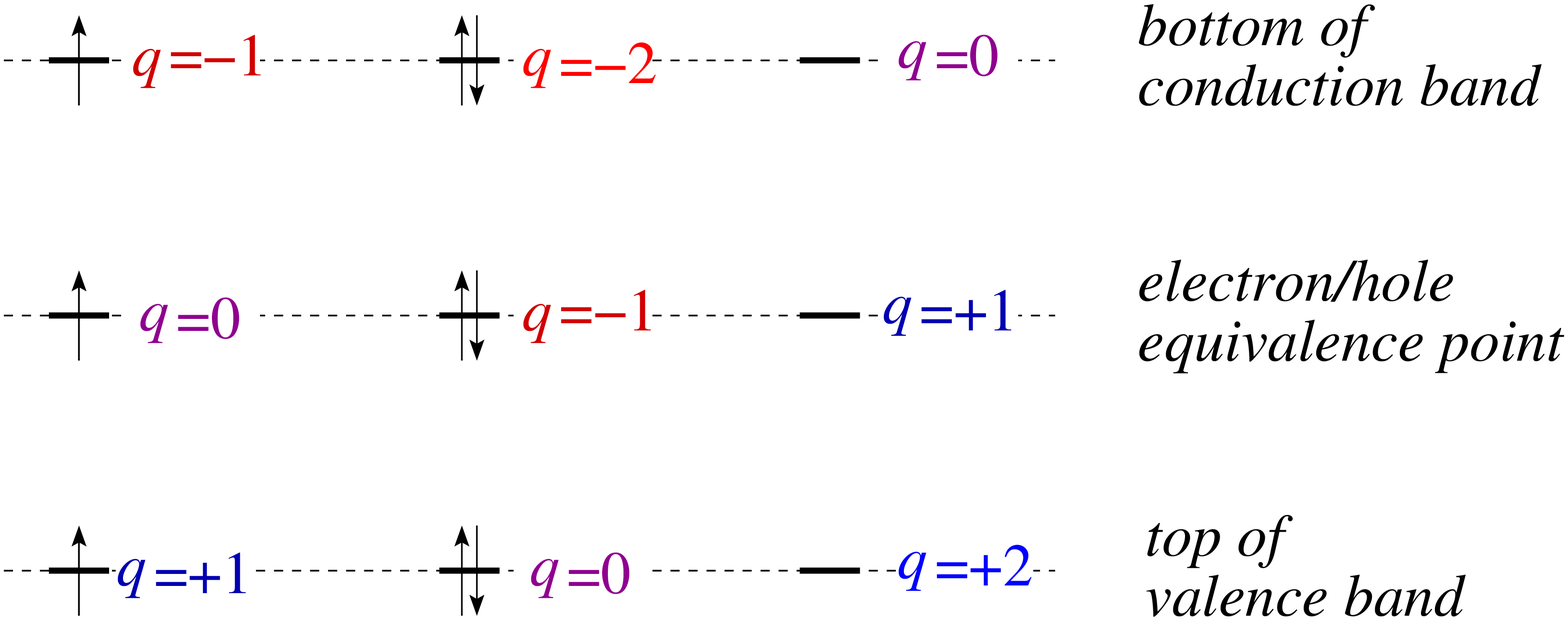}
  \caption{\label{ehduality} The charge of a state depending on its
    position in the forbidden gap and the filling.}
\end{figure}

In the chalcogenides, one expects that most of the midgap states would
be either fully occupied or empty: On the one hand, the chalcogenides
exhibit spatial variation in the electronegativity.  Suppose the
variation is on average equal to $\varepsilon$. Then a midgap state
based on the more electronegative site is typically lowered,
energy-wise, by the amount $\varepsilon/2$ relative to the middle of
the gap.  Likewise, defects centered on the less electronegative
element would be destabilized by the same amount.  In addition, one
expects that a singly occupied midgap state will be modestly
stabilized by binding an electron or hole~\cite{ZL_JCP}---consistent
with simple electron counting arguments.~\cite{ZLMicro2} These notions
are summarized in Fig.~\ref{intimate}(b). The apparent charge of the
defect will depend not only on the occupation of the midgap state but
also on its position in the band,~\cite{PhysRevLett.49.1455} see the
informal chart in Fig.~\ref{ehduality}.

A midgap state cannot be removed by elastic deformation but, instead,
only via annihilation with a defect of opposite malcoordination, which
resides on the same chain. For this reason, such oppositely
malcoordinated configurations are often called soliton and
anti-soliton. When the spatial separation between such oppositely
malcoordinated defects is small, the respective energy levels form a
resonance.  The lower level of the resonance becomes occupied, the
upper level vacant, as in Fig.~\ref{intimate}(c), where we also sketch
the corresponding structural motifs.  Once the defects fully
recombine, the bottom level merges with the valence band and the top
level with the conduction band. We note that odd-numbered cyclic ring
molecules at half-filling must host at least one defect that in
principle cannot be removed.  The latter will thus host a midgap state
which is singly-occupied. Finally, Fig.~\ref{intimate}(d)
schematically shows what happens when an electron or hole is added to
a chain and the system is allowed to relax to form a polaron. In
contrast with a continuum view, in which a polaron can be thought of
as a generic impurity-like state, the lattice deformation results in
the appearance of at least {\em two} midgap states. The resulting
configuration can be thought of as a charged soliton in a close
proximity with a neutral soliton,~\cite{RevModPhys.60.781} or a charge
carrier bound to an intimate soliton-antisoliton pair.

\begin{figure}[t]
  \centering
  \includegraphics[width= \figurewidth]{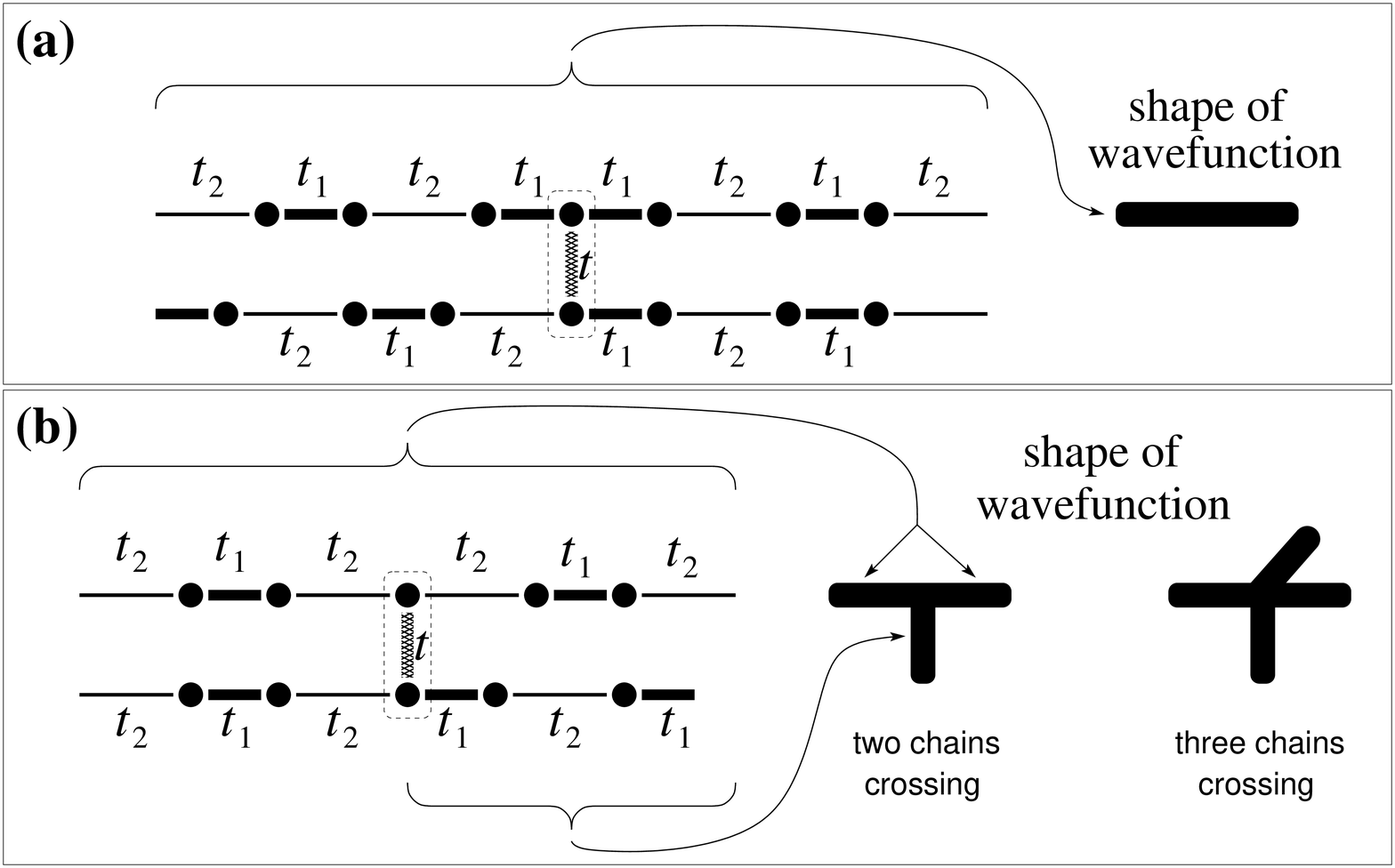}
  \caption{\label{chaincross} {\bf (a)} Graphical description of the
    tight-binding Hamiltonian used here to infer the electronic
    structure for two coupled chains, one hosting an overcoordinated
    atom and one perfectly dimerized. The sites connected by the
    transfer element $t$ are physically located on the same atom, as
    is signified by the dashed-line frame. On the r.h.s., we show the
    resulting shape of the wave function, according to the data in
    Fig.~\ref{over}. {\bf (b)} Same as (a), but the defected chain now
    hosts an under-coordinated atom. In this case, the wavefunction of
    the midgap state is shared between the chains, see
    Fig.~\ref{under}. In the case of two intersecting chains, one
    expects a T-shaped wavefunction, in which the angle between the
    two linear fragments is not necessarily $90^\circ$. For a greater
    number of intersecting chains, the shape could be described as
    urchin-shaped. Only two of the ``rays'' are expected to be
    approximately co-linear.}
\end{figure}

A distinct feature of the midgap states is their shape: Since the
wavefunctions of the topological states are hosted by chain-like
dimensional motifs, they are relatively localized in the direction
perpendicular to the motif, but could be substantially delocalized
along the motif itself. The delocalization along the chain is
surprisingly large given how deep the states are. This is because the
localization length is no longer determined by the depth of the state
but, instead, by the ratio of the transfer integrals of the strong and
weak bond.  The amount of delocalization is additionally increased
because the malcoordination is smoothly distributed over the chain,
typically over ten bond lengths or so, see the Supplementary Material
for more detail and Ref.~\onlinecite{ZLMicro2} for specific molecular
realizations. It was argued in that work that the effect of the solid
matrix housing the chain is, largely, to renormalize the on-site
energies and electronic transfer integrals along the chain. Here we
revisit this proposition by considering two or more chains crossing at
one site. One can think of one chain as made, for instance, primarily
of $pp\sigma$-bonded $p_x$ orbitals and the other chain of
$pp\sigma$-bonded $p_y$ orbitals.  Exactly one of the chains hosts a
malcoordination defect by construction. A graphical illustration of
this set-up is shown in Fig.~\ref{chaincross}, panels (a) and (b)
corresponding to over-coordination and under-coordination,
respectively. The chains are depicted as parallel to avoid crowding
the picture, however they are not parallel in the physical space. The
transfer integral between the two orbitals at the intersection between
the chains is denoted with $t$. These two orbitals belong to the same
atom; the non-zero overlap between the orbitals can come about because
the chains are not exactly perpendicular and/or because $sp$-mixing is
present. For simplicity, we allow the bond lengths to have only two
values. These values would correspond to the covalent and secondary
bond in a perfectly dimerized chain; the electronic hopping integrals
are $t_1$ and $t_2$ ($t_1 > t_2$) for the covalent and secondary bond,
respectively. Also for simplicity, we set all of the on-site energies
at the same value $\epsilon = 0$. The results can be straightforwardly
generalized for a non-vanishing, sign-alternating on-site
energy.~\cite{PhysRevLett.49.1455, ZLMicro2} The latter situation
would be directly relevant to linear chain-like motifs in which
chalcogen and pnictogen alternate in sequence. Finally, we set
transfer integrals for next-nearest and farther neighbors at zero.

\begin{figure}[t]
  \centering
  \includegraphics[width= .8 \figurewidth]{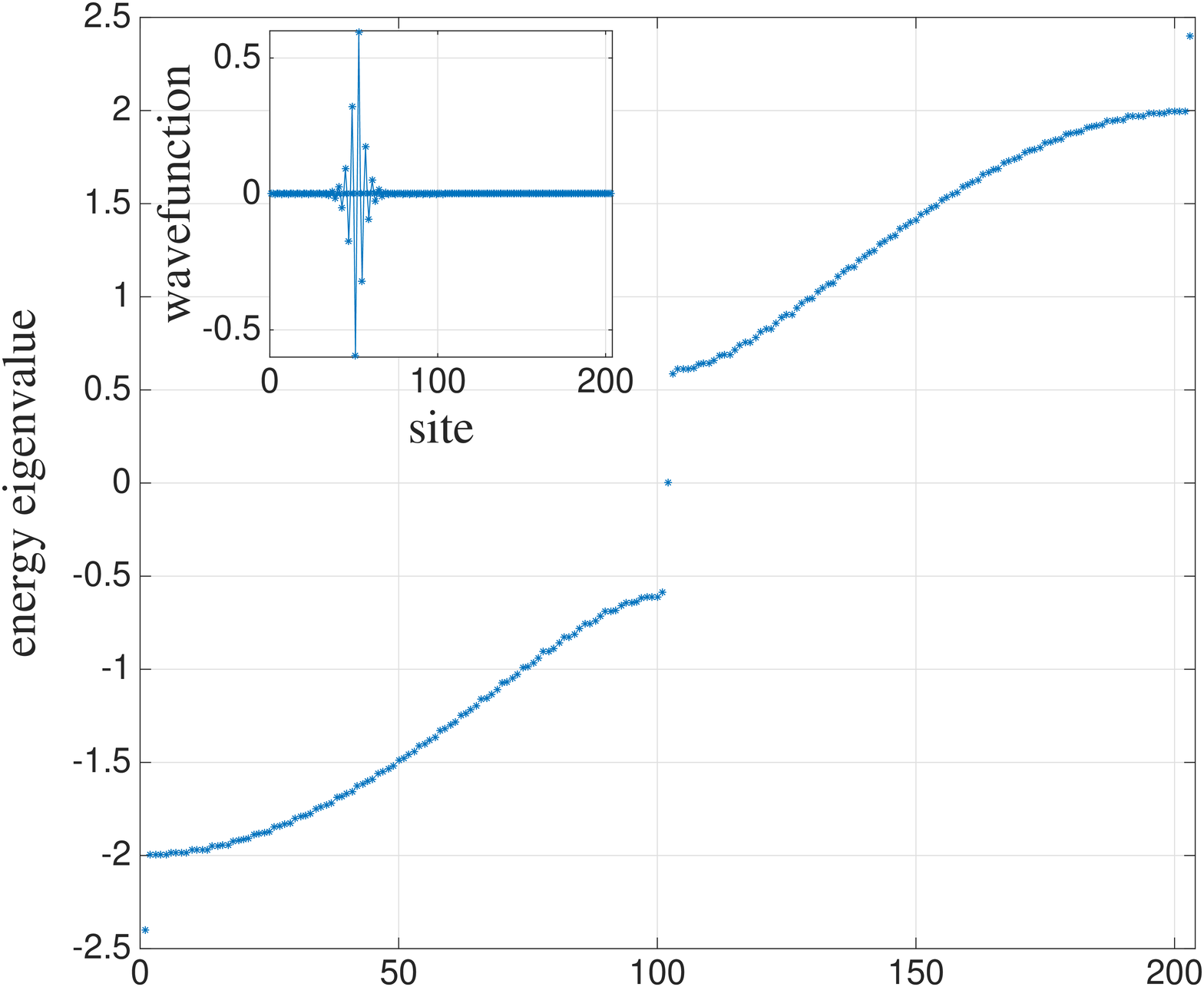}
  \caption{\label{over} The energy spectrum for the two intersecting
    chains from Fig.~\ref{chaincross}(a), the eigenvalues ordered in
    the ascending order. The lengths of the defected and defect-free
    chains are 103 and 100, respectively. $t_1 = - 1.3$, $t_2 = -
    0.7$, $t= - 1.0$.  The midgap state at $E=0$ is clearly seen, the
    corresponding wavefunction shown in the Inset, where sites 1
    through 103 comprise the defected chain and sites 104 through 203
    the defect-free chain. We observe that the midgap state based on
    overcoordinated atoms stays on the defected chain. The
    delocalization length is seen to be around twenty sites. Note the
    defect also gives rise to two bound states outside the bands,
    wavefunctions not shown, which is shared between the chains.}
\end{figure}

\begin{figure}[t]
  \centering
  \includegraphics[width= .8 \figurewidth]{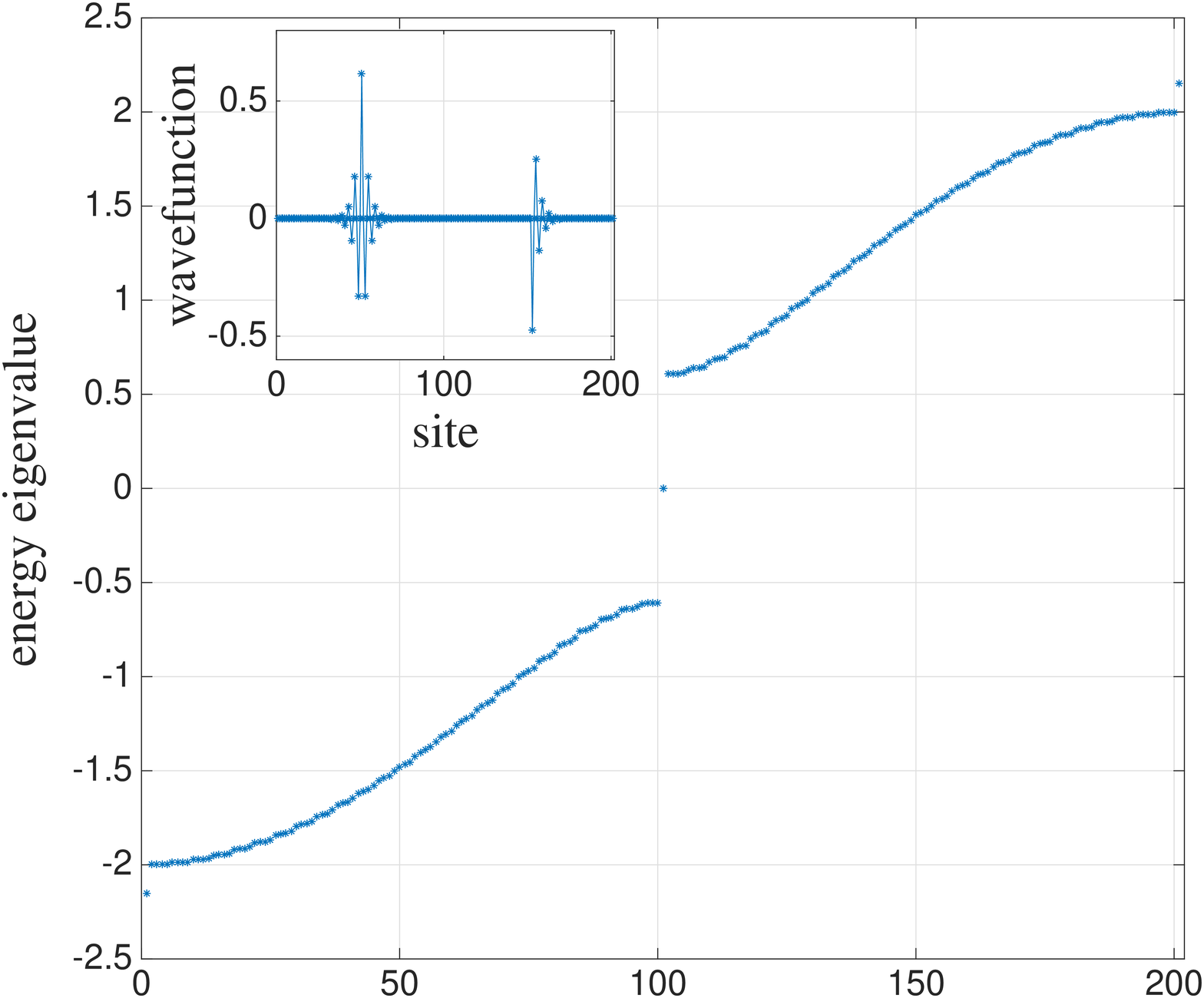}
  \caption{\label{under} Same as Fig.~\ref{over}, but for the
    under-coordinated situation from Fig.~\ref{chaincross}(b). The
    defected and defect-free chains contain 101 and 100 sites,
    respectively. In the inset, sites 1 through 101 comprise the
    defected chain and sites 102 through 201 the defect-free
    chain. Here we observe that the midgap state wavefunction is
    shared between the chains. The wavefunction penetrates the
    defect-free chain only on one side of the crossing point. }
\end{figure}

The resulting Hamiltonian is easily numerically diagonalized, the
corresponding spectra and the midgap wave-functions shown in
Figs.~\ref{over} and \ref{under}. We observe that in the case of
overcoordination, the wave-function is confined to the chain housing
the defect, consistent with the analysis of Zhugayevych and
Lubchenko.~\cite{ZLMicro1, ZLMicro2} In contrast, when the midgap
state is caused by under-coordination, a substantial part of the the
wave function now ``spills'' into the crossing, defect-free chains.
We notice that the wavefunction of the midgap state, on such a
defect-free chain, is non-vanishing only on one side of the
intersection. We have directly checked that the same conclusions apply
when three chains intersect. In cases when the wavefunction of the
midgap state is shared between a number of intersecting chains, its
shape can be thought of as a set of linear portions emanating from the
same lattice site; only two of these portions are expected to be
approximately co-linear.

\begin{figure}[t]
  \centering
  \includegraphics[width=.9 \figurewidth]{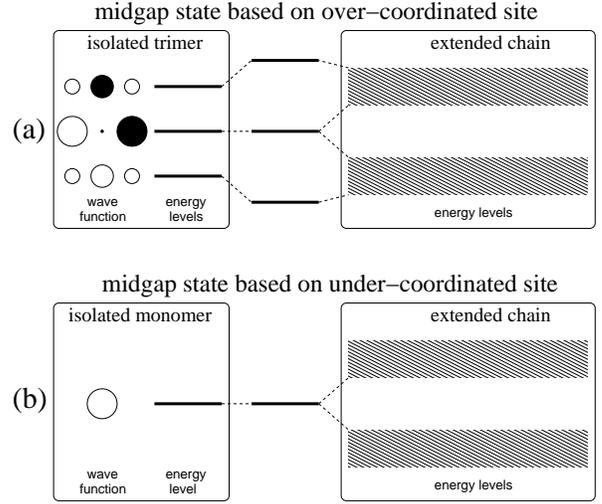}
  \caption{\label{informal} {\bf (a)} An informal molecular orbital
    (MO) diagram illustrating why the midgap state centered on an
    over-coordinated atom in an odd function with respect to
    reflection about the central site of the midgap state. The size
    and color of the circles on the l.h.s.  reflect, respectively, the
    magnitude and sign of the coefficient at the pertinent atomic
    orbital in the MOs. Clearly, the MO corresponding to the $E=0$
    state is odd and vanishes on the central site. {\bf (b)} Same as
    (a), but for an under-coordinated site. Here the midgap state's
    wavefunction is even. }
\end{figure}

These contrasting behaviors with respect to the wavefunction's shape
are discussed formally in the Supplemental Material, but can be
rationalized already using the following, informal line of reasoning,
which is graphically illustrated in Fig.~\ref{informal}. In the
absence of geometric optimization, which would distribute the
malcoordination over an extended portion of the chain, an {\em
  over}-coordinated center can be thought of as a trimer weakly
coupled to a perfectly dimerized chain via two weak bonds, as in
Fig.~\ref{informal}(a). The bottom and top level of the trimer will
be, respectively, stabilized and destabilized as a result of the
coupling. The non-bonding state of the trimer, on the other hand, will
remain in the middle of the gap by symmetry, since it is coupled
equally strongly to both the bottom and top band of the chain. The
corresponding wavefunction vanishes at the central site of the trimer,
either without or with coupling. 

In contrast, an {\em under}-coordinated center in a non-relaxed
geometry can be thought of as a single orbital at $E=0$ coupled to a
perfectly dimerized chain, as in Fig.~\ref{informal}(b). The wave
function of the resulting midgap state will be non-zero and, in fact,
will have its largest value at that orbital. When a defected chain is
coupled to a perfectly dimerized chain, the electron will be also able
to tunnel to some extent into the dimerized chain, but will do so only
on one side. This is because a state inside the forbidden gap must
have a wavefunction that is either exponentially increasing or
decreasing function of the coordinate. In a perfectly dimerized chain,
either behavior would have to be maintained throughout; the resulting
wavefunction would not be normalizable. This notion can be made
formal, see the Supplemental Information. There we also show that the
above conclusions still apply, even if one allows the chain to
geometrically optimize.

The question of what should happen to a midgap state when the chain is
enclosed in a solid matrix is much more complicated than the above
setup where only two or three chains intersect. The solid can be
thought of as a very large number of closed chains that cross at the
defect site; the resulting interference effects could be
significant. We have conducted preliminary tests on the effects of an
ordered solid matrix using small slabs of {\em crystalline}
As$_2$Se$_3$, see the Supplemental Material. We observed that even
after the sample is allowed to geometrically optimize, the midgap
state remains robustly within the gap and, in fact, moves closer to
the gap's center. The wavefunction of midgap state tends to peak at
the slab's surface while its shape becomes more anisotropic following
geometric optimization.

\begin{figure}[t!]
  \centering
  \includegraphics[width= \figurewidth]{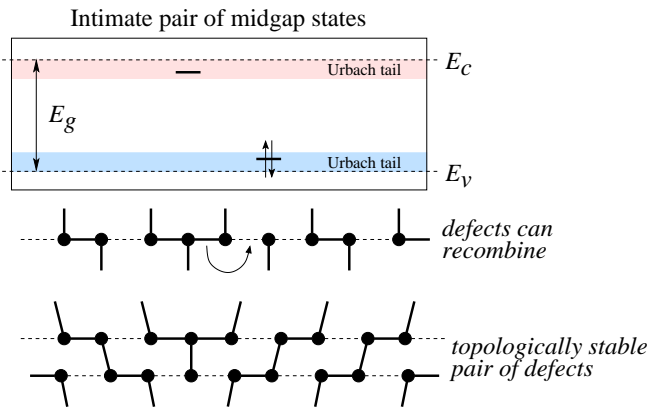}
  \caption{\label{intpair} A qualitative illustration of how two
    defects of opposite deviation from the correct coordination could
    or could not recombine if brought together. The recombination
    event, if any, would correspond to switching the bond as shown by
    the curved arrow. Upon recombination, the two levels would merge
    with the nearest band while the perfect alternation pattern of
    bond strength is restored. If infinitely apart, the defects would
    be both near the gap center, energy-wise, and singly occupied. }
\end{figure}

Finally we note that when the defects happen to be on distinct chains,
they may not be able to recombine but, instead, form a bound pair as
at the bottom of Fig.~\ref{intpair}, see a concrete molecular
realization in Fig.~10 of Ref.~\onlinecite{ZLMicro2}.

\section{The topological midgap states: Present samples}
\label{midgap2}

Motivated by the notions made in Section~\ref{midgap1}, we next
quantify both the extent and compactness of states within the gap and
its immediate vicinity. A concrete way to quantify the localization of
an orbital $\psi$ is to compute the so called inverse participation
ratio (IPR):
\begin{equation} \text{IPR} = \frac{\int d^3 \br \left|\psi
      (\br)\right|^4}{\left[\int d^3 \br \left|\psi
        (\br)\right|^2\right]^2},
\end{equation}
where the denominator is included in case the orbital is not
normalized to unity. A moment thought shows that the above expression
generically scales as the inverse volume of the region occupied by the
orbital. Thus the inverse participation ratio gives a volumetric
measure of localization of the respective wavefunction. The so
evaluated measure of localization is shown with the red dots in
Fig.~\ref{participation-ratio}. Consistent with expectation, the
states comprising the mobility band are delocalized over the whole
sample. There may be an ever so slight increase in localization going
toward the gap. This increase picks up closer to the edge of mobility
band so that by the end of the Urbach tail, the states are seen to
occupy a volume that is about one order of magnitude less than the
extended states. (This difference depends on the sample size, of
course.) We observe that the overall degree of localization of the
deep midgap states is greater still than that of the Urbach states,
but not dramatically so.

\begin{figure}[t!]
  \centering
  \includegraphics[width=1\figurewidth]{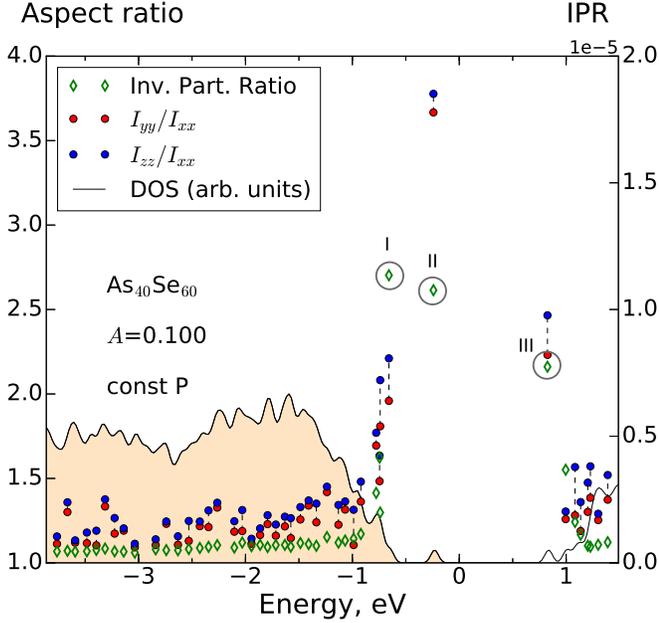}
  \caption{\label{inertia-ratio} Aspect ratios $I_{yy}/I_{xx}$ and
    $I_{zz}/I_{xx}$ and the corresponding inverse participation ratio
    (IRP) for select states in the band gap and its vicinity for a
    As$_{40}$Se$_{60}$ sample that hosts at least one midgap
    state. The thin solid line shows the electronic density of states,
    shading indicates filled states.  }
\end{figure}

To assess the {\em shape} of the wavefunctions, we compute its
``tensor of inertia'' according to:
\begin{equation}
  I_{ij} = \int r_i r_j d\rho,
\end{equation}
where $\rho$ is the charge density at a point $\br$. There is a
technical complication in computing the above quantity since the
sample is periodically continued, by construction, and so one must
make a decision as to the precise location of the repeat unit so that
the wavefunction is most connected and, hence, compact.  After
choosing the optimal location for the centroid of the wavefunction, we
compute the corresponding inertia tensor and bring it to a diagonal
form. We next sort the resulting principal moments of inertia in
ascending order and label them as follows: $I_{xx} \leq I_{yy} \leq
I_{zz}$. Since here we are interested only in the shape, not the
absolute extent of the wavefunction, we consider only the aspect
ratios of $I_{yy}/I_{xx}$ and $I_{zz}/I_{xx}$. According to the ZL
predictions, these ratios should significantly exceed unity at least
for the midgap states that are confined to one chain. One may also
generically expect the aspect ratio to be numerically close to one for
the extended states within the mobility bands.

\begin{figure}[t]
  \centering
  \includegraphics[width=1\figurewidth]{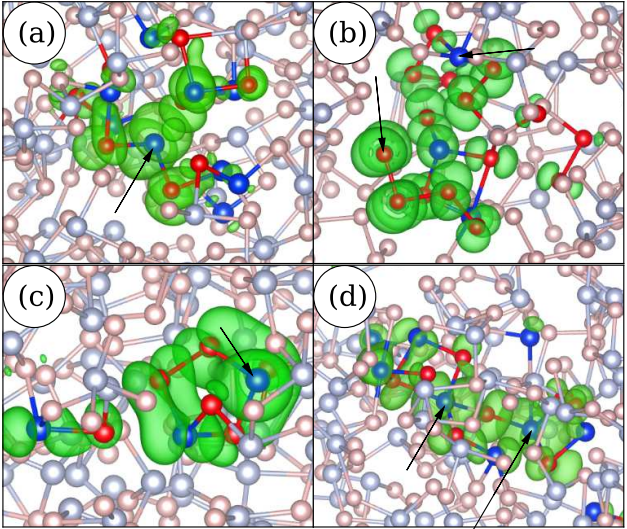}
  \caption{\label{defects} Charge density corresponding to the
    wavefunctions of the select deep midgap states for four samples
    marked by the arrows in Fig.~\ref{participation-ratio}. In each
    panel, we have added arrows pointing toward malcoordinated atoms.}
\end{figure}

These expectations are well borne out by our data, which are displayed
in Fig.~\ref{inertia-ratio}. We observe that the aspect ratio largely
echoes the participation ratio: The deeper inside the gap, the more
non-compact the wavefunctions tend to be. Again, the difference
between the deep-midgap and Urbach states is substantial but not
dramatic. It is instructive to visualize some of those deep midgap
states. We have selected four representative examples shown in
Fig.~\ref{defects}; the location of these states in the respective
electronic spectra can be found in Fig.~\ref{participation-ratio}.

We observe that at least in the cases encountered in the present
study, the wavefunctions of non-extended states are rather
anisotropic, and more so the deeper the corresponding state is into
the mobility gap. While this is consistent with the general analysis
of ZL,~\cite{ZLMicro2} the atomic motifs housing the wavefunctions in
the present study are significantly more complex than the standalone
molecular liner motifs generated by ZL.  In any event, all of the
motifs in Fig.~\ref{defects} contain at least one malcoordinated atom.
Specifically, motif (a) contains an under-coordinated arsenic, (b)
under-coordinated selenium and over-coordinated arsenic, (c) an
under-coordinated arsenic, and (d) two over-coordinated
arsenics. Consistent with the discussion in Section~\ref{midgap1}, the
midgap state wavefunction stemming from under-coordination in panel
(a) appears to branch out, while the defect in (d) contains
over-coordinated atoms and is elongated preferentially in one spatial
direction.  Defect (c) is notable in that the bulk of electron density
resides on a 5-member closed ring. As mentioned earlier, such
odd-numbered rings must host midgap states.

\begin{figure}[t]
  \centering
  \includegraphics[width= \figurewidth]{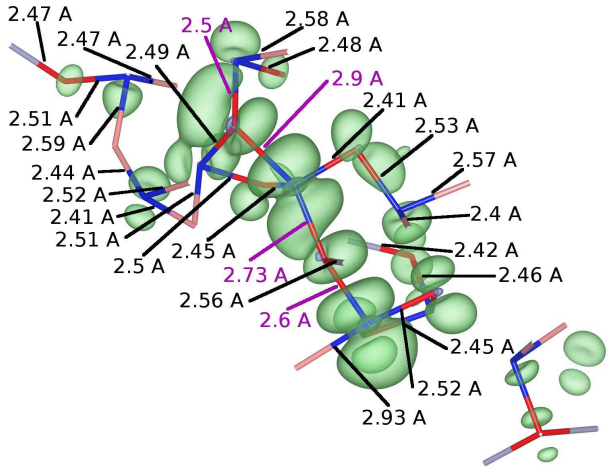}
  \caption{\label{linear} A different view of the midgap-state from
    Fig.~\ref{defects}(d).  The lengths of the bonds forming the chain
    motif that houses the bulk of midgap state are highlighted in
    purple. }
\end{figure}

We note that when quasi-linear, the midgap states recovered in the
present study are more complicated objects than what was foreseen by
ZL predictions.~\cite{ZLMicro2} To illustrate this notion, we provide
in Fig.~\ref{linear} a different view of the quasi-linear midgap
wavefunction from panel (d). In Fig.~\ref{linear}, we have highlighted
the bonds connecting the atoms that contribute significantly to the
wavefunction. Importantly, the bond lengths along the bulk portion of
the wavefunction, highlighted in purple, significantly exceed the
length of the covalent As-Se bond, i.e. 2.4~\AA.~\cite{LL1} That the
bond length near a topological defect should have an intermediate
value between than for the covalent and secondary bond is a hallmark
feature of the midgap states.~\cite{ZL_JCP, ZLMicro2} These bonds
clearly form a continuous, chain-like pattern. In addition, we observe
a ``satellite'' portion of the wavefunction running alongside its
primary portion. This satellite portion evidently has a substantial
contribution from lone pairs and likely stems from mixing between the
$p$-orbitals that are, respectively, parallel and perpendicular to the
line of the defect.

If standalone, a filled state that is close to the center of the gap
would be similar to the negatively charged topological state from
Fig.~\ref{intimate}(b). On the other hand, we observe that {\em
  vacant} midgap states are also generally present in the same sample
implying that the just mentioned filled state may, instead, be the
lower level of a resonance formed by two or more neutral midgap
states, as in Fig.~\ref{intimate}(c). There appears to be no
conclusive way to distinguish between those situations in the present
computational setup since in such modestly-size samples, it might be
difficult to produce even a relatively isolated, let alone truly
standalone defect.  The situation is however clearer with the filled
states close to the valence band, i.e., the Urbach states, since they
certainly have vacant counterparts close to the conduction band. Thus
in view of Fig.~\ref{intimate}(c), one may think of the Urbach states
as resulting from intimate soliton-antisoliton pairs. This notion adds
quite a bit of microscopic detail to the conventional idea of
Urbach-tail states as a generic consequence of disorder. Indeed, we
observe in Fig.~\ref{inertia-ratio} that the characteristics of a
filled state change only gradually with the energy of the state. Thus
the Urbach states share, to an extent, some of the properties with the
deep midgap states, such as the anisotropy in shape.

\begin{figure}[t!]
  \centering
  \includegraphics[width=.7\textwidth]{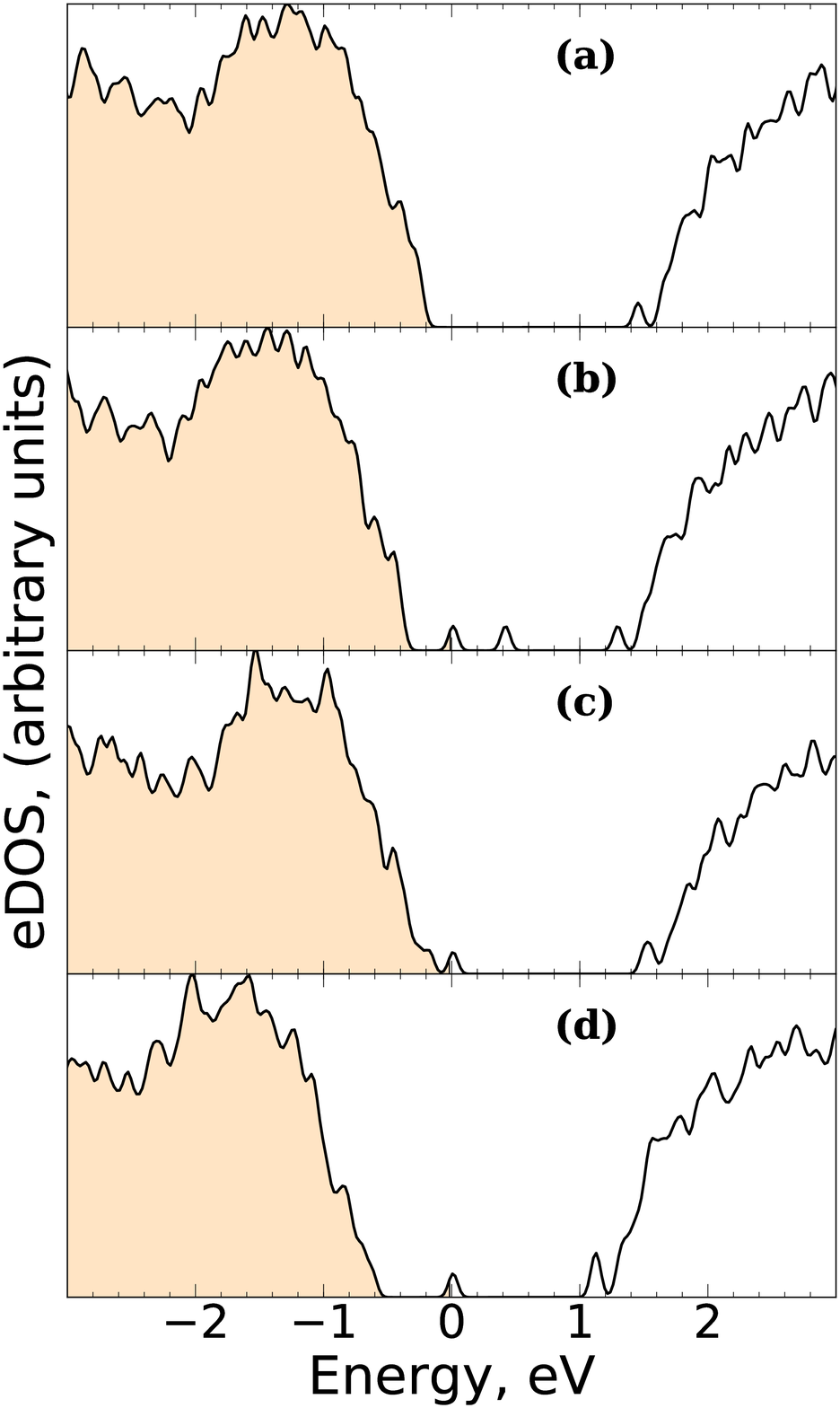}
  \caption{\label{figure1} (a) Electronic density of states (DOS) of
    an individual sample containing an even number of electrons. The
    rest of the panels show the DOS for the corresponding samples
    containing an odd number of electrons and prepared according to
    protocols I (b), II (c), and III (d). Shading indicates filled
    states. }
\end{figure}

We have observed above that very deep midgap states spontaneously
arise in some samples and match the characteristics of the topological
midgap states predicted by Zhugayevych and Lubchenko.  Next we study
what happens when the system is forced to have at least one {\em
  dangling} bond, which can be accomplished by using an odd number of
electrons. Because selenium and arsenic have an even and odd number of
electrons, respectively, we can ensure that the system has at least
one unpaired spin---and hence a dangling bond---by using an odd number
of arsenic atoms. This we accomplish by removing one of the arsenic
atoms from a sample that contains an even number of such atoms using
three distinct protocols: In protocol (I) we remove an arsenic atom
randomly from a sample that has been already geometrically
optimized. No further optimization is performed. In protocol (II), we
further optimize the sample obtained in protocol (I). In protocol
(III), one arsenic atom is removed already from the parent
structure. Only after this is the structure geometrically
optimized. The logic behind these protocols is as follows: In protocol
(I), we expect the defect to be as close as possible in character to a
vacancy.  In protocol (II), we allow this ``vacancy'' to relax but
subject to an environment that is already mechanically stable. In
protocol (III) the system is given the greatest amount of freedom to
relax afforded by the structure-building algorithm in Ref.~\cite{LL1}
The amount of residual strain in the structure is expected to be the
greatest in structures of type I and the least in structures of type
III.  Everywhere below, we limit ourselves to the stoichiometric
compound As$_{0.4}$Se$_{0.6}$. (The stoichiometry is obeyed only
approximately.)

The electronic density of states for individual realizations for all
three protocols is illustrated in Fig.~\ref{figure1}, alongside that
for the original structure that contained an even number of electrons.
We observe that the gross features of the electronic spectrum are not
sensitive to the detailed preparation protocol.

\begin{figure}[t]
  \centering
  \includegraphics[width=.9\textwidth]{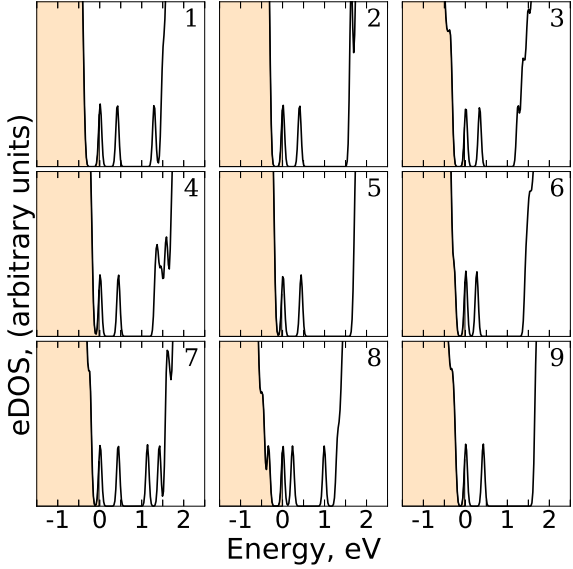}
  \caption{\label{figure2} Band structures of nine different samples
    with an odd number of electrons, as prepared using protocol I.}
\end{figure}

\begin{figure}[t]
  \centering
  \includegraphics[width=.9\textwidth]{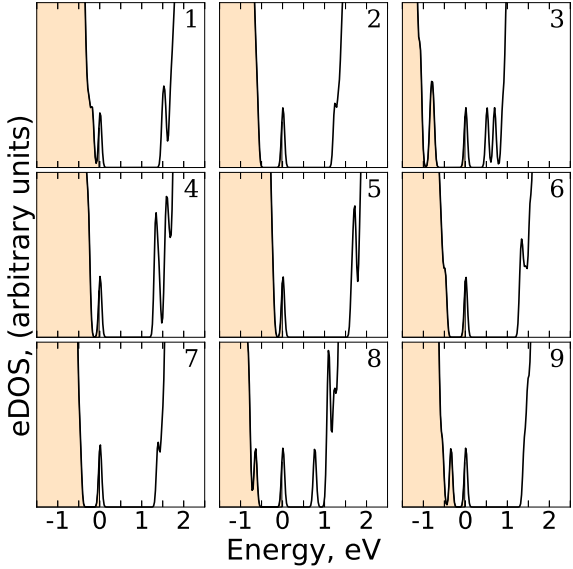}
  \caption{\label{figure3} Band structures of nine different samples
    prepared using protocol II. These correspond, panel-wise, to
    Fig.~\ref{figure2}.}
\end{figure}

\begin{figure}[t]
  \centering
  \includegraphics[width=.9\textwidth]{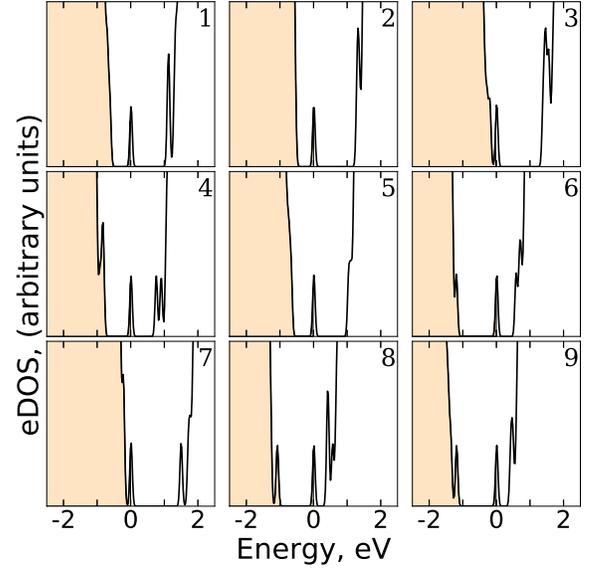}
  \caption{\label{figure4} Band structures of nine different samples
    prepared using protocol III.  These correspond, panel-wise, to
    Figs.~\ref{figure2} and ~\ref{figure3}.}
\end{figure}

We next focus on the electronic density of states within the mobility
gap and its immediate vicinity. Nine realizations for protocols I, II,
and III are shown, respectively in Figs.~\ref{figure2}, \ref{figure3},
and \ref{figure4}. In all figures, shading indicates filled
levels. Consistent with the expectation that samples of type I should
exhibit most strain, such samples show on average the greatest number
of midgap states. Yet even though samples of type II and III are
allowed to relax, some of them still host very deep-lying midgap
states. We observe that the midgap states in some samples of type I
are similar, at least superficially, to the hole-polaron configuration
in Fig.~\ref{intimate}(d), see for instance panel 8. This result is
perhaps not too surprising since the region centered on the removed
arsenic atom has a lower electron density or, equivalently, increased
hole density. In contrast, samples of type II seem to house primarily
neutral states from Fig.~\ref{intimate}(a). Samples of type (III)
exhibit such states as well and, in addition, states similar to the
electron-polaron states from Fig.~\ref{intimate}(d), as in panel 9.
While it seems reasonable that structures of type III would exhibit
the highest electron density of the three structure types, it is not
at all obvious why electron-polaron-like configurations should be so
stable as to be readily found already in the small ensemble structures
we have generated in the present study.

Finally we refer the reader to the Supplemental Material for the
information on the localization and anisotropy of the electronic
states for systems with an odd number of electrons, which is presented
there in the format of Fig.~\ref{inertia-ratio}. The main conclusion
is that the midgap states with systems with an odd-number of electrons
are entirely analogous to systems with filled states. In a
distinction, the midgap states in systems of type I are significantly
more localized than when the sample is allowed to relax. This is
expected since upon relaxation, the lone pair residing in the cavity
formed by removing an atom is able to mix more readily with the rest
of the orbitals.

\section{Summary}
\label{summary}

We have computationally-generated samples of glassy arsenic selenide,
for several stoichiometries, that appear to exhibit all putative types
of electronic states thought to exist in these materials. The gross
features of the electronic density of states do not vary from sample
to sample. The corresponding portions of the density of states are
attributed to the mobility bands; these states are extended and their
density of states is self-averaging.  In contrast, states near the
edge of the mobility bands strongly fluctuate in energy. The
statistics of this variation match well the distribution of the
venerable Urbach-tail states.

Most significantly, we recover a special set of electronic states
whose characteristics match those of the topological midgap states
predicted earlier by Zhugayevych and Lubchenko (ZL). These special
midgap states are very deep into the mobility gap and are not simply a
generic consequence of static disorder in the atomic
position. Instead, they stem from the vast degeneracy of the free
energy landscape of a glassy liquid. The midgap states have been
predicted~\cite{ZLMicro2} to reside on domain walls separating
distinct minima on the free energy landscape. We have shown that in
addition to the chain-like shape for the midgap states predicted in
Ref.~\cite{ZLMicro2}, more complicated, urchin-like and cyclic shapes
are possible, too.

Different charge states of the midgap states have been observed that
could be potentially identified with standalone midgap states or their
intimate pairs, and also with added charge carriers. Clearly, such
polaron-like states are significant in the context of the mechanism of
electrical conductance in amorphous chalcogenides. The current view is
that in the chalcogenides, electrical current is carried by small
polarons,~\cite{Emin_rev, Emin_revII} each of which can be thought of
as compound particle consisting of a charge carrier proper and the
polarization of the lattice, largely in the spirit of the Born model
of solvation of charge in a polar solvent.  In contrast with the
continuum Born picture, the present treatment directly identifies sets
of orbitals that house the charge carrier. Such sets greatly exceed in
size and complexity what would be expected of a small polaron. The
present results also indicate that the polarization of the lattice
involves very particular changes in local bonding such as emergence of
malcoordination.

The above findings were enabled by the availability of a
quantum-chemical approximation that recovers the width of the mobility
gap with reasonable accuracy, viz., a specific flavor of a hybrid DFT
approximation. Standard DFT approaches tend to underestimate the width
of the gap thus effectively concealing many of the features of the
electronic spectrum this work focused on.

We hope that a more conclusive identification of the presently
reported deep midgap states with the topological midgap states
predicted by ZL will be possible when we learn how to simulate
transitions between distinct free energy minima in a glassy
chalcogenide or similar compounds. When approached directly, such a
simulation is computationally very costly because of excessively high
relaxation barriers.~\cite{L_AP} Circumventing this complication is
work in progress.


\section{Acknowledgments}

This work has been supported by the National Science Foundation Grants
CHE-0956127, CHE-1465125, and the Welch Foundation Grant E-1765.  We
gratefully acknowledge the use of the Maxwell/Opuntia Cluster and the
untiring support from the Center for Advanced Computing and Data
Systems at the University of Houston.  Partial support for this work
was provided by resources of the uHPC cluster managed by the
University of Houston and acquired through NSF Award Number
ACI-1531814.

\bibliography{/Users/vas/Documents/tex/ACP/lowT,paper}

\begin{thebibliography}{84}%
\makeatletter
\providecommand \@ifxundefined [1]{%
 \@ifx{#1\undefined}
}%
\providecommand \@ifnum [1]{%
 \ifnum #1\expandafter \@firstoftwo
 \else \expandafter \@secondoftwo
 \fi
}%
\providecommand \@ifx [1]{%
 \ifx #1\expandafter \@firstoftwo
 \else \expandafter \@secondoftwo
 \fi
}%
\providecommand \natexlab [1]{#1}%
\providecommand \enquote  [1]{``#1''}%
\providecommand \bibnamefont  [1]{#1}%
\providecommand \bibfnamefont [1]{#1}%
\providecommand \citenamefont [1]{#1}%
\providecommand \href@noop [0]{\@secondoftwo}%
\providecommand \href [0]{\begingroup \@sanitize@url \@href}%
\providecommand \@href[1]{\@@startlink{#1}\@@href}%
\providecommand \@@href[1]{\endgroup#1\@@endlink}%
\providecommand \@sanitize@url [0]{\catcode `\\12\catcode `\$12\catcode
  `\&12\catcode `\#12\catcode `\^12\catcode `\_12\catcode `\%12\relax}%
\providecommand \@@startlink[1]{}%
\providecommand \@@endlink[0]{}%
\providecommand \url  [0]{\begingroup\@sanitize@url \@url }%
\providecommand \@url [1]{\endgroup\@href {#1}{\urlprefix }}%
\providecommand \urlprefix  [0]{URL }%
\providecommand \Eprint [0]{\href }%
\providecommand \doibase [0]{http://dx.doi.org/}%
\providecommand \selectlanguage [0]{\@gobble}%
\providecommand \bibinfo  [0]{\@secondoftwo}%
\providecommand \bibfield  [0]{\@secondoftwo}%
\providecommand \translation [1]{[#1]}%
\providecommand \BibitemOpen [0]{}%
\providecommand \bibitemStop [0]{}%
\providecommand \bibitemNoStop [0]{.\EOS\space}%
\providecommand \EOS [0]{\spacefactor3000\relax}%
\providecommand \BibitemShut  [1]{\csname bibitem#1\endcsname}%
\let\auto@bib@innerbib\@empty
\bibitem [{\citenamefont {Anderson}(1958)}]{AndersonLoc}%
  \BibitemOpen
  \bibfield  {author} {\bibinfo {author} {\bibfnamefont {P.~W.}\ \bibnamefont
  {Anderson}},\ }\href@noop {} {\bibfield  {journal} {\bibinfo  {journal}
  {Phys. Rev.}\ }\textbf {\bibinfo {volume} {109}},\ \bibinfo {pages} {1492}
  (\bibinfo {year} {1958})}\BibitemShut {NoStop}%
\bibitem [{\citenamefont {Mott}(1982)}]{Mott1982}%
  \BibitemOpen
  \bibfield  {author} {\bibinfo {author} {\bibfnamefont {N.}~\bibnamefont
  {Mott}},\ }\href@noop {} {\bibfield  {journal} {\bibinfo  {journal} {Proc. R.
  Soc. Lond.}\ }\textbf {\bibinfo {volume} {A 382}},\ \bibinfo {pages} {1}
  (\bibinfo {year} {1982})}\BibitemShut {NoStop}%
\bibitem [{\citenamefont {Cohen}\ \emph {et~al.}(1969)\citenamefont {Cohen},
  \citenamefont {Fritzsche},\ and\ \citenamefont {Ovshinsky}}]{CFO}%
  \BibitemOpen
  \bibfield  {author} {\bibinfo {author} {\bibfnamefont {M.~H.}\ \bibnamefont
  {Cohen}}, \bibinfo {author} {\bibfnamefont {H.}~\bibnamefont {Fritzsche}}, \
  and\ \bibinfo {author} {\bibfnamefont {S.~R.}\ \bibnamefont {Ovshinsky}},\
  }\href@noop {} {\bibfield  {journal} {\bibinfo  {journal} {Phys. Rev. Lett.}\
  }\textbf {\bibinfo {volume} {22}},\ \bibinfo {pages} {1065} (\bibinfo {year}
  {1969})}\BibitemShut {NoStop}%
\bibitem [{\citenamefont {Mott}(1990)}]{Mott1990}%
  \BibitemOpen
  \bibfield  {author} {\bibinfo {author} {\bibfnamefont {N.~F.}\ \bibnamefont
  {Mott}},\ }\href@noop {} {\emph {\bibinfo {title} {Metal-Insulator
  Transitions}}}\ (\bibinfo  {publisher} {Taylor and Francis},\ \bibinfo
  {address} {London},\ \bibinfo {year} {1990})\BibitemShut {NoStop}%
\bibitem [{\citenamefont {Emin}(1983{\natexlab{a}})}]{Emin_rev}%
  \BibitemOpen
  \bibfield  {author} {\bibinfo {author} {\bibfnamefont {D.}~\bibnamefont
  {Emin}},\ }\href@noop {} {\bibfield  {journal} {\bibinfo  {journal} {Comments
  Solid State Phys.}\ }\textbf {\bibinfo {volume} {11}},\ \bibinfo {pages} {35}
  (\bibinfo {year} {1983}{\natexlab{a}})}\BibitemShut {NoStop}%
\bibitem [{\citenamefont {Emin}(1983{\natexlab{b}})}]{Emin_revII}%
  \BibitemOpen
  \bibfield  {author} {\bibinfo {author} {\bibfnamefont {D.}~\bibnamefont
  {Emin}},\ }\href@noop {} {\bibfield  {journal} {\bibinfo  {journal} {Comments
  Solid State Phys.}\ }\textbf {\bibinfo {volume} {11}},\ \bibinfo {pages} {59}
  (\bibinfo {year} {1983}{\natexlab{b}})}\BibitemShut {NoStop}%
\bibitem [{\citenamefont {Urbach}(1953)}]{PhysRev.92.1324}%
  \BibitemOpen
  \bibfield  {author} {\bibinfo {author} {\bibfnamefont {F.}~\bibnamefont
  {Urbach}},\ }\href@noop {} {\bibfield  {journal} {\bibinfo  {journal} {Phys.
  Rev.}\ }\textbf {\bibinfo {volume} {92}},\ \bibinfo {pages} {1324} (\bibinfo
  {year} {1953})}\BibitemShut {NoStop}%
\bibitem [{\citenamefont {Toyozawa}(1961)}]{Toyozawa1961}%
  \BibitemOpen
  \bibfield  {author} {\bibinfo {author} {\bibfnamefont {Y.}~\bibnamefont
  {Toyozawa}},\ }\href@noop {} {\bibfield  {journal} {\bibinfo  {journal}
  {Progr. Theor. Phys.}\ }\textbf {\bibinfo {volume} {26}},\ \bibinfo {pages}
  {29} (\bibinfo {year} {1961})}\BibitemShut {NoStop}%
\bibitem [{\citenamefont {Dow}\ and\ \citenamefont
  {Redfield}(1972)}]{PhysRevB.5.594}%
  \BibitemOpen
  \bibfield  {author} {\bibinfo {author} {\bibfnamefont {J.~D.}\ \bibnamefont
  {Dow}}\ and\ \bibinfo {author} {\bibfnamefont {D.}~\bibnamefont {Redfield}},\
  }\href {\doibase 10.1103/PhysRevB.5.594} {\bibfield  {journal} {\bibinfo
  {journal} {Phys. Rev. B}\ }\textbf {\bibinfo {volume} {5}},\ \bibinfo {pages}
  {594} (\bibinfo {year} {1972})}\BibitemShut {NoStop}%
\bibitem [{\citenamefont {Mahan}(1966)}]{PhysRev.145.602}%
  \BibitemOpen
  \bibfield  {author} {\bibinfo {author} {\bibfnamefont {G.~D.}\ \bibnamefont
  {Mahan}},\ }\href {\doibase 10.1103/PhysRev.145.602} {\bibfield  {journal}
  {\bibinfo  {journal} {Phys. Rev.}\ }\textbf {\bibinfo {volume} {145}},\
  \bibinfo {pages} {602} (\bibinfo {year} {1966})}\BibitemShut {NoStop}%
\bibitem [{\citenamefont {Kostadinov}(1977)}]{Kostadinov}%
  \BibitemOpen
  \bibfield  {author} {\bibinfo {author} {\bibfnamefont {I.~Z.}\ \bibnamefont
  {Kostadinov}},\ }\href@noop {} {\bibfield  {journal} {\bibinfo  {journal} {J.
  Phys. C: Solid State Phys.}\ }\textbf {\bibinfo {volume} {10}},\ \bibinfo
  {pages} {L263} (\bibinfo {year} {1977})}\BibitemShut {NoStop}%
\bibitem [{\citenamefont {Brezin}\ and\ \citenamefont
  {Parisi}(1980)}]{0022-3719-13-12-005}%
  \BibitemOpen
  \bibfield  {author} {\bibinfo {author} {\bibfnamefont {E.}~\bibnamefont
  {Brezin}}\ and\ \bibinfo {author} {\bibfnamefont {G.}~\bibnamefont
  {Parisi}},\ }\href@noop {} {\bibfield  {journal} {\bibinfo  {journal} {J.
  Phys. C}\ }\textbf {\bibinfo {volume} {13}},\ \bibinfo {pages} {L307}
  (\bibinfo {year} {1980})}\BibitemShut {NoStop}%
\bibitem [{\citenamefont {Cardy}(1978)}]{0022-3719-11-8-006}%
  \BibitemOpen
  \bibfield  {author} {\bibinfo {author} {\bibfnamefont {J.~L.}\ \bibnamefont
  {Cardy}},\ }\href@noop {} {\bibfield  {journal} {\bibinfo  {journal} {J.
  Phys. C}\ }\textbf {\bibinfo {volume} {11}},\ \bibinfo {pages} {L321}
  (\bibinfo {year} {1978})}\BibitemShut {NoStop}%
\bibitem [{\citenamefont {John}\ \emph {et~al.}(1986)\citenamefont {John},
  \citenamefont {Soukoulis}, \citenamefont {Cohen},\ and\ \citenamefont
  {Economou}}]{PhysRevLett.57.1777}%
  \BibitemOpen
  \bibfield  {author} {\bibinfo {author} {\bibfnamefont {S.}~\bibnamefont
  {John}}, \bibinfo {author} {\bibfnamefont {C.}~\bibnamefont {Soukoulis}},
  \bibinfo {author} {\bibfnamefont {M.~H.}\ \bibnamefont {Cohen}}, \ and\
  \bibinfo {author} {\bibfnamefont {E.~N.}\ \bibnamefont {Economou}},\ }\href
  {\doibase 10.1103/PhysRevLett.57.1777} {\bibfield  {journal} {\bibinfo
  {journal} {Phys. Rev. Lett.}\ }\textbf {\bibinfo {volume} {57}},\ \bibinfo
  {pages} {1777} (\bibinfo {year} {1986})}\BibitemShut {NoStop}%
\bibitem [{\citenamefont {Zittartz}\ and\ \citenamefont
  {Langer}(1966)}]{PhysRev.148.741}%
  \BibitemOpen
  \bibfield  {author} {\bibinfo {author} {\bibfnamefont {J.}~\bibnamefont
  {Zittartz}}\ and\ \bibinfo {author} {\bibfnamefont {J.~S.}\ \bibnamefont
  {Langer}},\ }\href {\doibase 10.1103/PhysRev.148.741} {\bibfield  {journal}
  {\bibinfo  {journal} {Phys. Rev.}\ }\textbf {\bibinfo {volume} {148}},\
  \bibinfo {pages} {741} (\bibinfo {year} {1966})}\BibitemShut {NoStop}%
\bibitem [{\citenamefont {Economou}\ \emph {et~al.}(1970)\citenamefont
  {Economou}, \citenamefont {Kirkpatrick}, \citenamefont {Cohen},\ and\
  \citenamefont {Eggarter}}]{PhysRevLett.25.520}%
  \BibitemOpen
  \bibfield  {author} {\bibinfo {author} {\bibfnamefont {E.~N.}\ \bibnamefont
  {Economou}}, \bibinfo {author} {\bibfnamefont {S.}~\bibnamefont
  {Kirkpatrick}}, \bibinfo {author} {\bibfnamefont {M.~H.}\ \bibnamefont
  {Cohen}}, \ and\ \bibinfo {author} {\bibfnamefont {T.~P.}\ \bibnamefont
  {Eggarter}},\ }\href {\doibase 10.1103/PhysRevLett.25.520} {\bibfield
  {journal} {\bibinfo  {journal} {Phys. Rev. Lett.}\ }\textbf {\bibinfo
  {volume} {25}},\ \bibinfo {pages} {520} (\bibinfo {year} {1970})}\BibitemShut
  {NoStop}%
\bibitem [{\citenamefont {Lubchenko}\ and\ \citenamefont
  {Wolynes}(2004)}]{LW_aging}%
  \BibitemOpen
  \bibfield  {author} {\bibinfo {author} {\bibfnamefont {V.}~\bibnamefont
  {Lubchenko}}\ and\ \bibinfo {author} {\bibfnamefont {P.~G.}\ \bibnamefont
  {Wolynes}},\ }\href@noop {} {\bibfield  {journal} {\bibinfo  {journal} {J.
  Chem. Phys.}\ }\textbf {\bibinfo {volume} {121}},\ \bibinfo {pages} {2852}
  (\bibinfo {year} {2004})}\BibitemShut {NoStop}%
\bibitem [{\citenamefont {Lubchenko}\ and\ \citenamefont
  {Wolynes}(2018)}]{LWjamming}%
  \BibitemOpen
  \bibfield  {author} {\bibinfo {author} {\bibfnamefont {V.}~\bibnamefont
  {Lubchenko}}\ and\ \bibinfo {author} {\bibfnamefont {P.~G.}\ \bibnamefont
  {Wolynes}},\ }\href@noop {} {\bibfield  {journal} {\bibinfo  {journal} {J.
  Phys. Chem. B}\ }\textbf {\bibinfo {volume} {122}},\ \bibinfo {pages} {3280}
  (\bibinfo {year} {2018})}\BibitemShut {NoStop}%
\bibitem [{\citenamefont {Anderson}(1976)}]{PWA_negU2}%
  \BibitemOpen
  \bibfield  {author} {\bibinfo {author} {\bibfnamefont {P.~W.}\ \bibnamefont
  {Anderson}},\ }\href@noop {} {\bibfield  {journal} {\bibinfo  {journal} {J.
  Phys. (Paris)}\ }\textbf {\bibinfo {volume} {C4}},\ \bibinfo {pages} {339}
  (\bibinfo {year} {1976})}\BibitemShut {NoStop}%
\bibitem [{\citenamefont {Zhugayevych}\ and\ \citenamefont
  {Lubchenko}(2010{\natexlab{a}})}]{ZL_JCP}%
  \BibitemOpen
  \bibfield  {author} {\bibinfo {author} {\bibfnamefont {A.}~\bibnamefont
  {Zhugayevych}}\ and\ \bibinfo {author} {\bibfnamefont {V.}~\bibnamefont
  {Lubchenko}},\ }\href@noop {} {\bibfield  {journal} {\bibinfo  {journal} {J.
  Chem. Phys.}\ }\textbf {\bibinfo {volume} {132}},\ \bibinfo {pages} {044508}
  (\bibinfo {year} {2010}{\natexlab{a}})}\BibitemShut {NoStop}%
\bibitem [{\citenamefont {Biegelsen}\ and\ \citenamefont
  {Street}(1980)}]{BiegelsenStreet}%
  \BibitemOpen
  \bibfield  {author} {\bibinfo {author} {\bibfnamefont {D.~K.}\ \bibnamefont
  {Biegelsen}}\ and\ \bibinfo {author} {\bibfnamefont {R.~A.}\ \bibnamefont
  {Street}},\ }\href@noop {} {\bibfield  {journal} {\bibinfo  {journal} {Phys.
  Rev. Lett.}\ }\textbf {\bibinfo {volume} {44}},\ \bibinfo {pages} {803}
  (\bibinfo {year} {1980})}\BibitemShut {NoStop}%
\bibitem [{\citenamefont {Hautala}\ \emph {et~al.}(1988)\citenamefont
  {Hautala}, \citenamefont {Ohlsen},\ and\ \citenamefont
  {Taylor}}]{PhysRevB.38.11048}%
  \BibitemOpen
  \bibfield  {author} {\bibinfo {author} {\bibfnamefont {J.}~\bibnamefont
  {Hautala}}, \bibinfo {author} {\bibfnamefont {W.~D.}\ \bibnamefont {Ohlsen}},
  \ and\ \bibinfo {author} {\bibfnamefont {P.~C.}\ \bibnamefont {Taylor}},\
  }\href@noop {} {\bibfield  {journal} {\bibinfo  {journal} {Phys. Rev. B}\
  }\textbf {\bibinfo {volume} {38}},\ \bibinfo {pages} {11048} (\bibinfo {year}
  {1988})}\BibitemShut {NoStop}%
\bibitem [{\citenamefont {Shimakawa}\ \emph {et~al.}(1995)\citenamefont
  {Shimakawa}, \citenamefont {Kolobov},\ and\ \citenamefont
  {Elliott}}]{ShimakawaElliott}%
  \BibitemOpen
  \bibfield  {author} {\bibinfo {author} {\bibfnamefont {K.}~\bibnamefont
  {Shimakawa}}, \bibinfo {author} {\bibfnamefont {A.}~\bibnamefont {Kolobov}},
  \ and\ \bibinfo {author} {\bibfnamefont {S.~R.}\ \bibnamefont {Elliott}},\
  }\href@noop {} {\bibfield  {journal} {\bibinfo  {journal} {Adv. Phys.}\
  }\textbf {\bibinfo {volume} {44}},\ \bibinfo {pages} {475} (\bibinfo {year}
  {1995})}\BibitemShut {NoStop}%
\bibitem [{\citenamefont {Bishop}\ \emph {et~al.}(1977)\citenamefont {Bishop},
  \citenamefont {Strom},\ and\ \citenamefont {Taylor}}]{PhysRevB.15.2278}%
  \BibitemOpen
  \bibfield  {author} {\bibinfo {author} {\bibfnamefont {S.~G.}\ \bibnamefont
  {Bishop}}, \bibinfo {author} {\bibfnamefont {U.}~\bibnamefont {Strom}}, \
  and\ \bibinfo {author} {\bibfnamefont {P.~C.}\ \bibnamefont {Taylor}},\
  }\href@noop {} {\bibfield  {journal} {\bibinfo  {journal} {Phys. Rev. B}\
  }\textbf {\bibinfo {volume} {15}},\ \bibinfo {pages} {2278} (\bibinfo {year}
  {1977})}\BibitemShut {NoStop}%
\bibitem [{\citenamefont {Mollot}\ \emph {et~al.}(1980)\citenamefont {Mollot},
  \citenamefont {Cernogora},\ and\ \citenamefont {{Benoit \`{a} la
  Guillaume}}}]{Mollot1980}%
  \BibitemOpen
  \bibfield  {author} {\bibinfo {author} {\bibfnamefont {F.}~\bibnamefont
  {Mollot}}, \bibinfo {author} {\bibfnamefont {J.}~\bibnamefont {Cernogora}}, \
  and\ \bibinfo {author} {\bibfnamefont {C.}~\bibnamefont {{Benoit \`{a} la
  Guillaume}}},\ }\href@noop {} {\bibfield  {journal} {\bibinfo  {journal}
  {Phil. Mag. B}\ }\textbf {\bibinfo {volume} {42}},\ \bibinfo {pages} {643}
  (\bibinfo {year} {1980})}\BibitemShut {NoStop}%
\bibitem [{\citenamefont {Tada}\ and\ \citenamefont
  {Ninomiya}(1989{\natexlab{a}})}]{TadaNinomiya}%
  \BibitemOpen
  \bibfield  {author} {\bibinfo {author} {\bibfnamefont {T.}~\bibnamefont
  {Tada}}\ and\ \bibinfo {author} {\bibfnamefont {T.}~\bibnamefont
  {Ninomiya}},\ }\href@noop {} {\bibfield  {journal} {\bibinfo  {journal} {Sol.
  St. Comm.}\ }\textbf {\bibinfo {volume} {71}},\ \bibinfo {pages} {247}
  (\bibinfo {year} {1989}{\natexlab{a}})}\BibitemShut {NoStop}%
\bibitem [{\citenamefont {Tada}\ and\ \citenamefont
  {Ninomiya}(1989{\natexlab{b}})}]{TadaNinomiya2}%
  \BibitemOpen
  \bibfield  {author} {\bibinfo {author} {\bibfnamefont {T.}~\bibnamefont
  {Tada}}\ and\ \bibinfo {author} {\bibfnamefont {T.}~\bibnamefont
  {Ninomiya}},\ }\href@noop {} {\bibfield  {journal} {\bibinfo  {journal} {J.
  Non-Cryst. Sol.}\ }\textbf {\bibinfo {volume} {114}},\ \bibinfo {pages} {88}
  (\bibinfo {year} {1989}{\natexlab{b}})}\BibitemShut {NoStop}%
\bibitem [{\citenamefont {Tada}\ and\ \citenamefont
  {Ninomiya}(1989{\natexlab{c}})}]{TadaNinomiya3}%
  \BibitemOpen
  \bibfield  {author} {\bibinfo {author} {\bibfnamefont {T.}~\bibnamefont
  {Tada}}\ and\ \bibinfo {author} {\bibfnamefont {T.}~\bibnamefont
  {Ninomiya}},\ }\href@noop {} {\bibfield  {journal} {\bibinfo  {journal} {J.
  Non-Cryst. Sol.}\ }\textbf {\bibinfo {volume} {137\&138}},\ \bibinfo {pages}
  {997} (\bibinfo {year} {1989}{\natexlab{c}})}\BibitemShut {NoStop}%
\bibitem [{\citenamefont {Kolomiets}(1981)}]{Kolomiets1981}%
  \BibitemOpen
  \bibfield  {author} {\bibinfo {author} {\bibfnamefont {B.~T.}\ \bibnamefont
  {Kolomiets}},\ }\href@noop {} {\bibfield  {journal} {\bibinfo  {journal} {J.
  Phys. (Paris) C4}\ }\textbf {\bibinfo {volume} {42}},\ \bibinfo {pages} {887}
  (\bibinfo {year} {1981})}\BibitemShut {NoStop}%
\bibitem [{\citenamefont {Mott}(1993)}]{Mott1993}%
  \BibitemOpen
  \bibfield  {author} {\bibinfo {author} {\bibfnamefont {N.~F.}\ \bibnamefont
  {Mott}},\ }\href@noop {} {\emph {\bibinfo {title} {Conduction in
  Non-crystalline Materials}}}\ (\bibinfo  {publisher} {Clarendon Press},\
  \bibinfo {address} {Oxford},\ \bibinfo {year} {1993})\BibitemShut {NoStop}%
\bibitem [{\citenamefont {Anderson}(1975)}]{PWA_negU}%
  \BibitemOpen
  \bibfield  {author} {\bibinfo {author} {\bibfnamefont {P.~W.}\ \bibnamefont
  {Anderson}},\ }\href@noop {} {\bibfield  {journal} {\bibinfo  {journal}
  {Phys. Rev. Lett.}\ }\textbf {\bibinfo {volume} {34}},\ \bibinfo {pages}
  {953} (\bibinfo {year} {1975})}\BibitemShut {NoStop}%
\bibitem [{\citenamefont {Street}\ and\ \citenamefont
  {Mott}(1975)}]{PhysRevLett.35.1293}%
  \BibitemOpen
  \bibfield  {author} {\bibinfo {author} {\bibfnamefont {R.~A.}\ \bibnamefont
  {Street}}\ and\ \bibinfo {author} {\bibfnamefont {N.~F.}\ \bibnamefont
  {Mott}},\ }\href@noop {} {\bibfield  {journal} {\bibinfo  {journal} {Phys.
  Rev. Lett.}\ }\textbf {\bibinfo {volume} {35}},\ \bibinfo {pages} {1293}
  (\bibinfo {year} {1975})}\BibitemShut {NoStop}%
\bibitem [{\citenamefont {Kastner}\ \emph {et~al.}(1976)\citenamefont
  {Kastner}, \citenamefont {Adler},\ and\ \citenamefont {Fritzsche}}]{KAF}%
  \BibitemOpen
  \bibfield  {author} {\bibinfo {author} {\bibfnamefont {M.}~\bibnamefont
  {Kastner}}, \bibinfo {author} {\bibfnamefont {D.}~\bibnamefont {Adler}}, \
  and\ \bibinfo {author} {\bibfnamefont {H.}~\bibnamefont {Fritzsche}},\
  }\href@noop {} {\bibfield  {journal} {\bibinfo  {journal} {Phys. Rev. Lett.}\
  }\textbf {\bibinfo {volume} {37}},\ \bibinfo {pages} {1504} (\bibinfo {year}
  {1976})}\BibitemShut {NoStop}%
\bibitem [{\citenamefont {Vanderbilt}\ and\ \citenamefont
  {Joannopoulos}(1981)}]{PhysRevB.23.2596}%
  \BibitemOpen
  \bibfield  {author} {\bibinfo {author} {\bibfnamefont {D.}~\bibnamefont
  {Vanderbilt}}\ and\ \bibinfo {author} {\bibfnamefont {J.~D.}\ \bibnamefont
  {Joannopoulos}},\ }\href@noop {} {\bibfield  {journal} {\bibinfo  {journal}
  {Phys. Rev. B}\ }\textbf {\bibinfo {volume} {23}},\ \bibinfo {pages} {2596}
  (\bibinfo {year} {1981})}\BibitemShut {NoStop}%
\bibitem [{\citenamefont {Li}\ and\ \citenamefont
  {Drabold}(2000)}]{PhysRevLett.85.2785}%
  \BibitemOpen
  \bibfield  {author} {\bibinfo {author} {\bibfnamefont {J.}~\bibnamefont
  {Li}}\ and\ \bibinfo {author} {\bibfnamefont {D.~A.}\ \bibnamefont
  {Drabold}},\ }\href {\doibase 10.1103/PhysRevLett.85.2785} {\bibfield
  {journal} {\bibinfo  {journal} {Phys. Rev. Lett.}\ }\textbf {\bibinfo
  {volume} {85}},\ \bibinfo {pages} {2785} (\bibinfo {year}
  {2000})}\BibitemShut {NoStop}%
\bibitem [{\citenamefont {Zhugayevych}\ and\ \citenamefont
  {Lubchenko}(2010{\natexlab{b}})}]{ZLMicro2}%
  \BibitemOpen
  \bibfield  {author} {\bibinfo {author} {\bibfnamefont {A.}~\bibnamefont
  {Zhugayevych}}\ and\ \bibinfo {author} {\bibfnamefont {V.}~\bibnamefont
  {Lubchenko}},\ }\href@noop {} {\bibfield  {journal} {\bibinfo  {journal} {J.
  Chem. Phys.}\ }\textbf {\bibinfo {volume} {133}},\ \bibinfo {pages} {234504}
  (\bibinfo {year} {2010}{\natexlab{b}})}\BibitemShut {NoStop}%
\bibitem [{\citenamefont {Pyykk\"{o}}(1997)}]{Pyykko}%
  \BibitemOpen
  \bibfield  {author} {\bibinfo {author} {\bibfnamefont {P.}~\bibnamefont
  {Pyykk\"{o}}},\ }\href@noop {} {\bibfield  {journal} {\bibinfo  {journal}
  {Chem. Rev.}\ }\textbf {\bibinfo {volume} {97}},\ \bibinfo {pages} {597}
  (\bibinfo {year} {1997})}\BibitemShut {NoStop}%
\bibitem [{\citenamefont {Xia}\ and\ \citenamefont {Wolynes}(2000)}]{XW}%
  \BibitemOpen
  \bibfield  {author} {\bibinfo {author} {\bibfnamefont {X.}~\bibnamefont
  {Xia}}\ and\ \bibinfo {author} {\bibfnamefont {P.~G.}\ \bibnamefont
  {Wolynes}},\ }\href@noop {} {\bibfield  {journal} {\bibinfo  {journal} {Proc.
  Natl. Acad. Sci. U.~S.~A.}\ }\textbf {\bibinfo {volume} {97}},\ \bibinfo
  {pages} {2990} (\bibinfo {year} {2000})}\BibitemShut {NoStop}%
\bibitem [{\citenamefont {Lubchenko}\ and\ \citenamefont {Wolynes}(2001)}]{LW}%
  \BibitemOpen
  \bibfield  {author} {\bibinfo {author} {\bibfnamefont {V.}~\bibnamefont
  {Lubchenko}}\ and\ \bibinfo {author} {\bibfnamefont {P.~G.}\ \bibnamefont
  {Wolynes}},\ }\href@noop {} {\bibfield  {journal} {\bibinfo  {journal} {Phys.
  Rev. Lett.}\ }\textbf {\bibinfo {volume} {87}},\ \bibinfo {pages} {195901}
  (\bibinfo {year} {2001})}\BibitemShut {NoStop}%
\bibitem [{\citenamefont {Lubchenko}\ and\ \citenamefont
  {Wolynes}(2007)}]{LW_ARPC}%
  \BibitemOpen
  \bibfield  {author} {\bibinfo {author} {\bibfnamefont {V.}~\bibnamefont
  {Lubchenko}}\ and\ \bibinfo {author} {\bibfnamefont {P.~G.}\ \bibnamefont
  {Wolynes}},\ }\href@noop {} {\bibfield  {journal} {\bibinfo  {journal} {Annu.
  Rev. Phys. Chem.}\ }\textbf {\bibinfo {volume} {58}},\ \bibinfo {pages} {235}
  (\bibinfo {year} {2007})}\BibitemShut {NoStop}%
\bibitem [{\citenamefont {Lubchenko}(2015)}]{L_AP}%
  \BibitemOpen
  \bibfield  {author} {\bibinfo {author} {\bibfnamefont {V.}~\bibnamefont
  {Lubchenko}},\ }\href@noop {} {\bibfield  {journal} {\bibinfo  {journal}
  {Adv. Phys.}\ }\textbf {\bibinfo {volume} {64}},\ \bibinfo {pages} {283}
  (\bibinfo {year} {2015})}\BibitemShut {NoStop}%
\bibitem [{\citenamefont {Lubchenko}(2006)}]{L_Lindemann}%
  \BibitemOpen
  \bibfield  {author} {\bibinfo {author} {\bibfnamefont {V.}~\bibnamefont
  {Lubchenko}},\ }\href@noop {} {\bibfield  {journal} {\bibinfo  {journal} {J.
  Phys. Chem. B}\ }\textbf {\bibinfo {volume} {110}},\ \bibinfo {pages} {18779}
  (\bibinfo {year} {2006})}\BibitemShut {NoStop}%
\bibitem [{\citenamefont {Rabochiy}\ and\ \citenamefont
  {Lubchenko}(2012)}]{RL_Tcr}%
  \BibitemOpen
  \bibfield  {author} {\bibinfo {author} {\bibfnamefont {P.}~\bibnamefont
  {Rabochiy}}\ and\ \bibinfo {author} {\bibfnamefont {V.}~\bibnamefont
  {Lubchenko}},\ }\href@noop {} {\bibfield  {journal} {\bibinfo  {journal} {J.
  Phys. Chem. B}\ }\textbf {\bibinfo {volume} {116}},\ \bibinfo {pages} {5729}
  (\bibinfo {year} {2012})}\BibitemShut {NoStop}%
\bibitem [{\citenamefont {Heeger}\ \emph {et~al.}(1988)\citenamefont {Heeger},
  \citenamefont {Kivelson}, \citenamefont {Schrieffer},\ and\ \citenamefont
  {Su}}]{RevModPhys.60.781}%
  \BibitemOpen
  \bibfield  {author} {\bibinfo {author} {\bibfnamefont {A.~J.}\ \bibnamefont
  {Heeger}}, \bibinfo {author} {\bibfnamefont {S.}~\bibnamefont {Kivelson}},
  \bibinfo {author} {\bibfnamefont {J.~R.}\ \bibnamefont {Schrieffer}}, \ and\
  \bibinfo {author} {\bibfnamefont {W.~P.}\ \bibnamefont {Su}},\ }\href@noop {}
  {\bibfield  {journal} {\bibinfo  {journal} {Rev. Mod. Phys.}\ }\textbf
  {\bibinfo {volume} {60}},\ \bibinfo {pages} {781} (\bibinfo {year}
  {1988})}\BibitemShut {NoStop}%
\bibitem [{\citenamefont {Vanderbilt}\ and\ \citenamefont
  {Joannopoulos}(1980)}]{PhysRevB.22.2927}%
  \BibitemOpen
  \bibfield  {author} {\bibinfo {author} {\bibfnamefont {D.}~\bibnamefont
  {Vanderbilt}}\ and\ \bibinfo {author} {\bibfnamefont {J.~D.}\ \bibnamefont
  {Joannopoulos}},\ }\href@noop {} {\bibfield  {journal} {\bibinfo  {journal}
  {Phys. Rev. B}\ }\textbf {\bibinfo {volume} {22}},\ \bibinfo {pages} {2927}
  (\bibinfo {year} {1980})}\BibitemShut {NoStop}%
\bibitem [{\citenamefont {Zhugayevych}\ and\ \citenamefont
  {Lubchenko}(2010{\natexlab{c}})}]{ZLMicro1}%
  \BibitemOpen
  \bibfield  {author} {\bibinfo {author} {\bibfnamefont {A.}~\bibnamefont
  {Zhugayevych}}\ and\ \bibinfo {author} {\bibfnamefont {V.}~\bibnamefont
  {Lubchenko}},\ }\href@noop {} {\bibfield  {journal} {\bibinfo  {journal} {J.
  Chem. Phys.}\ }\textbf {\bibinfo {volume} {133}},\ \bibinfo {pages} {234503}
  (\bibinfo {year} {2010}{\natexlab{c}})}\BibitemShut {NoStop}%
\bibitem [{\citenamefont {Golden}(2016)}]{GoldenThesis}%
  \BibitemOpen
  \bibfield  {author} {\bibinfo {author} {\bibfnamefont {J.~C.}\ \bibnamefont
  {Golden}},\ }\emph {\bibinfo {title} {SYMMETRY BREAKING IN CHEMICAL
  INTERACTIONS}},\ \href@noop {} {Ph.D. thesis},\ \bibinfo  {school}
  {University of Houston} (\bibinfo {year} {2016})\BibitemShut {NoStop}%
\bibitem [{\citenamefont {Lukyanov}\ and\ \citenamefont
  {Lubchenko}(2017)}]{LL1}%
  \BibitemOpen
  \bibfield  {author} {\bibinfo {author} {\bibfnamefont {A.}~\bibnamefont
  {Lukyanov}}\ and\ \bibinfo {author} {\bibfnamefont {V.}~\bibnamefont
  {Lubchenko}},\ }\href@noop {} {\bibfield  {journal} {\bibinfo  {journal} {J.
  Chem. Phys.}\ }\textbf {\bibinfo {volume} {147}},\ \bibinfo {pages} {114505}
  (\bibinfo {year} {2017})}\BibitemShut {NoStop}%
\bibitem [{\citenamefont {Parisi}\ and\ \citenamefont
  {Zamponi}(2010)}]{RevModPhys.82.789}%
  \BibitemOpen
  \bibfield  {author} {\bibinfo {author} {\bibfnamefont {G.}~\bibnamefont
  {Parisi}}\ and\ \bibinfo {author} {\bibfnamefont {F.}~\bibnamefont
  {Zamponi}},\ }\href@noop {} {\bibfield  {journal} {\bibinfo  {journal} {Rev.
  Mod. Phys.}\ }\textbf {\bibinfo {volume} {82}},\ \bibinfo {pages} {789}
  (\bibinfo {year} {2010})}\BibitemShut {NoStop}%
\bibitem [{\citenamefont {Deschamps}\ \emph {et~al.}(2015)\citenamefont
  {Deschamps}, \citenamefont {Genevois}, \citenamefont {Cui}, \citenamefont
  {Roiland}, \citenamefont {LePoll\`es}, \citenamefont {Furet}, \citenamefont
  {Massiot},\ and\ \citenamefont {Bureau}}]{doi:10.1021/acs.jpcc.5b02423}%
  \BibitemOpen
  \bibfield  {author} {\bibinfo {author} {\bibfnamefont {M.}~\bibnamefont
  {Deschamps}}, \bibinfo {author} {\bibfnamefont {C.}~\bibnamefont {Genevois}},
  \bibinfo {author} {\bibfnamefont {S.}~\bibnamefont {Cui}}, \bibinfo {author}
  {\bibfnamefont {C.}~\bibnamefont {Roiland}}, \bibinfo {author} {\bibfnamefont
  {L.}~\bibnamefont {LePoll\`es}}, \bibinfo {author} {\bibfnamefont
  {E.}~\bibnamefont {Furet}}, \bibinfo {author} {\bibfnamefont
  {D.}~\bibnamefont {Massiot}}, \ and\ \bibinfo {author} {\bibfnamefont
  {B.}~\bibnamefont {Bureau}},\ }\href@noop {} {\bibfield  {journal} {\bibinfo
  {journal} {J. Phys. Chem. B}\ }\textbf {\bibinfo {volume} {119}},\ \bibinfo
  {pages} {11852} (\bibinfo {year} {2015})}\BibitemShut {NoStop}%
\bibitem [{\citenamefont {Elliott}(1991)}]{ElliottNature1991}%
  \BibitemOpen
  \bibfield  {author} {\bibinfo {author} {\bibfnamefont {S.~R.}\ \bibnamefont
  {Elliott}},\ }\href@noop {} {\bibfield  {journal} {\bibinfo  {journal}
  {Nature}\ }\textbf {\bibinfo {volume} {354}},\ \bibinfo {pages} {445}
  (\bibinfo {year} {1991})}\BibitemShut {NoStop}%
\bibitem [{\citenamefont {Salmon}\ \emph {et~al.}(2005)\citenamefont {Salmon},
  \citenamefont {Martin}, \citenamefont {Mason},\ and\ \citenamefont
  {Cuello}}]{Salmon_ZnCl2}%
  \BibitemOpen
  \bibfield  {author} {\bibinfo {author} {\bibfnamefont {P.~S.}\ \bibnamefont
  {Salmon}}, \bibinfo {author} {\bibfnamefont {R.~A.}\ \bibnamefont {Martin}},
  \bibinfo {author} {\bibfnamefont {P.~E.}\ \bibnamefont {Mason}}, \ and\
  \bibinfo {author} {\bibfnamefont {G.~J.}\ \bibnamefont {Cuello}},\
  }\href@noop {} {\bibfield  {journal} {\bibinfo  {journal} {Nature}\ }\textbf
  {\bibinfo {volume} {435}},\ \bibinfo {pages} {75} (\bibinfo {year}
  {2005})}\BibitemShut {NoStop}%
\bibitem [{\citenamefont {Bauchy}\ \emph {et~al.}(2014)\citenamefont {Bauchy},
  \citenamefont {Kachmar},\ and\ \citenamefont {Micoulaut}}]{fpmdMic}%
  \BibitemOpen
  \bibfield  {author} {\bibinfo {author} {\bibfnamefont {M.}~\bibnamefont
  {Bauchy}}, \bibinfo {author} {\bibfnamefont {A.}~\bibnamefont {Kachmar}}, \
  and\ \bibinfo {author} {\bibfnamefont {M.}~\bibnamefont {Micoulaut}},\
  }\href@noop {} {\bibfield  {journal} {\bibinfo  {journal} {J. Chem. Phys.}\
  }\textbf {\bibinfo {volume} {141}},\ \bibinfo {pages} {194506} (\bibinfo
  {year} {2014})}\BibitemShut {NoStop}%
\bibitem [{\citenamefont {Perdew}\ and\ \citenamefont
  {Levy}(1983)}]{bandGap-problem1}%
  \BibitemOpen
  \bibfield  {author} {\bibinfo {author} {\bibfnamefont {J.}~\bibnamefont
  {Perdew}}\ and\ \bibinfo {author} {\bibfnamefont {M.}~\bibnamefont {Levy}},\
  }\href@noop {} {\bibfield  {journal} {\bibinfo  {journal} {Physical Review
  Letters}\ }\textbf {\bibinfo {volume} {51}},\ \bibinfo {pages} {1884}
  (\bibinfo {year} {1983})}\BibitemShut {NoStop}%
\bibitem [{\citenamefont {Sham}\ and\ \citenamefont
  {Schl{\"u}tter}(1983)}]{bandGap-problem2}%
  \BibitemOpen
  \bibfield  {author} {\bibinfo {author} {\bibfnamefont {L.}~\bibnamefont
  {Sham}}\ and\ \bibinfo {author} {\bibfnamefont {M.}~\bibnamefont
  {Schl{\"u}tter}},\ }\href@noop {} {\bibfield  {journal} {\bibinfo  {journal}
  {Physical Review Letters}\ }\textbf {\bibinfo {volume} {51}},\ \bibinfo
  {pages} {1888} (\bibinfo {year} {1983})}\BibitemShut {NoStop}%
\bibitem [{\citenamefont {Brazovskii}\ and\ \citenamefont
  {Kirova}(1981)}]{ISI:A1981MD41000002}%
  \BibitemOpen
  \bibfield  {author} {\bibinfo {author} {\bibfnamefont {S.}~\bibnamefont
  {Brazovskii}}\ and\ \bibinfo {author} {\bibfnamefont {N.}~\bibnamefont
  {Kirova}},\ }\href@noop {} {\bibfield  {journal} {\bibinfo  {journal} {JETP
  Lett.}\ }\textbf {\bibinfo {volume} {33}},\ \bibinfo {pages} {4} (\bibinfo
  {year} {1981})}\BibitemShut {NoStop}%
\bibitem [{\citenamefont {Kresse}\ and\ \citenamefont {Hafner}(1993)}]{vasp1}%
  \BibitemOpen
  \bibfield  {author} {\bibinfo {author} {\bibfnamefont {G.}~\bibnamefont
  {Kresse}}\ and\ \bibinfo {author} {\bibfnamefont {J.}~\bibnamefont
  {Hafner}},\ }\href@noop {} {\bibfield  {journal} {\bibinfo  {journal} {Phys.
  Rev. B}\ }\textbf {\bibinfo {volume} {47}},\ \bibinfo {pages} {558} (\bibinfo
  {year} {1993})}\BibitemShut {NoStop}%
\bibitem [{\citenamefont {Kresse}\ and\ \citenamefont {Hafner}(1994)}]{vasp2}%
  \BibitemOpen
  \bibfield  {author} {\bibinfo {author} {\bibfnamefont {G.}~\bibnamefont
  {Kresse}}\ and\ \bibinfo {author} {\bibfnamefont {J.}~\bibnamefont
  {Hafner}},\ }\href@noop {} {\bibfield  {journal} {\bibinfo  {journal} {Phys.
  Rev. B}\ }\textbf {\bibinfo {volume} {49}},\ \bibinfo {pages} {14251}
  (\bibinfo {year} {1994})}\BibitemShut {NoStop}%
\bibitem [{\citenamefont {Kresse}\ and\ \citenamefont
  {Furthm{\"u}ller}(1996{\natexlab{a}})}]{vasp3}%
  \BibitemOpen
  \bibfield  {author} {\bibinfo {author} {\bibfnamefont {G.}~\bibnamefont
  {Kresse}}\ and\ \bibinfo {author} {\bibfnamefont {J.}~\bibnamefont
  {Furthm{\"u}ller}},\ }\href@noop {} {\bibfield  {journal} {\bibinfo
  {journal} {Comput. Mat. Sci.}\ }\textbf {\bibinfo {volume} {6}},\ \bibinfo
  {pages} {15} (\bibinfo {year} {1996}{\natexlab{a}})}\BibitemShut {NoStop}%
\bibitem [{\citenamefont {Kresse}\ and\ \citenamefont
  {Furthm{\"u}ller}(1996{\natexlab{b}})}]{vasp4}%
  \BibitemOpen
  \bibfield  {author} {\bibinfo {author} {\bibfnamefont {G.}~\bibnamefont
  {Kresse}}\ and\ \bibinfo {author} {\bibfnamefont {J.}~\bibnamefont
  {Furthm{\"u}ller}},\ }\href@noop {} {\bibfield  {journal} {\bibinfo
  {journal} {Phys. Rev. B}\ }\textbf {\bibinfo {volume} {54}},\ \bibinfo
  {pages} {1169} (\bibinfo {year} {1996}{\natexlab{b}})}\BibitemShut {NoStop}%
\bibitem [{\citenamefont {Perdew}\ \emph {et~al.}(1992)\citenamefont {Perdew},
  \citenamefont {Chevary}, \citenamefont {Vosko}, \citenamefont {Jackson},
  \citenamefont {Pederson}, \citenamefont {Singh},\ and\ \citenamefont
  {Fiolhais}}]{vasp-gga-pw91-1}%
  \BibitemOpen
  \bibfield  {author} {\bibinfo {author} {\bibfnamefont {J.~P.}\ \bibnamefont
  {Perdew}}, \bibinfo {author} {\bibfnamefont {J.~A.}\ \bibnamefont {Chevary}},
  \bibinfo {author} {\bibfnamefont {S.~H.}\ \bibnamefont {Vosko}}, \bibinfo
  {author} {\bibfnamefont {K.~A.}\ \bibnamefont {Jackson}}, \bibinfo {author}
  {\bibfnamefont {M.~R.}\ \bibnamefont {Pederson}}, \bibinfo {author}
  {\bibfnamefont {D.~J.}\ \bibnamefont {Singh}}, \ and\ \bibinfo {author}
  {\bibfnamefont {C.}~\bibnamefont {Fiolhais}},\ }\href@noop {} {\bibfield
  {journal} {\bibinfo  {journal} {Phys. Rev. B}\ }\textbf {\bibinfo {volume}
  {46}},\ \bibinfo {pages} {6671} (\bibinfo {year} {1992})}\BibitemShut
  {NoStop}%
\bibitem [{\citenamefont {Perdew}\ \emph {et~al.}(1993)\citenamefont {Perdew},
  \citenamefont {Chevary}, \citenamefont {Vosko}, \citenamefont {Jackson},
  \citenamefont {Pederson}, \citenamefont {Singh},\ and\ \citenamefont
  {Fiolhais}}]{vasp-gga-pw91-2}%
  \BibitemOpen
  \bibfield  {author} {\bibinfo {author} {\bibfnamefont {J.~P.}\ \bibnamefont
  {Perdew}}, \bibinfo {author} {\bibfnamefont {J.~A.}\ \bibnamefont {Chevary}},
  \bibinfo {author} {\bibfnamefont {S.~H.}\ \bibnamefont {Vosko}}, \bibinfo
  {author} {\bibfnamefont {K.~A.}\ \bibnamefont {Jackson}}, \bibinfo {author}
  {\bibfnamefont {M.~R.}\ \bibnamefont {Pederson}}, \bibinfo {author}
  {\bibfnamefont {D.~J.}\ \bibnamefont {Singh}}, \ and\ \bibinfo {author}
  {\bibfnamefont {C.}~\bibnamefont {Fiolhais}},\ }\href@noop {} {\bibfield
  {journal} {\bibinfo  {journal} {Phys. Rev. B}\ }\textbf {\bibinfo {volume}
  {48}},\ \bibinfo {pages} {4978} (\bibinfo {year} {1993})}\BibitemShut
  {NoStop}%
\bibitem [{\citenamefont {Kozyukhin}\ \emph {et~al.}(2011)\citenamefont
  {Kozyukhin}, \citenamefont {Golovchak}, \citenamefont {Kovalskiy},
  \citenamefont {Shpotyuk},\ and\ \citenamefont {Jain}}]{kozyukhin-2011-edos}%
  \BibitemOpen
  \bibfield  {author} {\bibinfo {author} {\bibfnamefont {S.}~\bibnamefont
  {Kozyukhin}}, \bibinfo {author} {\bibfnamefont {R.}~\bibnamefont
  {Golovchak}}, \bibinfo {author} {\bibfnamefont {A.}~\bibnamefont
  {Kovalskiy}}, \bibinfo {author} {\bibfnamefont {O.}~\bibnamefont {Shpotyuk}},
  \ and\ \bibinfo {author} {\bibfnamefont {H.}~\bibnamefont {Jain}},\
  }\href@noop {} {\bibfield  {journal} {\bibinfo  {journal} {Физика и
  техника полупроводников}\ }\textbf {\bibinfo {volume}
  {45}},\ \bibinfo {pages} {433} (\bibinfo {year} {2011})}\BibitemShut
  {NoStop}%
\bibitem [{\citenamefont {Golovchak}\ \emph {et~al.}(2007)\citenamefont
  {Golovchak}, \citenamefont {Kovalskiy}, \citenamefont {Miller}, \citenamefont
  {Jain},\ and\ \citenamefont {Shpotyuk}}]{golovchak-2007-edos}%
  \BibitemOpen
  \bibfield  {author} {\bibinfo {author} {\bibfnamefont {R.}~\bibnamefont
  {Golovchak}}, \bibinfo {author} {\bibfnamefont {A.}~\bibnamefont
  {Kovalskiy}}, \bibinfo {author} {\bibfnamefont {A.~C.}\ \bibnamefont
  {Miller}}, \bibinfo {author} {\bibfnamefont {H.}~\bibnamefont {Jain}}, \ and\
  \bibinfo {author} {\bibfnamefont {O.}~\bibnamefont {Shpotyuk}},\ }\href@noop
  {} {\bibfield  {journal} {\bibinfo  {journal} {Phys. Rev. B}\ }\textbf
  {\bibinfo {volume} {76}},\ \bibinfo {pages} {125208} (\bibinfo {year}
  {2007})}\BibitemShut {NoStop}%
\bibitem [{\citenamefont {Bishop}\ and\ \citenamefont
  {Shevchik}(1975)}]{bishop-1975-edos}%
  \BibitemOpen
  \bibfield  {author} {\bibinfo {author} {\bibfnamefont {S.}~\bibnamefont
  {Bishop}}\ and\ \bibinfo {author} {\bibfnamefont {N.}~\bibnamefont
  {Shevchik}},\ }\href@noop {} {\bibfield  {journal} {\bibinfo  {journal}
  {Phys. Rev. B}\ }\textbf {\bibinfo {volume} {12}},\ \bibinfo {pages} {1567}
  (\bibinfo {year} {1975})}\BibitemShut {NoStop}%
\bibitem [{\citenamefont {Anderson}(1978)}]{RevModPhys.50.191}%
  \BibitemOpen
  \bibfield  {author} {\bibinfo {author} {\bibfnamefont {P.~W.}\ \bibnamefont
  {Anderson}},\ }\href@noop {} {\bibfield  {journal} {\bibinfo  {journal} {Rev.
  Mod. Phys.}\ }\textbf {\bibinfo {volume} {50}},\ \bibinfo {pages} {191}
  (\bibinfo {year} {1978})}\BibitemShut {NoStop}%
\bibitem [{\citenamefont {Seo}\ and\ \citenamefont
  {Hoffmann}(1999)}]{SeoHoffmann1999}%
  \BibitemOpen
  \bibfield  {author} {\bibinfo {author} {\bibfnamefont {D.}~\bibnamefont
  {Seo}}\ and\ \bibinfo {author} {\bibfnamefont {R.}~\bibnamefont {Hoffmann}},\
  }\href@noop {} {\bibfield  {journal} {\bibinfo  {journal} {J. Sol. State
  Chem.}\ }\textbf {\bibinfo {volume} {147}},\ \bibinfo {pages} {26} (\bibinfo
  {year} {1999})}\BibitemShut {NoStop}%
\bibitem [{\citenamefont {Li}\ \emph {et~al.}(2002)\citenamefont {Li},
  \citenamefont {Drabold}, \citenamefont {Krishnaswami}, \citenamefont {Chen},\
  and\ \citenamefont {Jain}}]{Drabold-2002-edos-theory}%
  \BibitemOpen
  \bibfield  {author} {\bibinfo {author} {\bibfnamefont {J.}~\bibnamefont
  {Li}}, \bibinfo {author} {\bibfnamefont {D.}~\bibnamefont {Drabold}},
  \bibinfo {author} {\bibfnamefont {S.}~\bibnamefont {Krishnaswami}}, \bibinfo
  {author} {\bibfnamefont {G.}~\bibnamefont {Chen}}, \ and\ \bibinfo {author}
  {\bibfnamefont {H.}~\bibnamefont {Jain}},\ }\href@noop {} {\bibfield
  {journal} {\bibinfo  {journal} {Phys. Rev. Lett.}\ }\textbf {\bibinfo
  {volume} {88}},\ \bibinfo {pages} {046803} (\bibinfo {year}
  {2002})}\BibitemShut {NoStop}%
\bibitem [{\citenamefont {Slusher}\ \emph {et~al.}(2004)\citenamefont
  {Slusher}, \citenamefont {Lenz}, \citenamefont {Hodelin}, \citenamefont
  {Sanghera}, \citenamefont {Shaw},\ and\ \citenamefont
  {Aggarwal}}]{Slusher-gapAs40Se60}%
  \BibitemOpen
  \bibfield  {author} {\bibinfo {author} {\bibfnamefont {R.~E.}\ \bibnamefont
  {Slusher}}, \bibinfo {author} {\bibfnamefont {G.}~\bibnamefont {Lenz}},
  \bibinfo {author} {\bibfnamefont {J.}~\bibnamefont {Hodelin}}, \bibinfo
  {author} {\bibfnamefont {J.}~\bibnamefont {Sanghera}}, \bibinfo {author}
  {\bibfnamefont {L.~B.}\ \bibnamefont {Shaw}}, \ and\ \bibinfo {author}
  {\bibfnamefont {I.~D.}\ \bibnamefont {Aggarwal}},\ }\href@noop {} {\bibfield
  {journal} {\bibinfo  {journal} {J. Opt. Soc. Am. B}\ }\textbf {\bibinfo
  {volume} {21}},\ \bibinfo {pages} {1146} (\bibinfo {year}
  {2004})}\BibitemShut {NoStop}%
\bibitem [{\citenamefont {Dahshan}\ \emph {et~al.}(2008)\citenamefont
  {Dahshan}, \citenamefont {Amer},\ and\ \citenamefont
  {Aly}}]{Dahshan-gapAs20Se80}%
  \BibitemOpen
  \bibfield  {author} {\bibinfo {author} {\bibfnamefont {A.}~\bibnamefont
  {Dahshan}}, \bibinfo {author} {\bibfnamefont {H.~H.}\ \bibnamefont {Amer}}, \
  and\ \bibinfo {author} {\bibfnamefont {K.~A.}\ \bibnamefont {Aly}},\
  }\href@noop {} {\bibfield  {journal} {\bibinfo  {journal} {Journal of Physics
  D: Applied Physics}\ }\textbf {\bibinfo {volume} {41}},\ \bibinfo {pages}
  {215401} (\bibinfo {year} {2008})}\BibitemShut {NoStop}%
\bibitem [{\citenamefont {Ming-Lei}\ \emph {et~al.}(2014)\citenamefont
  {Ming-Lei}, \citenamefont {Feng}, \citenamefont {Wen-Hou},\ and\
  \citenamefont {Zhi-Yong}}]{Fang-gapAs20-40-50}%
  \BibitemOpen
  \bibfield  {author} {\bibinfo {author} {\bibfnamefont {F.}~\bibnamefont
  {Ming-Lei}}, \bibinfo {author} {\bibfnamefont {X.}~\bibnamefont {Feng}},
  \bibinfo {author} {\bibfnamefont {W.}~\bibnamefont {Wen-Hou}}, \ and\
  \bibinfo {author} {\bibfnamefont {Y.}~\bibnamefont {Zhi-Yong}},\ }\href@noop
  {} {\bibfield  {journal} {\bibinfo  {journal} {Chinese Physics Letters}\
  }\textbf {\bibinfo {volume} {31}},\ \bibinfo {pages} {066101} (\bibinfo
  {year} {2014})}\BibitemShut {NoStop}%
\bibitem [{\citenamefont {Behera}\ \emph
  {et~al.}(2017{\natexlab{a}})\citenamefont {Behera}, \citenamefont {Behera},\
  and\ \citenamefont {Naik}}]{Behera-gapAs40Se60}%
  \BibitemOpen
  \bibfield  {author} {\bibinfo {author} {\bibfnamefont {M.}~\bibnamefont
  {Behera}}, \bibinfo {author} {\bibfnamefont {S.}~\bibnamefont {Behera}}, \
  and\ \bibinfo {author} {\bibfnamefont {R.}~\bibnamefont {Naik}},\ }\href@noop
  {} {\bibfield  {journal} {\bibinfo  {journal} {RSC Adv.}\ }\textbf {\bibinfo
  {volume} {7}},\ \bibinfo {pages} {18428} (\bibinfo {year}
  {2017}{\natexlab{a}})}\BibitemShut {NoStop}%
\bibitem [{\citenamefont {Felty}\ \emph {et~al.}(1967)\citenamefont {Felty},
  \citenamefont {Lucovsky},\ and\ \citenamefont {Myers}}]{FELTY1967555}%
  \BibitemOpen
  \bibfield  {author} {\bibinfo {author} {\bibfnamefont {E.}~\bibnamefont
  {Felty}}, \bibinfo {author} {\bibfnamefont {G.}~\bibnamefont {Lucovsky}}, \
  and\ \bibinfo {author} {\bibfnamefont {M.}~\bibnamefont {Myers}},\
  }\href@noop {} {\bibfield  {journal} {\bibinfo  {journal} {Solid State
  Communications}\ }\textbf {\bibinfo {volume} {5}},\ \bibinfo {pages} {555}
  (\bibinfo {year} {1967})}\BibitemShut {NoStop}%
\bibitem [{\citenamefont {Behera}\ \emph
  {et~al.}(2017{\natexlab{b}})\citenamefont {Behera}, \citenamefont {Naik},
  \citenamefont {Panda},\ and\ \citenamefont {Naik}}]{Behera-gapAs50Se50}%
  \BibitemOpen
  \bibfield  {author} {\bibinfo {author} {\bibfnamefont {M.}~\bibnamefont
  {Behera}}, \bibinfo {author} {\bibfnamefont {P.}~\bibnamefont {Naik}},
  \bibinfo {author} {\bibfnamefont {R.}~\bibnamefont {Panda}}, \ and\ \bibinfo
  {author} {\bibfnamefont {R.}~\bibnamefont {Naik}},\ }\href@noop {} {\bibfield
   {journal} {\bibinfo  {journal} {AIP Conference Proceedings}\ }\textbf
  {\bibinfo {volume} {1832}},\ \bibinfo {pages} {070009} (\bibinfo {year}
  {2017}{\natexlab{b}})}\BibitemShut {NoStop}%
\bibitem [{\citenamefont {N\v{e}mec}\ \emph {et~al.}(2005)\citenamefont
  {N\v{e}mec}, \citenamefont {Jedelsk\'{y}}, \citenamefont {Frumar},
  \citenamefont {\v{S}t\'{a}bl},\ and\ \citenamefont
  {\v{C}erno\v{s}ek}}]{Nemec-gapAs50-60}%
  \BibitemOpen
  \bibfield  {author} {\bibinfo {author} {\bibfnamefont {P.}~\bibnamefont
  {N\v{e}mec}}, \bibinfo {author} {\bibfnamefont {J.}~\bibnamefont
  {Jedelsk\'{y}}}, \bibinfo {author} {\bibfnamefont {M.}~\bibnamefont
  {Frumar}}, \bibinfo {author} {\bibfnamefont {M.}~\bibnamefont
  {\v{S}t\'{a}bl}}, \ and\ \bibinfo {author} {\bibfnamefont {Z.}~\bibnamefont
  {\v{C}erno\v{s}ek}},\ }\href@noop {} {\bibfield  {journal} {\bibinfo
  {journal} {Thin Solid Films}\ }\textbf {\bibinfo {volume} {484}},\ \bibinfo
  {pages} {140} (\bibinfo {year} {2005})}\BibitemShut {NoStop}%
\bibitem [{\citenamefont {Tauc}(1974)}]{Tauc1974}%
  \BibitemOpen
  \bibfield  {author} {\bibinfo {author} {\bibfnamefont {J.}~\bibnamefont
  {Tauc}},\ }\enquote {\bibinfo {title} {Optical properties of amorphous
  semiconductors},}\ in\ \href@noop {} {\emph {\bibinfo {booktitle} {{Amorphous
  and Liquid Semiconductors}}}},\ \bibinfo {editor} {edited by\ \bibinfo
  {editor} {\bibfnamefont {J.}~\bibnamefont {Tauc}}}\ (\bibinfo {year} {1974})\
  pp.\ \bibinfo {pages} {159--220}\BibitemShut {NoStop}%
\bibitem [{\citenamefont {Kittel}(1956)}]{Kittel}%
  \BibitemOpen
  \bibfield  {author} {\bibinfo {author} {\bibfnamefont {C.}~\bibnamefont
  {Kittel}},\ }\href@noop {} {\emph {\bibinfo {title} {Introduction to Solid
  State Physics}}}\ (\bibinfo  {publisher} {John Wiley \& Sons, {\it Inc.}},\
  \bibinfo {year} {1956})\BibitemShut {NoStop}%
\bibitem [{\citenamefont {Street}\ \emph {et~al.}(1978)\citenamefont {Street},
  \citenamefont {Nemanich},\ and\ \citenamefont
  {Connell}}]{BandGap-composition}%
  \BibitemOpen
  \bibfield  {author} {\bibinfo {author} {\bibfnamefont {R.~A.}\ \bibnamefont
  {Street}}, \bibinfo {author} {\bibfnamefont {R.~J.}\ \bibnamefont
  {Nemanich}}, \ and\ \bibinfo {author} {\bibfnamefont {G.~A.~N.}\ \bibnamefont
  {Connell}},\ }\href@noop {} {\bibfield  {journal} {\bibinfo  {journal} {Phys.
  Rev. B}\ }\textbf {\bibinfo {volume} {18}},\ \bibinfo {pages} {6915}
  (\bibinfo {year} {1978})}\BibitemShut {NoStop}%
\bibitem [{\citenamefont {Pfeiffer}\ \emph {et~al.}(1991)\citenamefont
  {Pfeiffer}, \citenamefont {Paesler},\ and\ \citenamefont
  {Agarwal}}]{Pfeiffer}%
  \BibitemOpen
  \bibfield  {author} {\bibinfo {author} {\bibfnamefont {G.}~\bibnamefont
  {Pfeiffer}}, \bibinfo {author} {\bibfnamefont {M.~A.}\ \bibnamefont
  {Paesler}}, \ and\ \bibinfo {author} {\bibfnamefont {S.~C.}\ \bibnamefont
  {Agarwal}},\ }\href@noop {} {\bibfield  {journal} {\bibinfo  {journal} {J.
  Non-Cryst. Sol.}\ }\textbf {\bibinfo {volume} {130}},\ \bibinfo {pages} {111}
  (\bibinfo {year} {1991})}\BibitemShut {NoStop}%
\bibitem [{\citenamefont {Harea}\ \emph {et~al.}(2003)\citenamefont {Harea},
  \citenamefont {Vasilev}, \citenamefont {Colomeico},\ and\ \citenamefont
  {Iovu}}]{Harea-2003-urbach-energy}%
  \BibitemOpen
  \bibfield  {author} {\bibinfo {author} {\bibfnamefont {D.~V.}\ \bibnamefont
  {Harea}}, \bibinfo {author} {\bibfnamefont {I.~A.}\ \bibnamefont {Vasilev}},
  \bibinfo {author} {\bibfnamefont {E.~P.}\ \bibnamefont {Colomeico}}, \ and\
  \bibinfo {author} {\bibfnamefont {M.~S.}\ \bibnamefont {Iovu}},\ }\href@noop
  {} {\bibfield  {journal} {\bibinfo  {journal} {J. Optoelectron. Adv. M.}\
  }\textbf {\bibinfo {volume} {5}},\ \bibinfo {pages} {1115} (\bibinfo {year}
  {2003})}\BibitemShut {NoStop}%
\bibitem [{\citenamefont {Rice}\ and\ \citenamefont
  {Mele}(1982)}]{PhysRevLett.49.1455}%
  \BibitemOpen
  \bibfield  {author} {\bibinfo {author} {\bibfnamefont {M.~J.}\ \bibnamefont
  {Rice}}\ and\ \bibinfo {author} {\bibfnamefont {E.~J.}\ \bibnamefont
  {Mele}},\ }\href {\doibase 10.1103/PhysRevLett.49.1455} {\bibfield  {journal}
  {\bibinfo  {journal} {Phys. Rev. Lett.}\ }\textbf {\bibinfo {volume} {49}},\
  \bibinfo {pages} {1455} (\bibinfo {year} {1982})}\BibitemShut {NoStop}%
\bibitem [{\citenamefont {Pearson}(1895)}]{Pearson1895-corrcoef}%
  \BibitemOpen
  \bibfield  {author} {\bibinfo {author} {\bibfnamefont {K.}~\bibnamefont
  {Pearson}},\ }\href@noop {} {\bibfield  {journal} {\bibinfo  {journal}
  {Proceedings of the Royal Society of London}\ }\textbf {\bibinfo {volume}
  {58}},\ \bibinfo {pages} {240} (\bibinfo {year} {1895})}\BibitemShut
  {NoStop}%
\bibitem [{\citenamefont {Takayama}\ \emph {et~al.}(1980)\citenamefont
  {Takayama}, \citenamefont {Lin-Liu},\ and\ \citenamefont
  {Maki}}]{PhysRevB.21.2388}%
  \BibitemOpen
  \bibfield  {author} {\bibinfo {author} {\bibfnamefont {H.}~\bibnamefont
  {Takayama}}, \bibinfo {author} {\bibfnamefont {Y.~R.}\ \bibnamefont
  {Lin-Liu}}, \ and\ \bibinfo {author} {\bibfnamefont {K.}~\bibnamefont
  {Maki}},\ }\href {\doibase 10.1103/PhysRevB.21.2388} {\bibfield  {journal}
  {\bibinfo  {journal} {Phys. Rev. B}\ }\textbf {\bibinfo {volume} {21}},\
  \bibinfo {pages} {2388} (\bibinfo {year} {1980})}\BibitemShut {NoStop}%
\bibitem [{\citenamefont {Stewart}()}]{mopac}%
  \BibitemOpen
  \bibfield  {author} {\bibinfo {author} {\bibfnamefont {J.~J.~P.}\
  \bibnamefont {Stewart}},\ }\href@noop {} {\enquote {\bibinfo {title}
  {{Stewart Computational Chemistry, Colorado Springs, CO, USA}},}\ }\bibinfo
  {howpublished} {HTTP://OpenMOPAC.net}\BibitemShut {NoStop}%
\end{thebibliography}%

\clearpage

\setcounter{section}{0}
\setcounter{equation}{0}
\setcounter{figure}{0}
\setcounter{table}{0}
\setcounter{page}{1}
\makeatletter
\renewcommand{\theequation}{S\arabic{equation}}
\renewcommand{\thefigure}{S\arabic{figure}}
\renewcommand{\thetable}{S\arabic{table}}
\renewcommand{\bibnumfmt}[1]{[S#1]}

\begin{widetext}

  \begin{center} {\bf \large {\em Supplemental Material}\/: Structural
      origin of the midgap electronic states and the Urbach tail in
      pnictogen-chalcogenide glasses} \medskip

    {\large Alexey Lukyanov$^{1}$, Jon C. Golden$^{1}$, and Vassiliy
      Lubchenko$^{1, 2}$} \medskip

    {\normalsize $^1$Department of Chemistry, University of Houston,
      Houston, TX 77204-5003

     $^2$Department of Physics, University of Houston, Houston, TX
     77204-5005}

  \end{center}
  
\end{widetext}

\section{On the choice of the quantum-chemical approximation}

The band gaps produced by a number of distinct approximations are
listed in Table~\ref{band-gaps-for-different-functionals} for several
optimized structures for the stoichiometric compound
As$_{40}$Se$_{60}$. The structures differ by the amount of vacancies
that must be introduced in the parent structure to achieve the desired
stoichiometry. The amount of vacancies also affects the mobility of
the atoms during the optimization.  Values $A < 0.1$ imply vacancies
must be introduced in the chalcogen sites, $A > 1$ at the pnictogen
sites.  We found in Ref.~\cite{LL1} that the optimized structures are
not overly sensitive to the value of $A$.  Likewise, we find here that
the band gap values are also quite robust.

  \begin{table}
    \begin{tabular}{|c|c|c|}
      \hline
      Functional         & Method 1 (DOS$^2$) & Method 2 (STDev) \\ \hline
      B3LYP              & 1.94               & 1.82     \\ 
      B3PW91             & 1.94               & 1.80    \\
      HSE03              & 1.34               & 1.24    \\
      HSE06              & 1.49               & 1.40    \\
      PW91               & 0.85               & 0.84         \\ 
      PBE                & 0.87               & 0.85     \\ \hline
      Experiment         & \multicolumn{2}{|c|}{1.78\cite{Slusher-gapAs40Se60},
                                                1.76\cite{Behera-gapAs40Se60},
                                                1.74\cite{FELTY1967555},
                                                1.64\cite{Fang-gapAs20-40-50}} \\
      \hline
    \end{tabular}
    \caption{\label{band-gaps-for-different-functionals} Table of the band gaps
      calculated using two different techniques: in Method 1 we assume that 
      conduction and valence bands have the functional forms $\propto \sqrt{E-E_c}$ 
      and $\propto \sqrt{E_v-E}$ respectively; in Method 2 we calculate
      a gap between the second energy levels within the valence and conduction 
      bands counting away from the gap. These methods were applied to the
      As$_{40}$Se$_{60}$ compound, parent structures were generated at 
      $A=0.1$, the mobility gap was estimated based for a density of states
      averaged over a number of samples. (The averaging is over 10  samples
      for B3LYP and PW91, over 5 samples for the rest of the approximations.)
    }
  \end{table}

  According to Table~\ref{band-gaps-for-different-functionals}, the
  band gaps obtained with B3LYP and B3PW91 hybrid functionals are
  significantly more consistent with observation. Because of this
  circumstance and the availability of the B3LYP functional in the
  default distribution of VASP, we have chosen B3LYP for the rest of
  the calculations. We have tested for the effects of varying the
  plane wave energy cut-off, within the range 200-400~eV, on the
  quality of the spectra. Larger values provide for better accuracy
  but incur greater computational cost. We have found that the value
  300~eV provides adequate accuracy in the full spectral range of the
  DOS, as assessed using the Bivariate (Pearson)
  correlation~\cite{Pearson1895-corrcoef} for each pair of spectra.
  All further simulations were performed at the $\Gamma$ point in the
  first Brillouin zone with the energy cut-off set to 300 eV and the
  threshold area parameter $A=0.100$ unless specifically noted
  otherwise.

\begin{figure}[H]
  \centering
  \includegraphics[width=.8 \figurewidth]{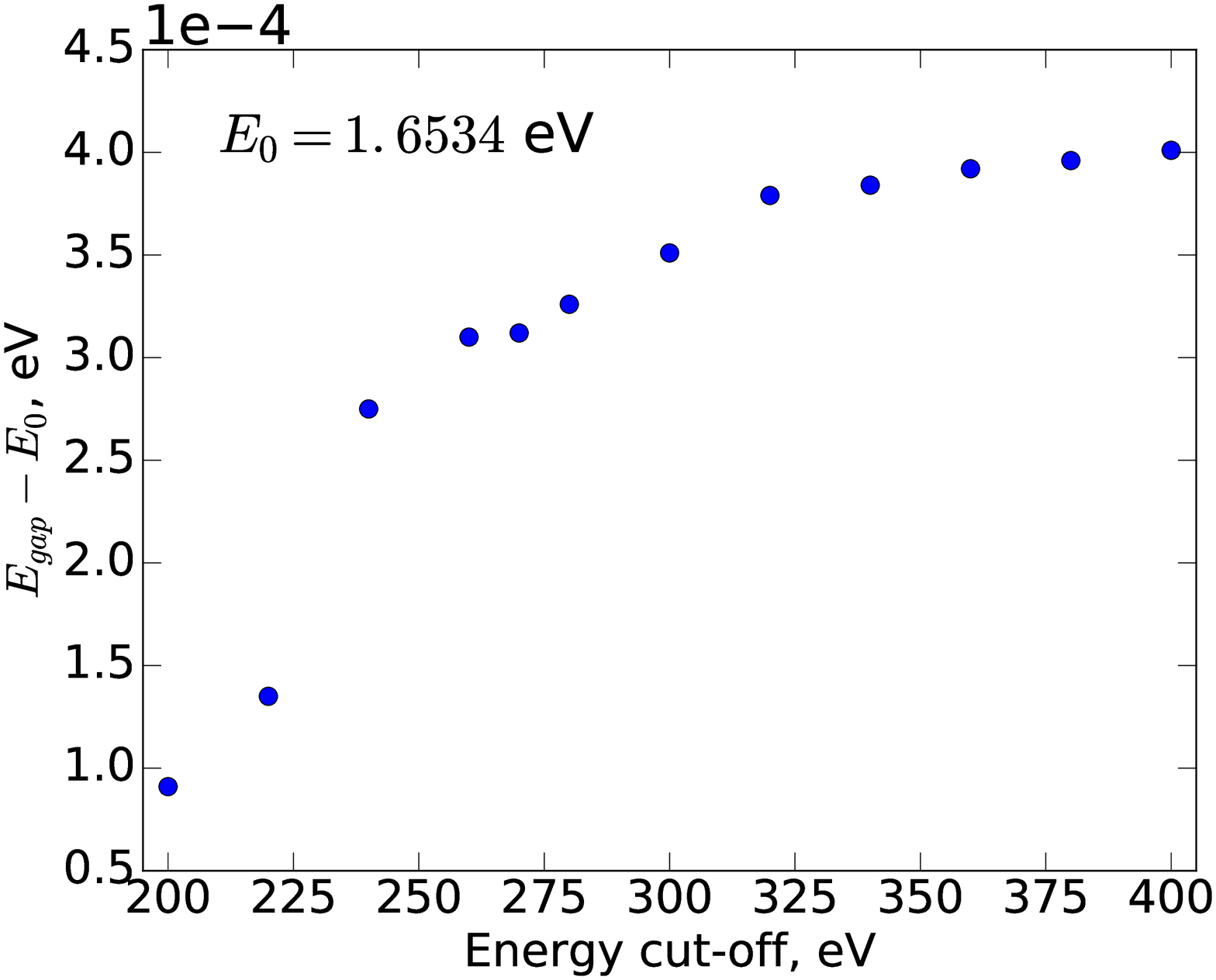}
  \caption{\label{encut-var-band-gap} Relative change in the band gap
    value with variation of the cutoff energy. Simulations performed
    on As$_{40}$Se$_{60}$ samples using the B3LYP hybrid functional
    for the electron exchange-correlation energy part.  }
\end{figure}

\begin{figure}[H]
  \centering
  \includegraphics[width=.8 \figurewidth]{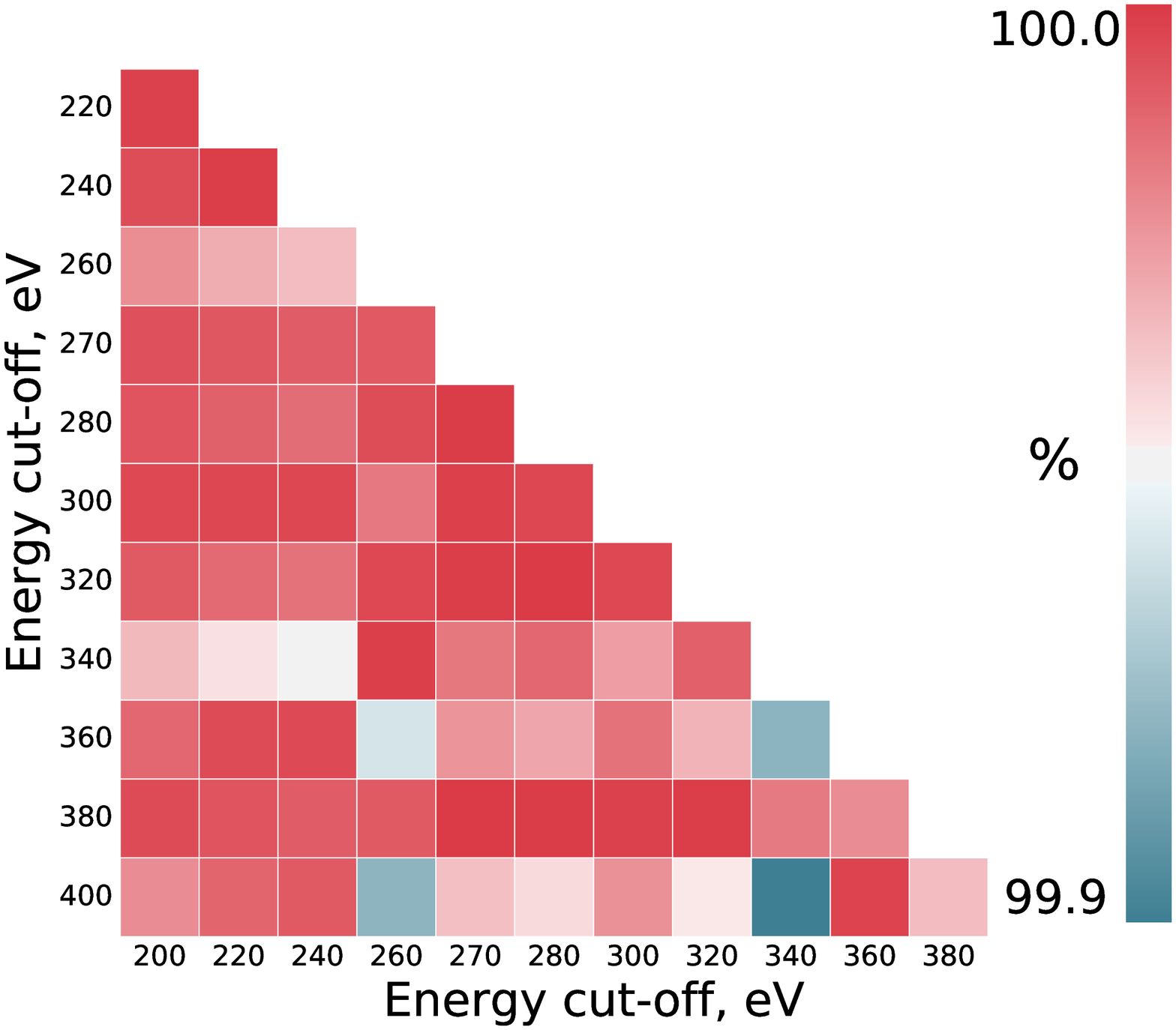}
  \caption{\label{encut-var-corr-map} Bivariate (Pearson) correlation
    map of the electronic density of states (eDOS) for the simulations
    performed for different values of the energy cut-off on an
    As$_{40}$Se$_{60}$ sample (B3LYP).  }
\end{figure}

Within chosen cut-off energy limits the band gap changes within
0.02\%, a negligibly small value (Fig.~\ref{encut-var-band-gap}).  The
densities of states were compared pairwise. For each pair the
Pearson's correlation coefficient\cite{Pearson1895-corrcoef} was
calculated. In Fig.~\ref{encut-var-corr-map} the correlation matrix
demonstrates a correlation over 99\% between different simulations,
based on which we can make a conclusion that eDOS is not overly
sensitive to varying the energy cut-off.

\section{Individual electronic spectra for other stoichiometries}

\begin{figure}[H]
  \centering
  \includegraphics[width=.9\figurewidth]{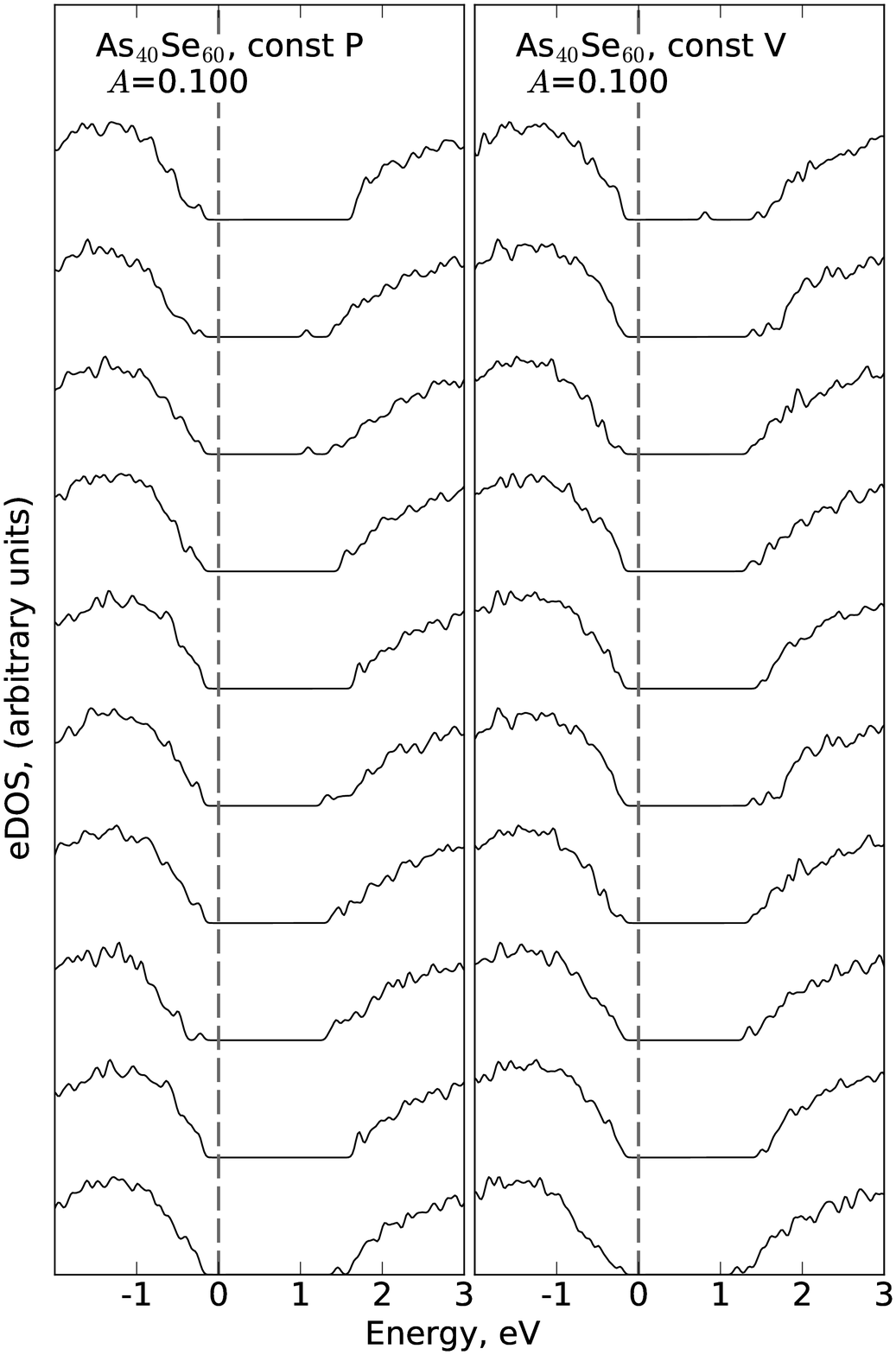}
  \caption{\label{band-gaps-As40Se60} Electronic density of states for
    five distinct amorphous samples of the compound
    As$_{40}$Se$_{60}$, optimized at constant pressure and volume,
    respectively. C.f.  Fig.~\ref{band-gaps-As20Se80} of the main
    text.}
\end{figure}

\begin{figure}[H]
  \centering
  \includegraphics[width=.9\figurewidth]{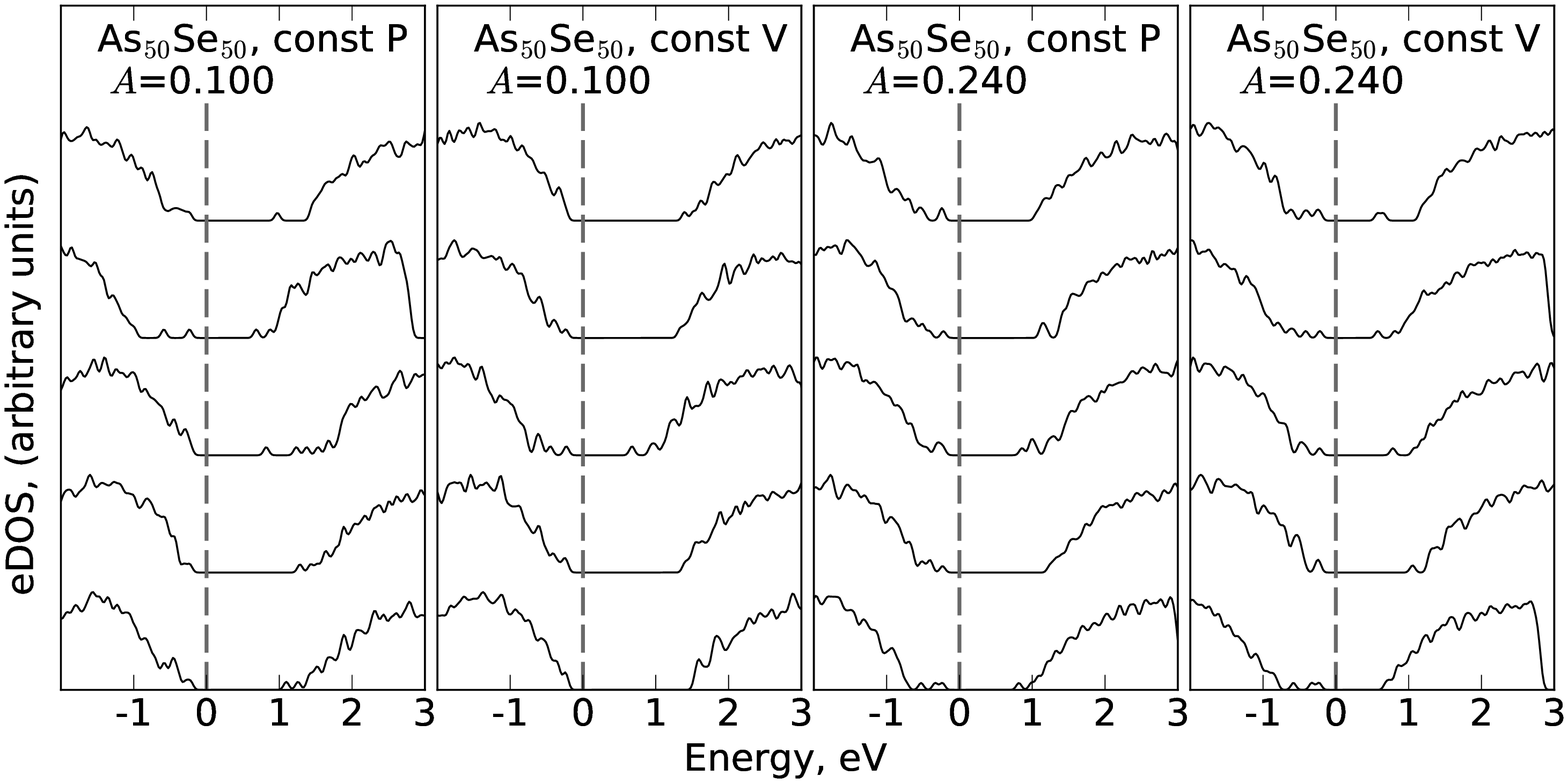}
  \caption{\label{band-gaps-As50Se50} Electronic density of states for
    five distinct amorphous samples of the compound for
    As$_{50}$Se$_{50}$ and two distinct values of the parameter $A$,
    optimized at constant pressure and volume, respectively. C.f.
    Fig.~\ref{band-gaps-As20Se80} of the main text and
    Fig.~\ref{band-gaps-As40Se60}.}
\end{figure}

\section{Comparison with electronic spectra obtained in earlier
  studies}

\begin{figure}[H]
  \centering
  \includegraphics[width=1\figurewidth]{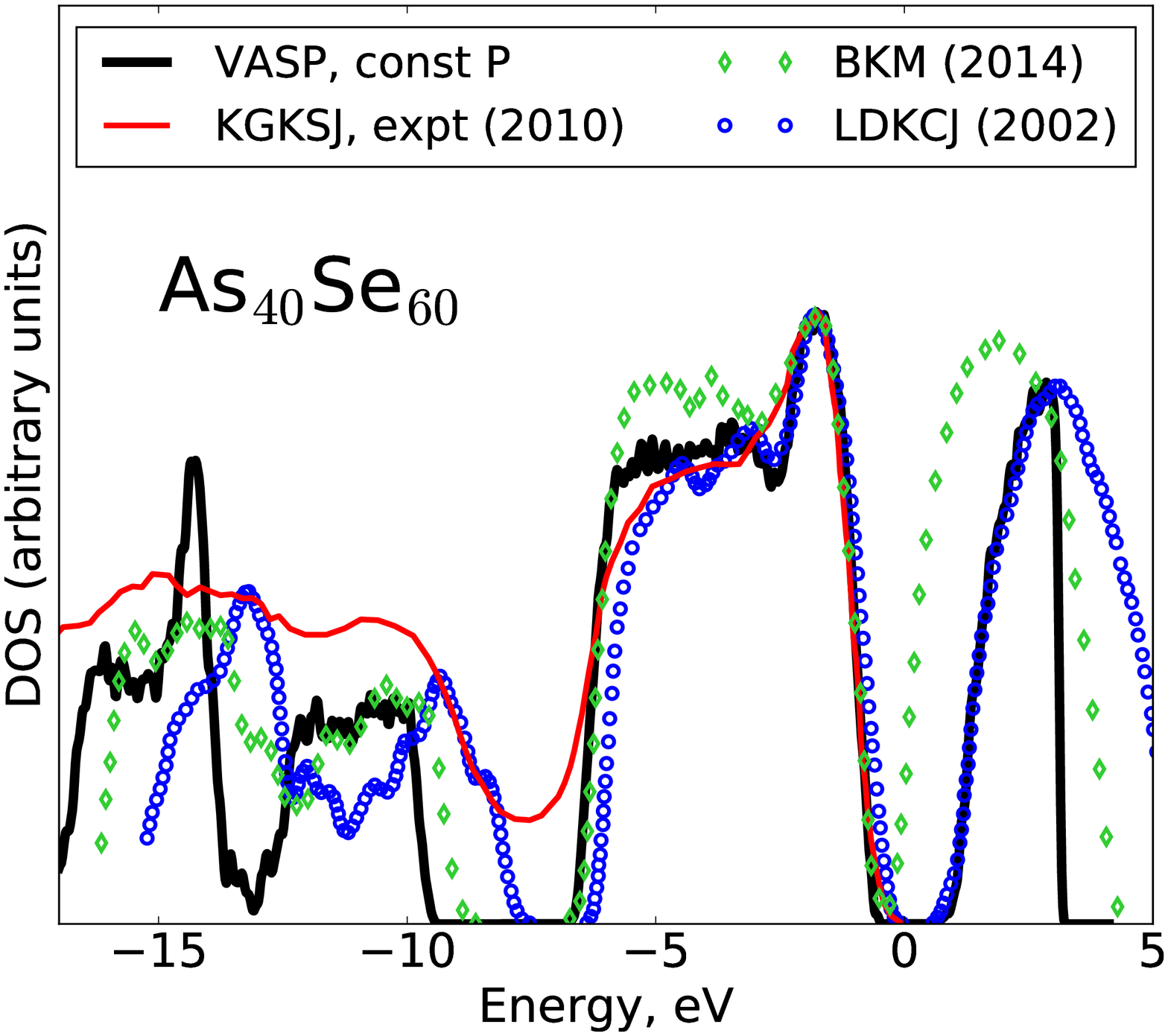}
  \caption{\label{compEarlierTheories} Predicted total electronic
    density of states (eDOS) for the As$_{40}$Se$_{60}$ glass,
    const-$P$, compared with predictions due to Bauchy et
    al. (BKM).~\cite{fpmdMic} Experimental eDOS inferred from X-ray
    photo-emission spectroscopy studies due to Kozyukhin {\em at al.}
    (KGKSJ)\cite{kozyukhin-2011-edos}. }
\end{figure}

\begin{figure}[H]
  \centering
  \includegraphics[width=1\figurewidth]{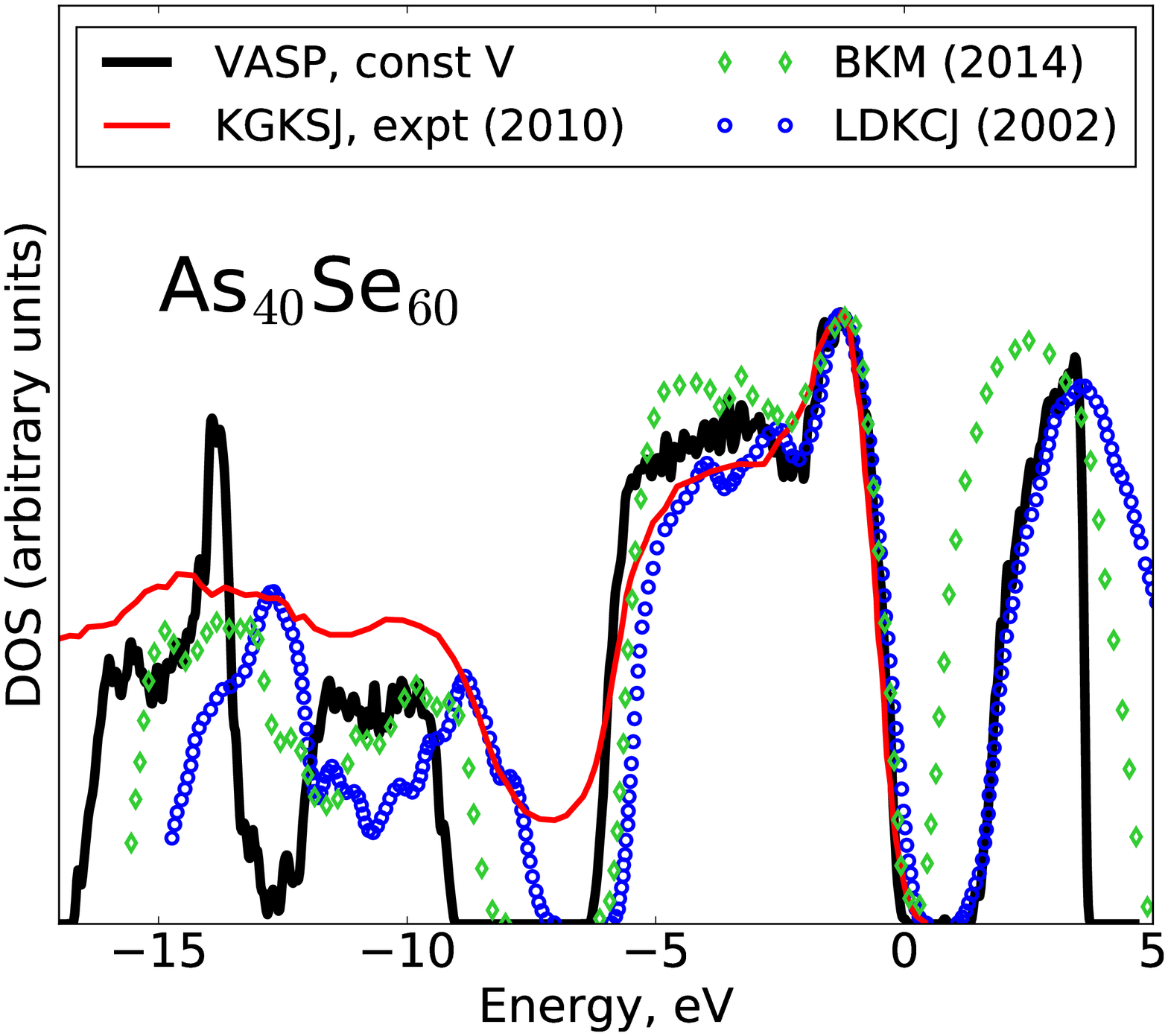}
  \caption{\label{compEarlierTheories1} Same as
    Fig.~\ref{compEarlierTheories} but at constant volume. }
\end{figure}

\begin{figure}[H]
  \centering
  \includegraphics[width=1\figurewidth]{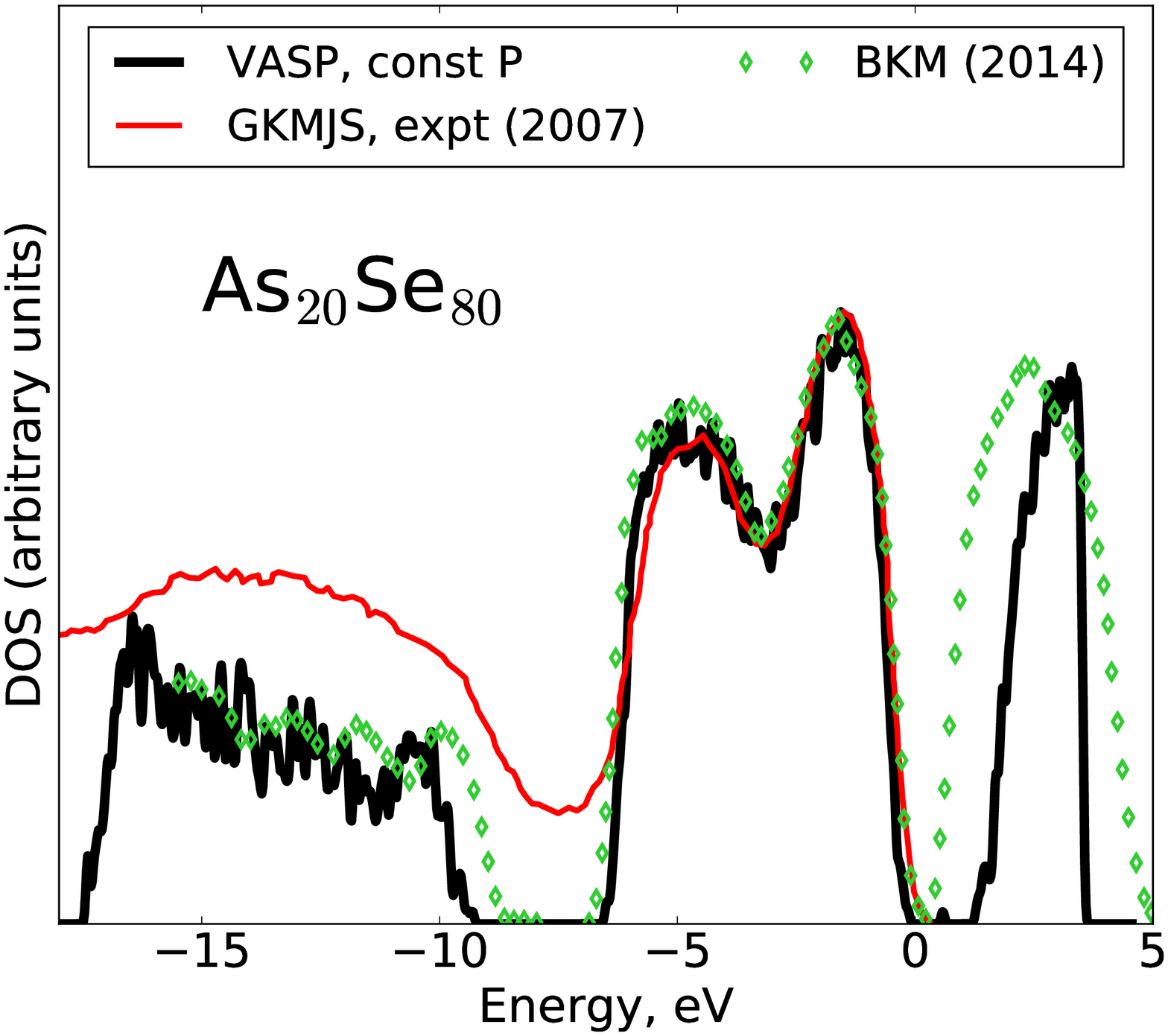}
  \caption{\label{compEarlierTheories2} Predicted total electronic
    density of states (eDOS) for the As$_{20}$Se$_{80}$ glass,
    const-$P$, compared with predictions due to Bauchy et
    al. (BKM).~\cite{fpmdMic} Experimental eDOS due to Golovchak {\em
      et al.}  (GKMJS).~\cite{golovchak-2007-edos} }
\end{figure}

\begin{figure}[H]
  \centering
  \includegraphics[width=1\figurewidth]{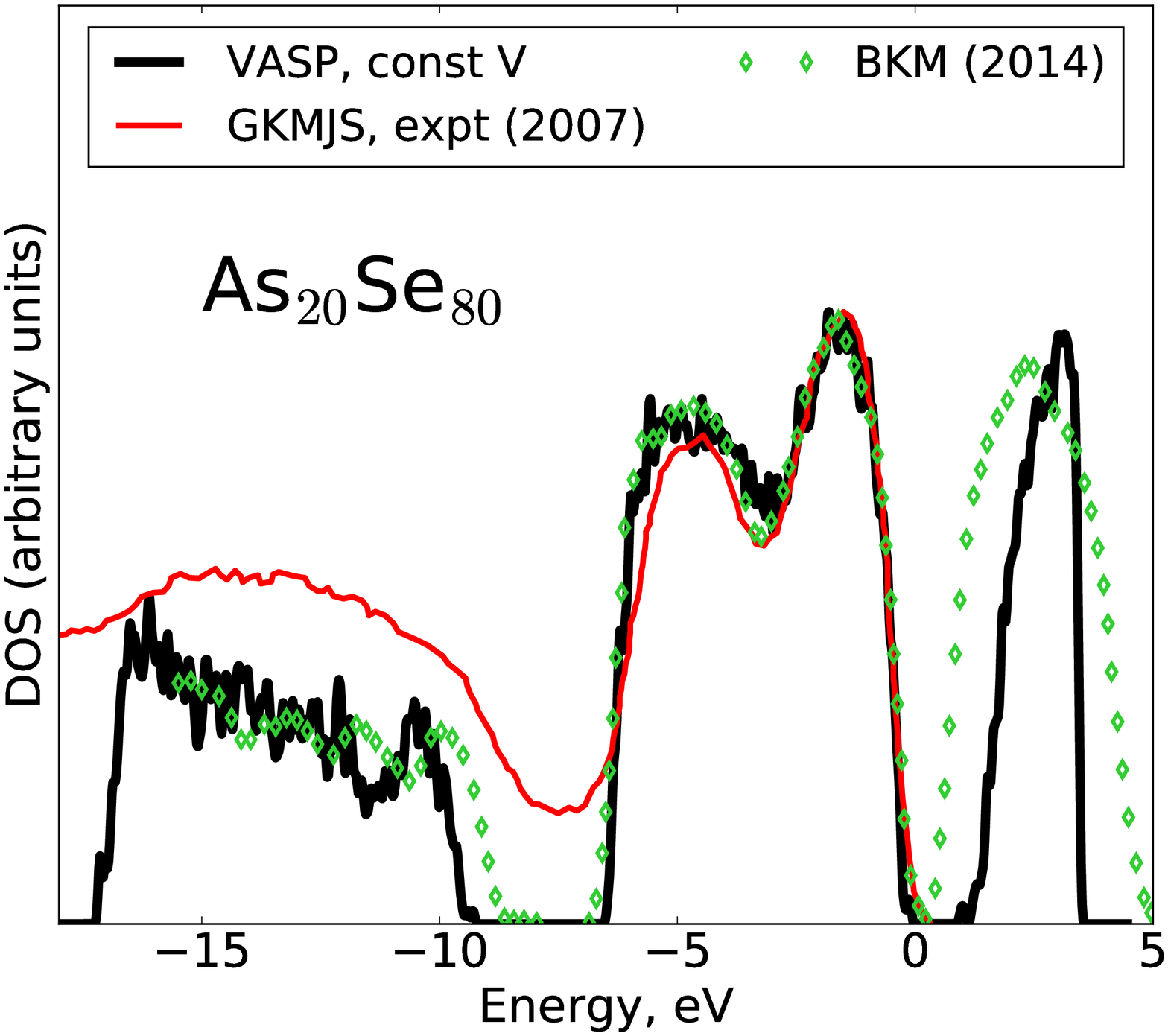}
  \caption{\label{compEarlierTheories3} Same as
    Fig.~\ref{compEarlierTheories2} but at constant volume. }
\end{figure}

\begin{figure}[H]
  \centering
  \includegraphics[width=1\figurewidth]{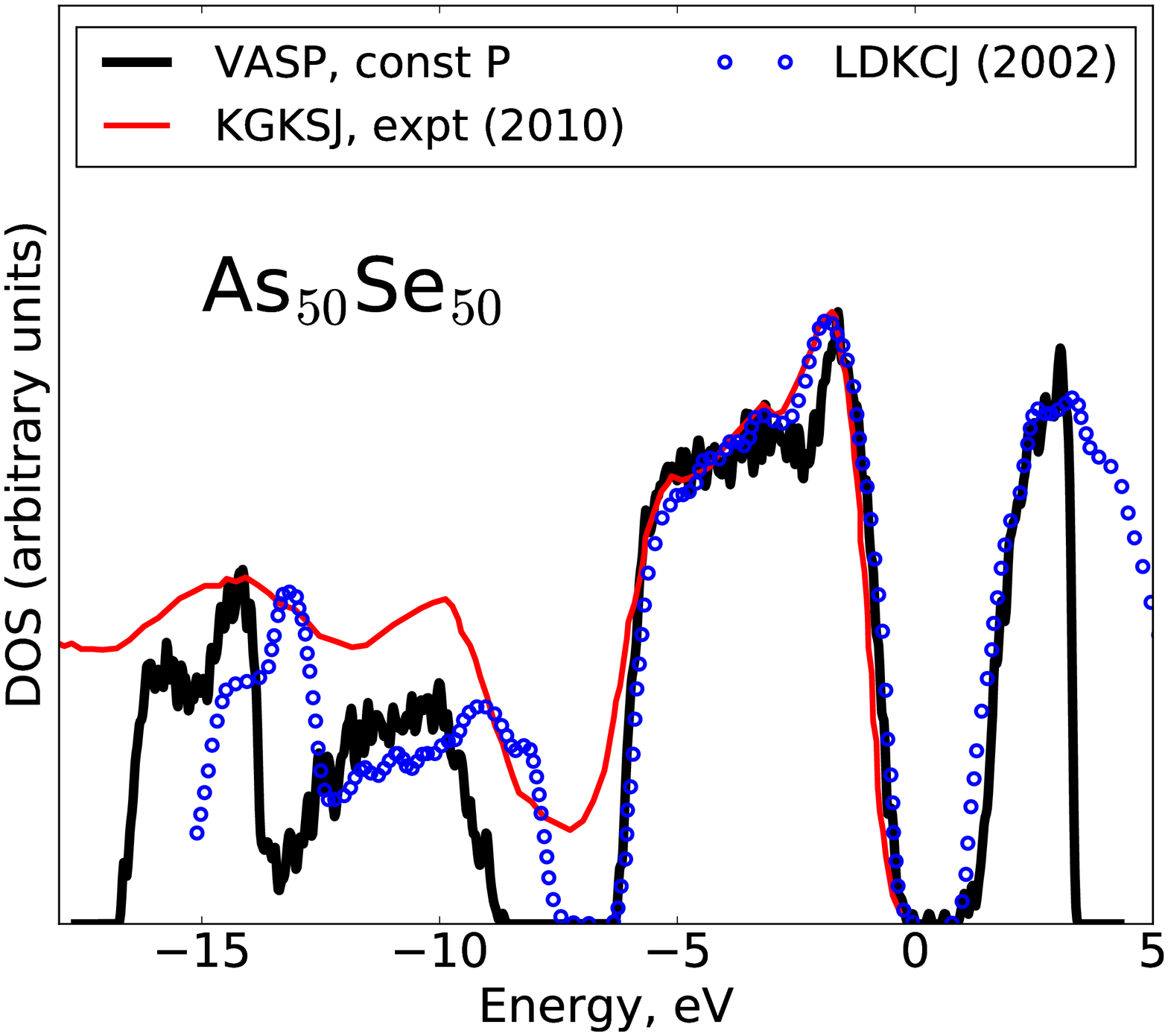}
  \caption{\label{compEarlierTheories4} Predicted total electronic
    density of states (eDOS) for the As$_{50}$Se$_{50}$ glass,
    const-$P$, compared with predictions due to Li et
    al. (LDKCJ).~\cite{fpmdMic} Experimental eDOS inferred from X-ray
    photo-emission spectroscopy studies by Kozyukhin {\em at al.}
    (KGKSJ).\cite{kozyukhin-2011-edos} }
\end{figure}

\begin{figure}[H]
  \centering
  \includegraphics[width=1\figurewidth]{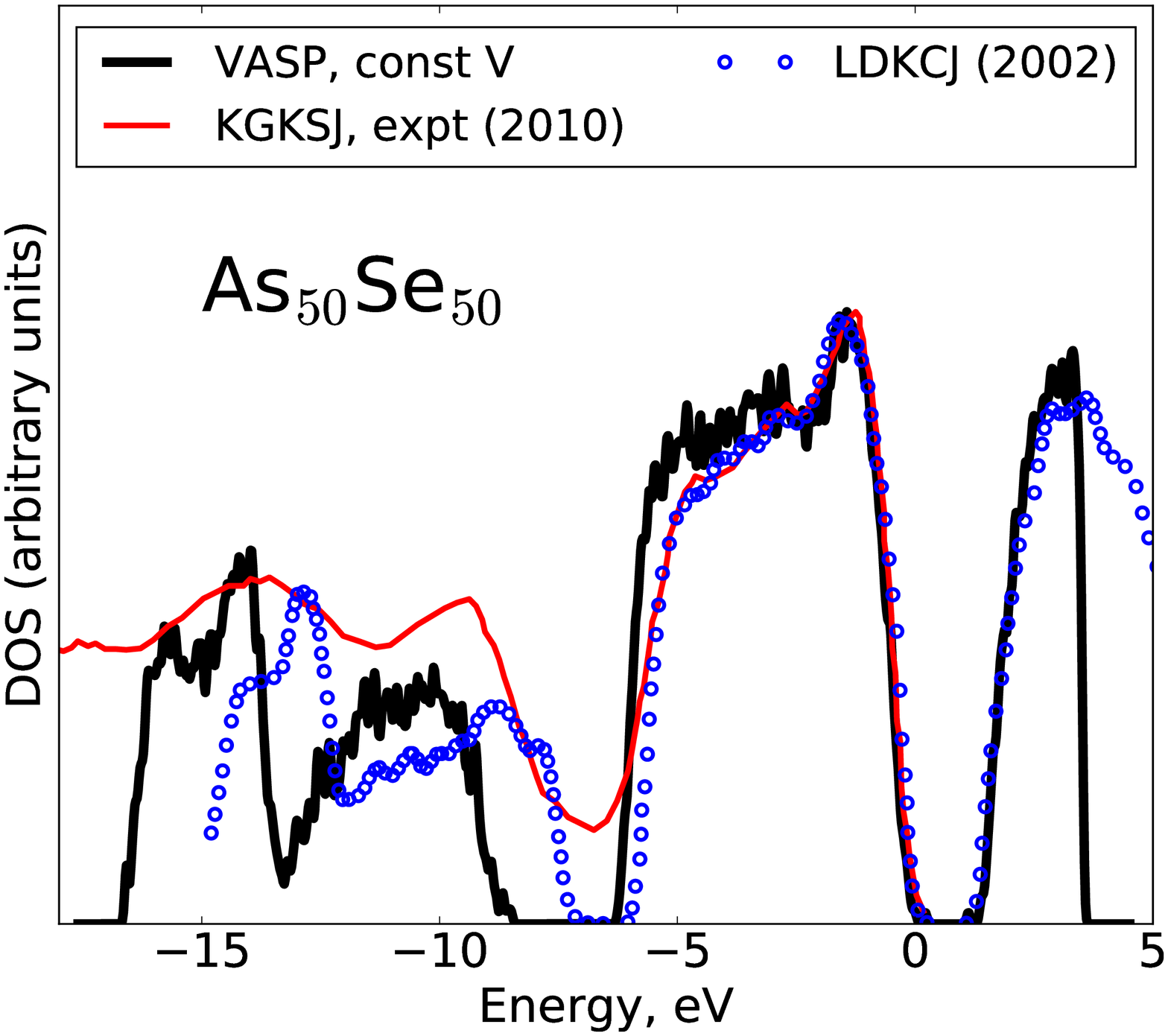}
  \caption{\label{compEarlierTheories5} Same as
    Fig.~\ref{compEarlierTheories4} but at constant volume. }
\end{figure}

\section{Statistics of energy levels}

\begin{figure}[H]
  \centering
  \includegraphics[width=0.7 \figurewidth]{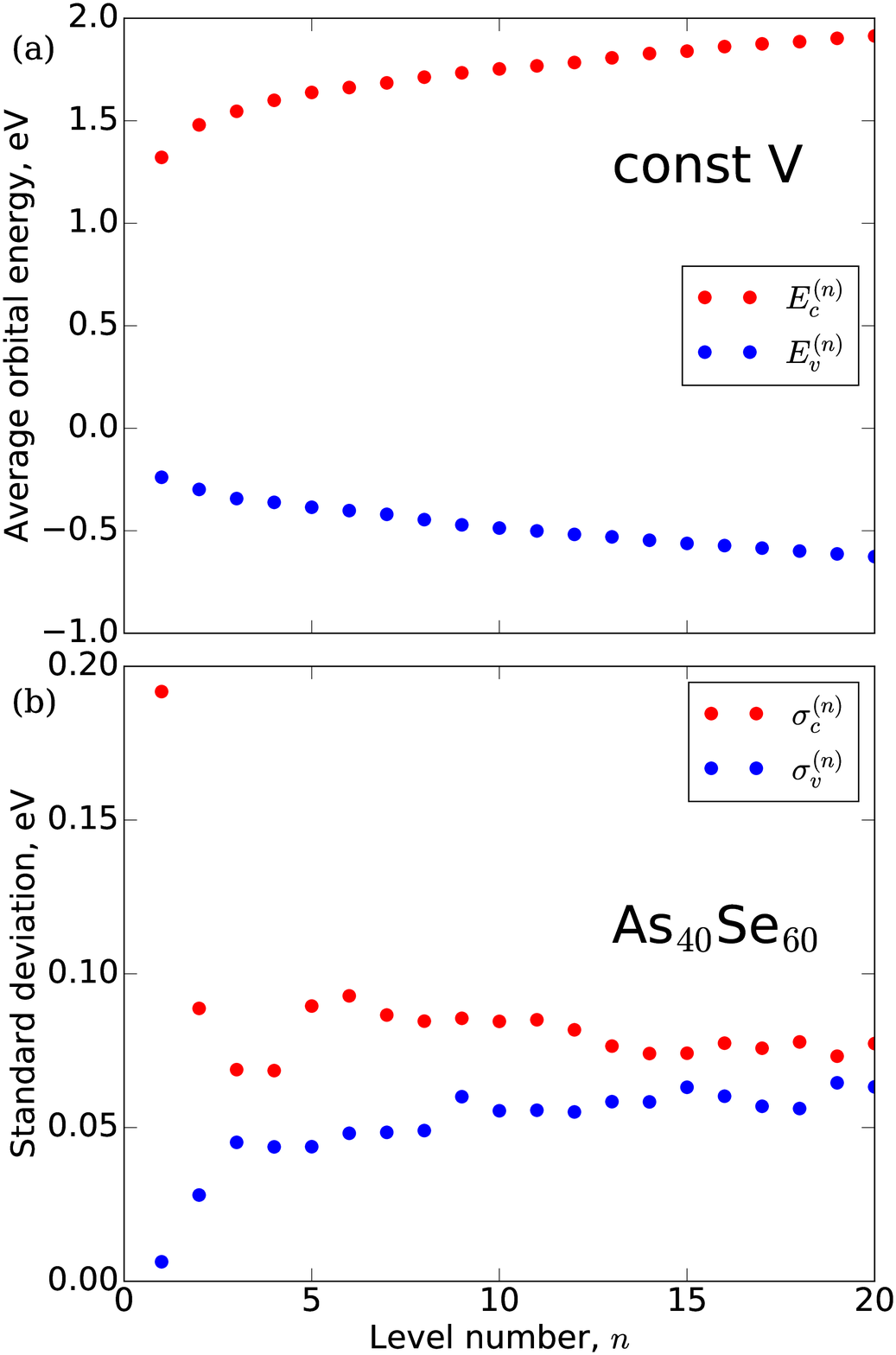}
  \caption{\label{EnSigma1} Same as Fig.~\ref{EnSigma} of the main
    text, but at constant volume. }
\end{figure}

\begin{figure}[H]
  \centering
  \includegraphics[width=0.7 \figurewidth]{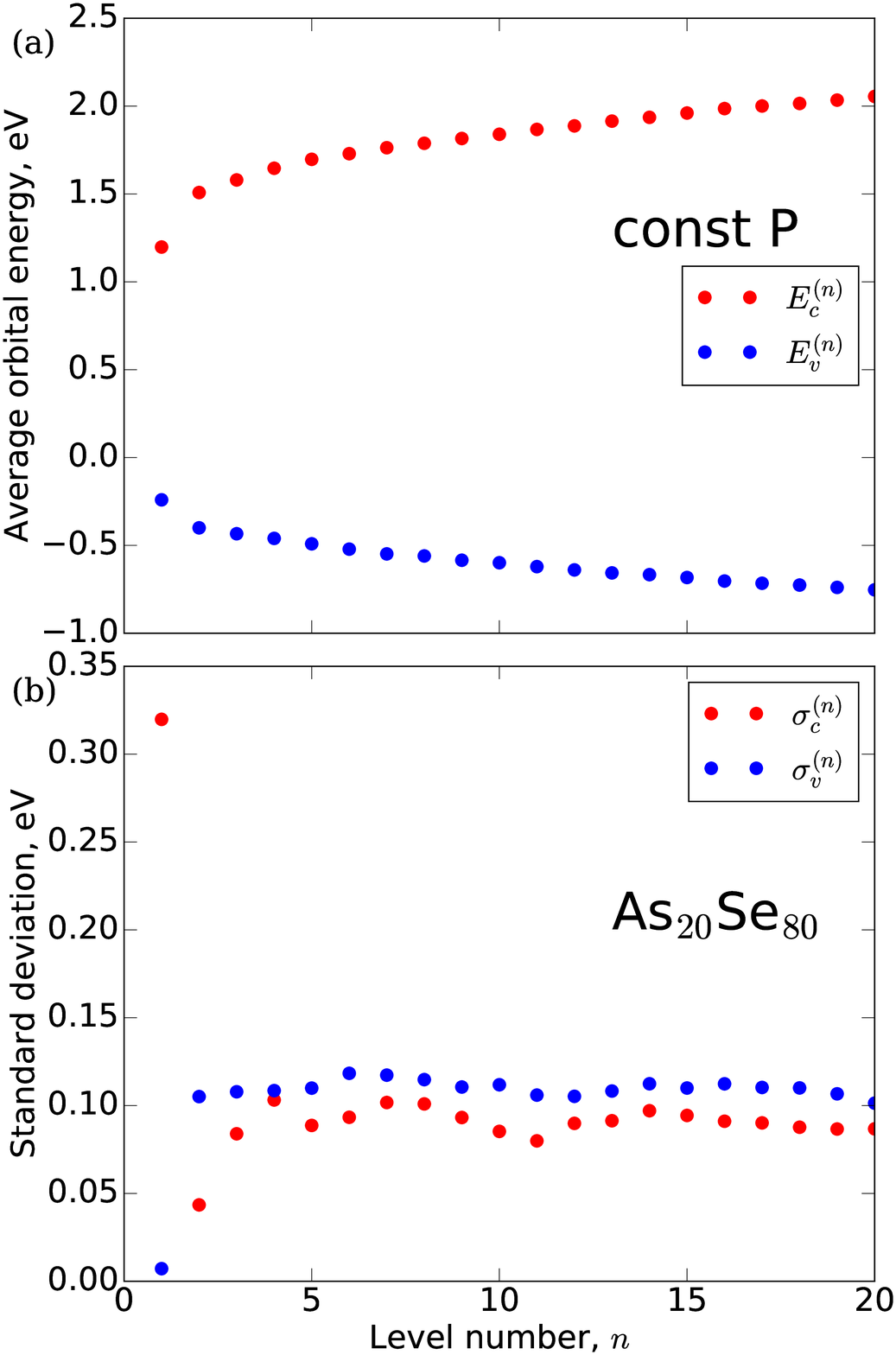}
  \caption{\label{EnSigma2} (a) The average energies $E_v^{(n)}$ and
    $E_c^{(n)}$ of the $n$-term as counted starting, respectively,
    from the HOMO and LUMO respectively. (b) The corresponding
    standard deviation. As$_{20}$Se$_{80}$, const-$P$.}
\end{figure}

\begin{figure}[H]
  \centering
  \includegraphics[width=0.7 \figurewidth]{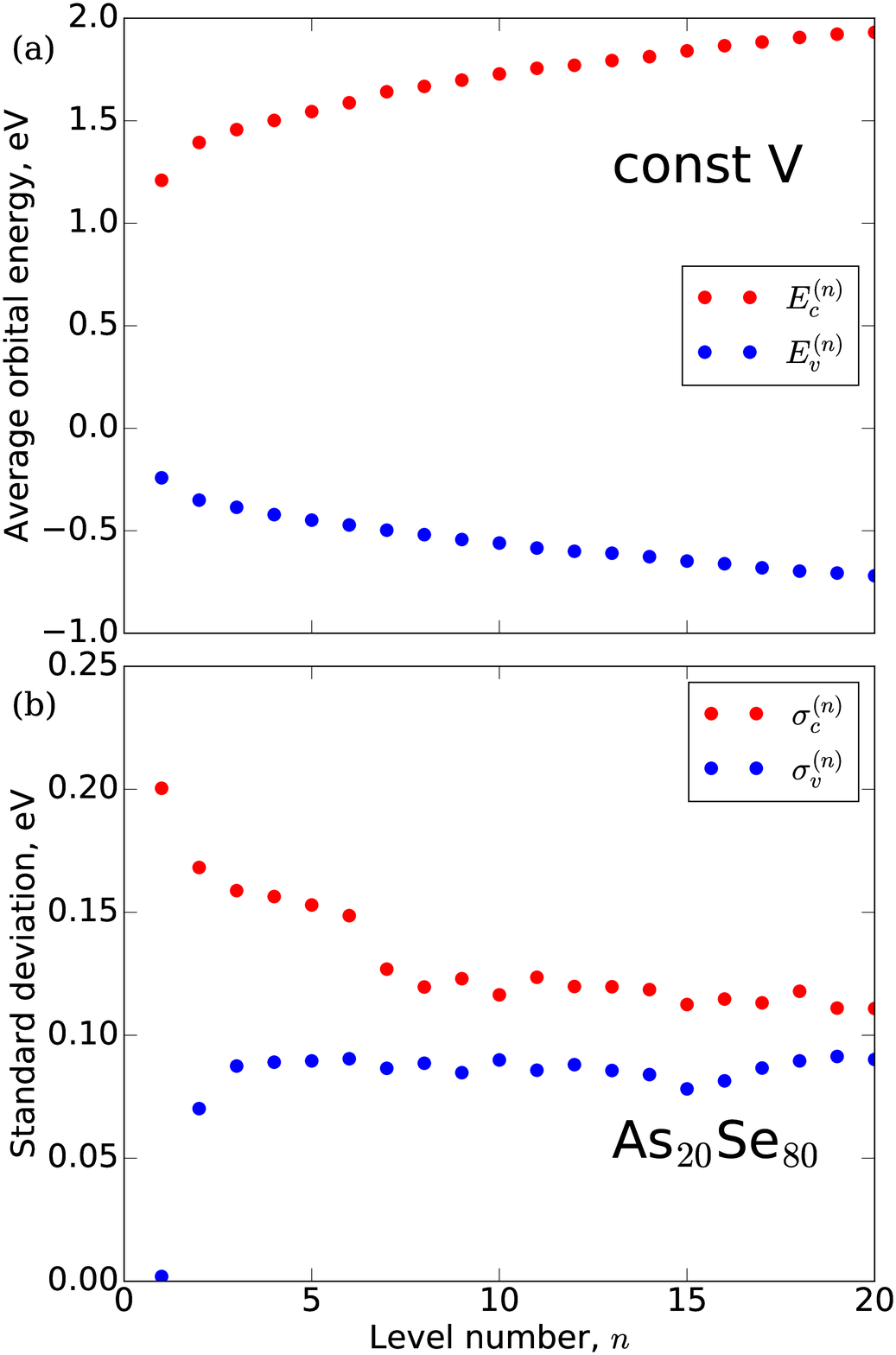}
  \caption{\label{EnSigma3} Same as Fig.~\ref{EnSigma2}, but at
    constant volume. }
\end{figure}

\begin{figure}[H]
  \centering
  \includegraphics[width=0.7 \figurewidth]{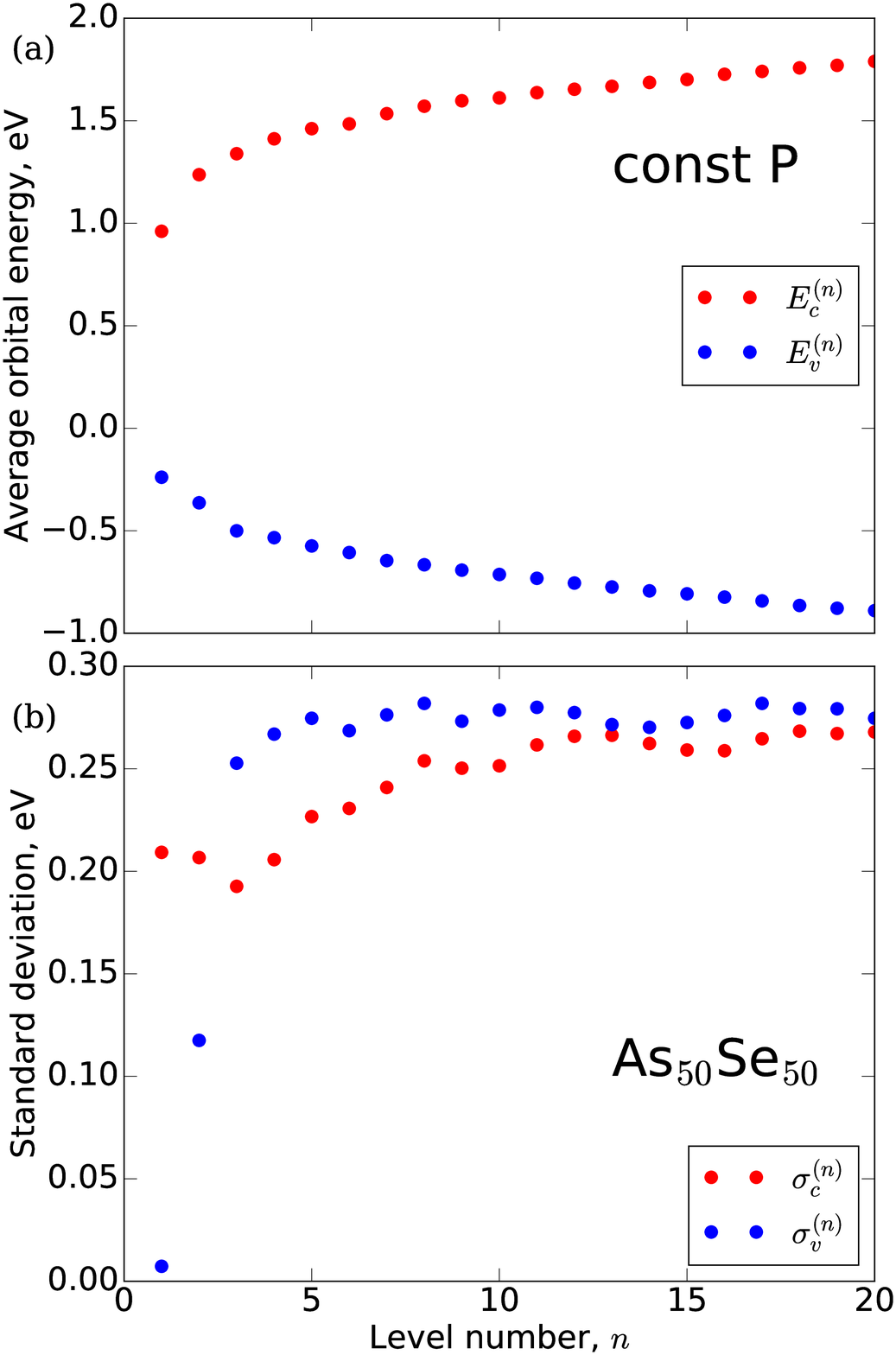}
  \caption{\label{EnSigma4} (a) The average energies $E_v^{(n)}$ and
    $E_c^{(n)}$ of the $n$-term as counted starting, respectively,
    from the HOMO and LUMO respectively. (b) The corresponding
    standard deviation. As$_{50}$Se$_{50}$, const-$P$.}
\end{figure}

\begin{figure}[H]
  \centering
  \includegraphics[width=0.8 \figurewidth]{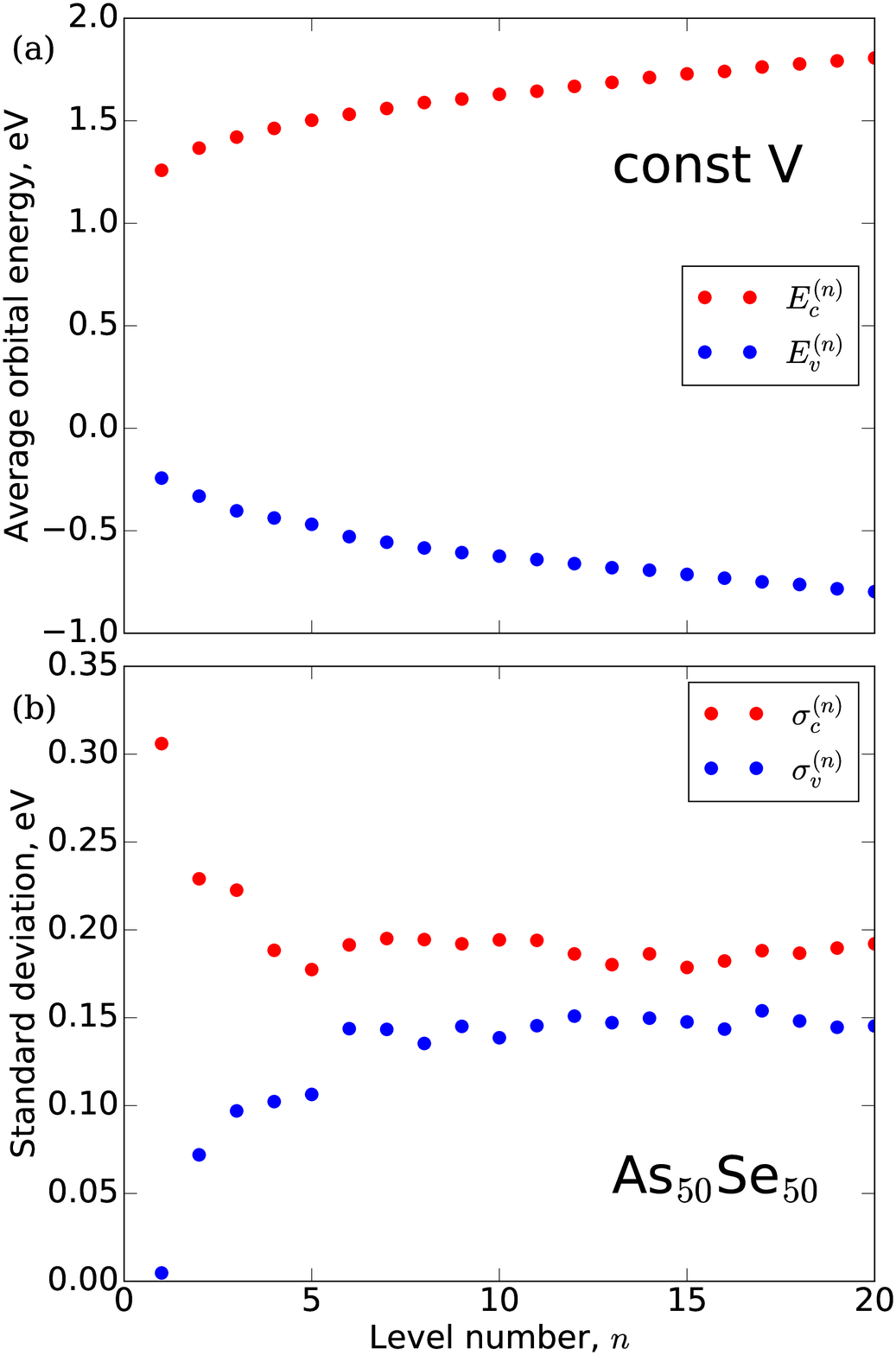}
  \caption{\label{EnSigma5} Same as Fig.~\ref{EnSigma4}, but at
    constant volume. }
\end{figure}

\section{Comparison of electronic spectra for samples obtained using
  different values of the threshold parameter $A$}

\begin{figure}[H]
  \centering
  \includegraphics[width=1\figurewidth]{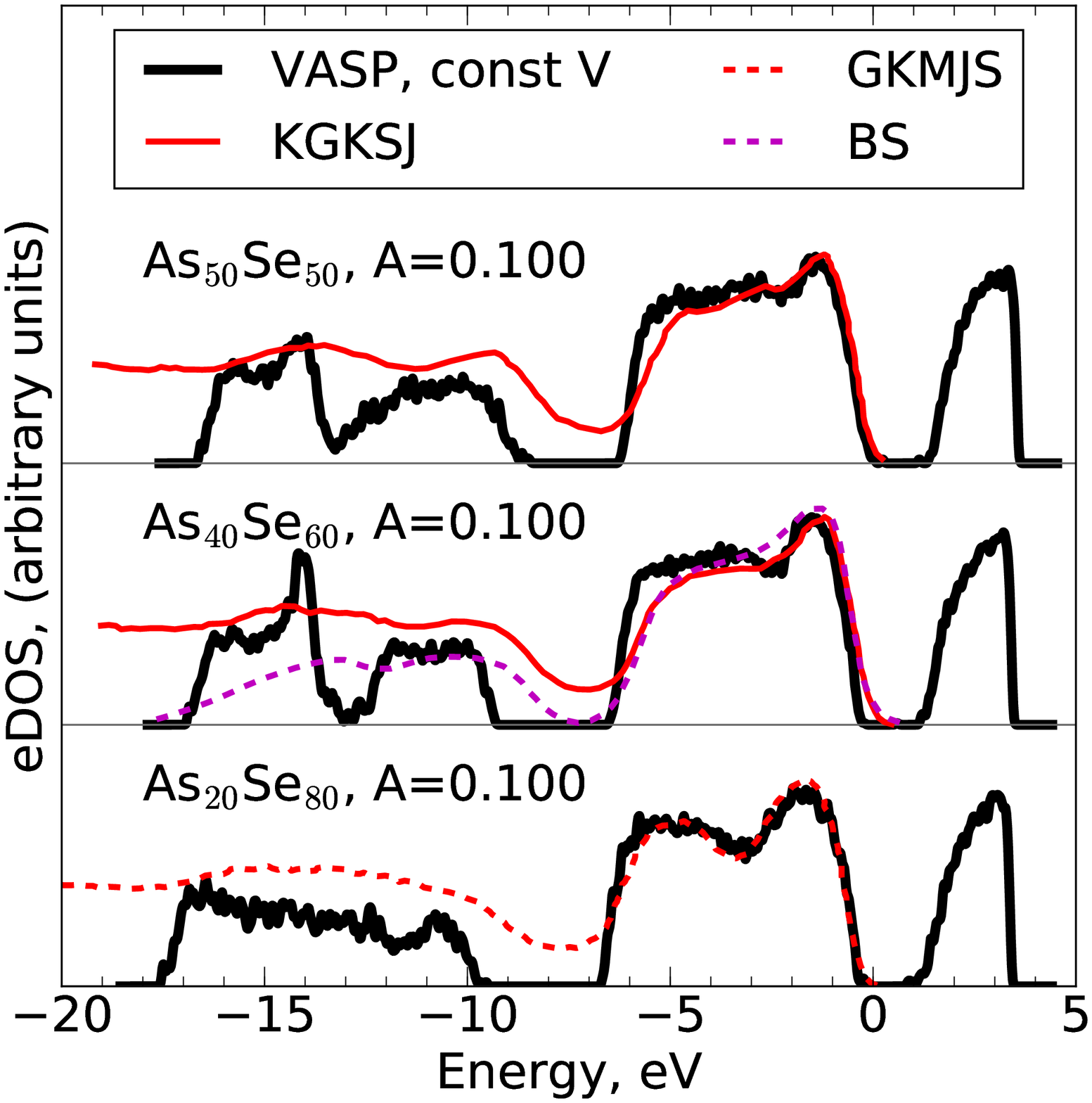}
  \caption{\label{edos-avg-stack-const-V-1} Averaged electronic
    density of states for $V$=const simulations at $A=0.100$. }
\end{figure}

\begin{figure}[H]
  \centering
  \includegraphics[width=1\figurewidth]{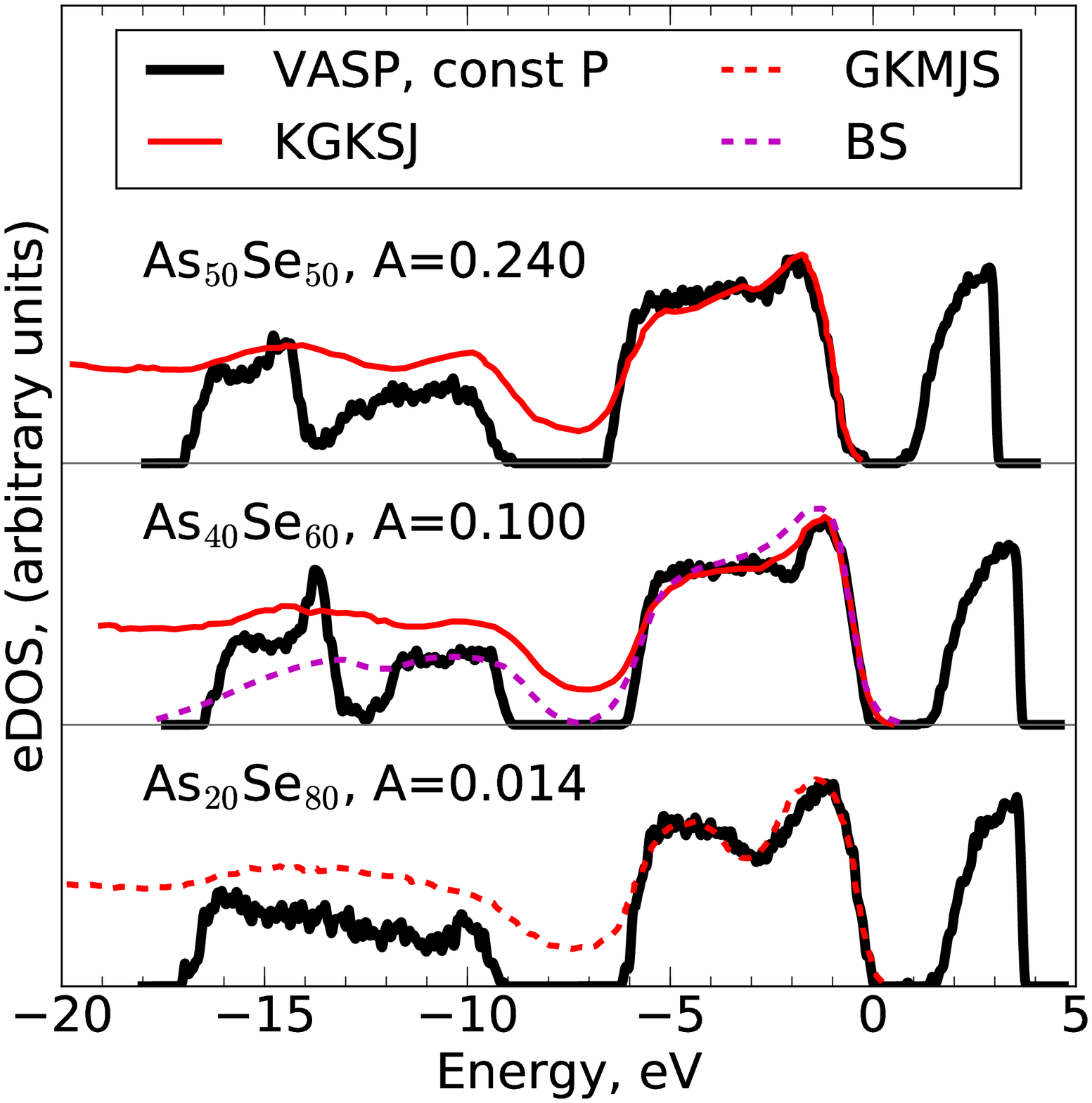}
  \caption{\label{edos-avg-stack-const-P-2} Averaged eDOS for
    const-$P$ simulations for different values of $A$, all which
    correspond to a low concentration of vacancies in the parent
    structure. Experimental data have been shifted to match the
    position of the valence band.  }
\end{figure}

\begin{figure}[H]
  \centering
  \includegraphics[width=1\figurewidth]{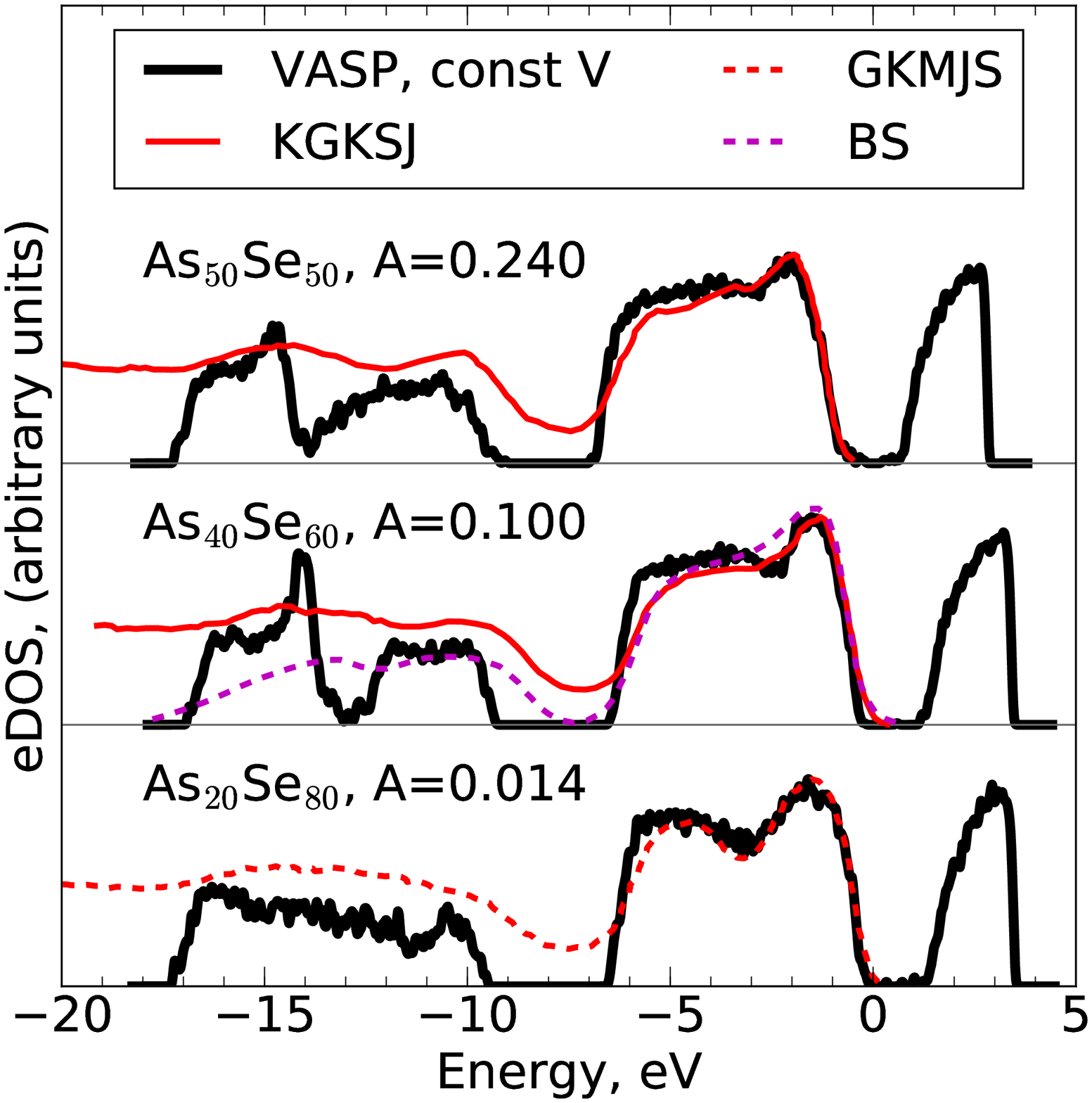}
  \caption{\label{edos-avg-stack-const-V-2} Same as
    Fig.~\ref{edos-avg-stack-const-P-2} but at constant volume. }
\end{figure}

\section{Basic properties of the wavefunctions of the topological
  midgap states, also in the presence of cross-linking with perfectly
  dimerized chains}

\begin{figure}[t]
  \centering
  \includegraphics[width=.9 \figurewidth]{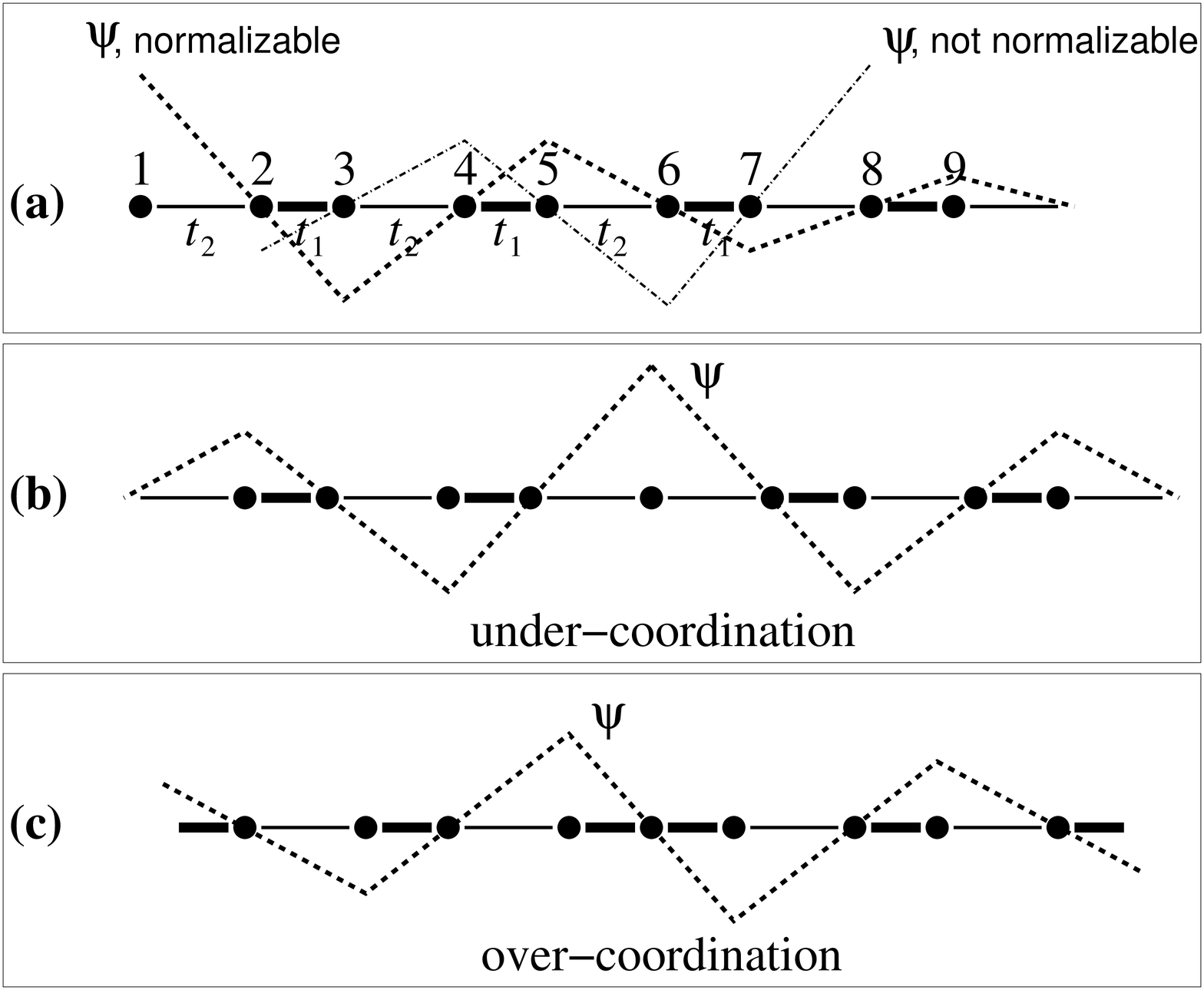}
  \caption{\label{terminal} {\bf (a)} An extended chain ending with
    the weaker bond. The numerals number the sites, $t_1$ and $t_2$
    denote the transfer integral for the stronger and weaker bond,
    respectively. The dashed line shows that the wavefunction of the
    midgap, edge state at $E=0$ decays exponentially into the
    bulk. The dashed-dotted line demonstrates that if the chain
    terminated with the stronger bond, the wavefunction would not be
    normalizable thus indicating that there would not be midgap edge
    state. {\bf (b)} shows how the wavefunctions for the midgap state
    based on an under-coordinated site can be obtained using the setup
    from (a). {\bf (c)} Same as (b) but for the topological midgap
    state resulting from over-coordination.  }
\end{figure}

To set the stage, we consider an extended, perfectly dimerized chain
terminating with the weaker bond on one end, as in
Fig.~\ref{terminal}(a). In the latter figure, $t_1$ and $t_2$ denote
the electronic transfer integrals for the stronger and weaker bond,
respectively: $t_1 > t_2$.  Everywhere below, we assume the on-site
energies are equal to zero, for simplicity. The argument can be
extended straightforwardly for a non-vanishing, sign-alternating
on-site energy.~\cite{PhysRevLett.49.1455, ZLMicro2} Such a situation
would be directly relevant to chain-like motifs in which chalcogen and
pnictogen alternate in sequence.

Irrespective of whether the terminal bond is weak or strong, the chain
has a bulk spectrum consisting of two bands, the outer edges of the
bands at $\pm |t_1 + t_2|$, the band-gap edges at $\pm |t_1 - t_2|$.
There is also a midgap, ``edge'' state for the configuration shown in
Fig.~\ref{terminal}(a) exactly in the middle of the forbidden gap. To
see this, we write down the stationary Schr\"odinger equation for the
components $\psi_n$ of the wave function, where the subscript $n$
labels the sites of the chain:
\begin{align}  - t_2 \psi_2 = E \psi_1 \nonumber \\
  - t_2 \psi_1 - t_1 \psi_3 = E \psi_2 \label{terminalEqns} \\
  - t_1 \psi_2  - t_2 \psi_4 = E \psi_3 \nonumber \\
  \ldots \nonumber
\end{align}

Clearly, Eqs.~(\ref{terminalEqns}) allow for a midgap solution exactly
in the middle of the gap:
\begin{eqnarray}  E &=& 0 \nonumber \\
  \psi_{2n} &=& 0 \label{terminalEqns1} \\
  \psi_{2n+1} &=& - \frac{t_2}{t_1} \psi_{2n-1}. \nonumber
\end{eqnarray}
The wavefunction vanishes on all even-numbered sites, while on the
odd-numbered sites, it alternates in sign while decreasing in
magnitude by a factor of $|t_2/t_1|$ per each pair of sites as one
moves away from the terminal site.  At the latter site, the
wavefunction has its largest value. Thus the wave function will
decrease exponentially as a function of the distance $|x|$ away from
the terminus:
\begin{equation} \psi(x) \propto e^{-|x| [\ln(t_1/t_2)]/2a},
\end{equation}
where $a$ is the average spacing between the sites. (We note a
peculiar feature of Eq.~(\ref{terminalEqns1}): If the transfer
integral is inversely proportional to the distance between the sites,
the wavefunction as a function of the physical coordinate consists of
a set of straight lines crossing the origin at the even-numbered sites
and thus can be easily drawn by hand, see the dashed line in
Fig.~\ref{terminal}(a).)

Using the same logic, one can show that such a midgap edge state would
not exist if the terminal bond were the strong one because this would
entail an exponential increase of the wavefunction toward the bulk of
the chain and, consequently, lack of normalizability for the
wavefunction. Again, this is straightforwardly evidenced graphically,
see the dashed-dotted line in Fig.~\ref{terminal}(a).

Non-withstanding their apparent simplicity, Eqs.~(\ref{terminalEqns})
demonstrate that the $E=0$ midgap state will persist even if the bond
strength varies somewhat in space, as long as the local value of the
$|t_2/t_1|$ ratio tends to a steady value less than one in the bulk in
the chain. (The latter condition is necessary to have well defined
bulk bands in the first place.)  The wavefunction will remain zero on
the even-numbered sites.  To avoid confusion we note that depending on
the detailed dependence of the local value of $|t_2/t_1|$ ratio on the
coordinate, other midgap states may be
present.~\cite{PhysRevB.21.2388}

Now, the above setup can be used to appreciate that the vicinity of an
undercoordinated site, in an otherwise perfectly dimerized extended
chain, will host a $E=0$ midgap state, as in
Fig.~\ref{terminal}(b). Hereby, the central site of the defect will
correspond to site 1 from Fig.~\ref{terminal}(a) while the chain
itself and midgap wavefunction to the left of site 1 will be the exact
mirror images of the respective entities from the r.h.s. of site 1.
Thus the wave function is even with respect to the reflection about
the central site and vanishes on sites that a separated by an odd
number of bonds from the central site.  Likewise, the vicinity of an
{\em over}-coordinated site will also host a midgap state, as in
Fig.~\ref{terminal}(c).  The center of the defect now corresponds to
site 2 from Fig.~\ref{terminal}(a). The corresponding wave function is
odd with respect to the reflection about the central site; it vanishes
on the central site and on sites that are separated by an even number
of bonds from the central site. These properties of the midgap states
are of course well known, see Ref.~\onlinecite{RevModPhys.60.781} and
references therein. According to those earlier works, if let
geometrically optimize, the chain will relax so as to get rid of all
midgap states other than the one in the
middle,~\cite{PhysRevB.21.2388} which we have seen is robust with
respect to vibrational deformations of the chain. This robustness
comes about for very general, topological
reasons.~\cite{RevModPhys.60.781}

\begin{figure}[t]
  \centering
  \includegraphics[width=.9 \figurewidth]{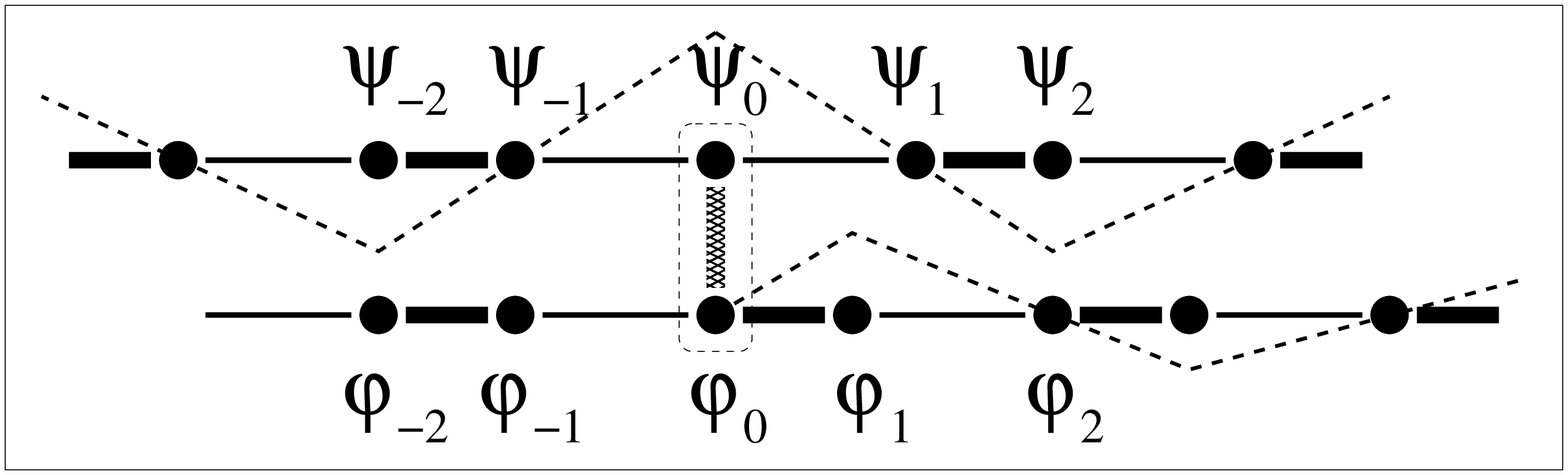}
  \caption{\label{coupled} Graphical description of the tight-binding
    Hamiltonian used here to infer the electronic structure for two
    coupled chains, one hosting an under-coordinated atom and one
    perfectly dimerized. The sites contained withing the dashed-line
    frame are physically located on the same atom, see also
    Fig.~\ref{chaincross} of the main text.  The zigzag-like dashed
    line shows a specific realization of the wavefunction of the $E=0$
    midgap state, i.e., the value of the coefficients $\psi$ and
    $\varphi$ at the respective sites.}
\end{figure}

Next, let us couple a chain containing an under-coordinated center
with a perfectly dimerized chain, the corresponding transfer integral
set equal to $t$, see Fig.~\ref{chaincross} of the main text and
Fig.~\ref{coupled}. Note that now we number the sites on the chains so
that the sites participating in the inter-chain coupling---one of them
hosting the defect---are labeled ``$0$''. The wave functions on the
defected and perfectly dimerized chain are labeled by the letters
$\psi$ and $\varphi$, respectively. It will suffice to write out only
two entries of the Schr\"odinger equation
\begin{align}  - t_2 \psi_{-1} - t_2 \psi_1 - t \varphi_0 = E \psi_0 \label{SE1} \\
  - t_2 \varphi_{-1} - t_1 \varphi_1 - t \psi_0 = E \varphi_0. \label{SE2}
\end{align}
Eq.~(\ref{SE1}) and the rest of the entries pertaining to the defected
chain still allow for a midgap solution at $E=0$, where the
wavefunction is an even function vanishing on odd-numbered sites, so
long as $\varphi_0 = 0$. Eq.~(\ref{SE2}), on the other hand, allows
for a solution at $E=0$ such that $\varphi_n = 0$, $n < 0$, while
\begin{equation} \varphi_1 = -\frac{t}{t_1} \psi_0. \label{spill}
\end{equation}
The rest of the positively-numbered $\varphi$'s obey the same
equations as the quantities $\psi$ in
Eq.~(\ref{terminalEqns1}). Clearly, the positively-numbered segment of
the dimerized chain is analogous to the setup in
Fig.~\ref{terminal}(a), since $\varphi_0 = 0$ while $\varphi_1 \ne
0$. The electron occupying the midgap state in the defected chain
tunnels within the positively-numbered side of the dimerized chain in
the same fashion as the electron occupying the terminal site in
Fig.~\ref{terminal}(a) tunnels toward the bulk of the chain in that
figure. The extent of ``spillage'' of the midgap state wavefunction
from the defected to the defect-free chain will be determined by the
strength $t$ of the inter-chain coupling, according to
Eq.~(\ref{spill}), and could be significant. Physically, the coupling
could be realized, for instance, through $sp$-mixing, which would
involve neighboring atoms and/or if the chains are not strictly
perpendicular. In the latter case, the coupling $t$ goes roughly $t_1
\cos \alpha$, where $\alpha$ is the angle between the chains. Finally,
similar logic can be used to show that a wavefunction centered on an
over-coordinated site will not ``spill'' into the crossing
chain. Indeed, coupling two chains does not change the symmetry of the
Hamiltonian with respect to reflection about the respective sites
through which the chains are coupled. Thus Eq.~(\ref{SE1}) still
yields a midgap state at $E=0$ whereby $\psi_{-1} = - \psi_1$, $\psi_0
= 0$, and thus $\varphi_0 = 0$. Consequently, if the dimerized chain
were to co-host a $E=0$ midgap state, the corresponding wavefunction
would have to be an exponential function of the coordinate throughout:
$\varphi_{2n+2} = - \frac{t_2}{t_1} \varphi_{2n}$, c.f. the bottom
equation in (\ref{terminalEqns1}). Since such a wavefunction is not
normalizable, we conclude that the $E=0$ midgap state based on an
over-coordinated site would be confined to the chain along which the
malcoordination takes place.

\section{Midgap states induced in crystalline slabs}

\begin{figure}[H]
  \centering
  \includegraphics[width=.9\figurewidth]{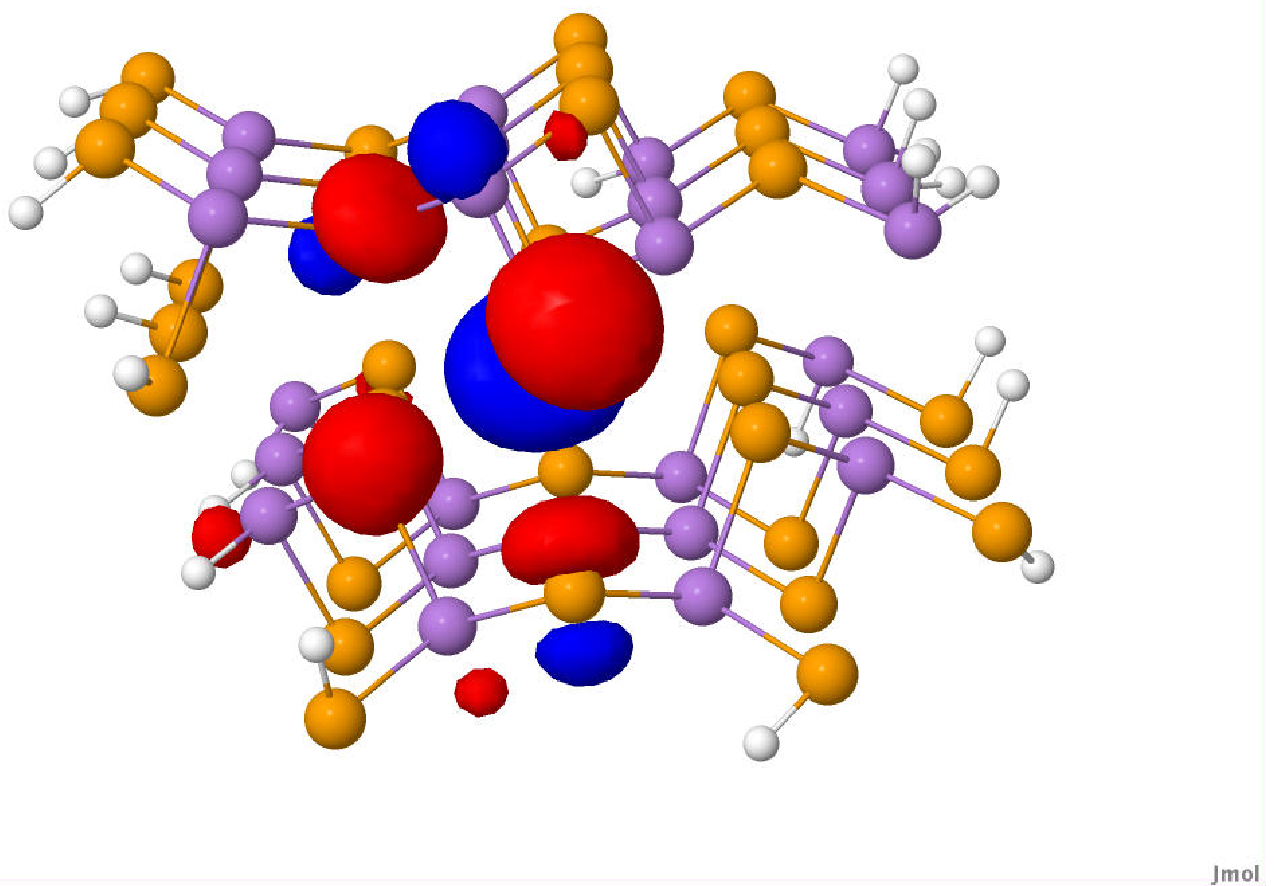}
  \caption{\label{unrelaxed} The structure and electronic spectrum of
    a slab of crystalline As$_2$Se$_3$ passivated by hydrogens. The
    passivation is to achieve the proper bond number (3 for As and 2
    for Se) everywhere except in one place on the surface so as to
    have exactly one over-coordinated atom.  The structure is not
    optimized. The midgap level is seen deep inside the gap but
    substantially away from the gap center. The corresponding
    wave-function is shown as the blobs, the two colors denoting the
    signs of the wavefunction. The quantum chemistry software:
    MOPAC2016, Version: 18.117L, James J. P. Stewart.~\cite{mopac} We
    note that by construction, MOPAC greatly overestimates the
    magnitude of the forbidden gap, see a related discussion in
    Refs.~\onlinecite{ZLMicro1, ZLMicro2}.}
\end{figure}

\begin{figure}[H]
  \centering
  \includegraphics[width=.9\figurewidth]{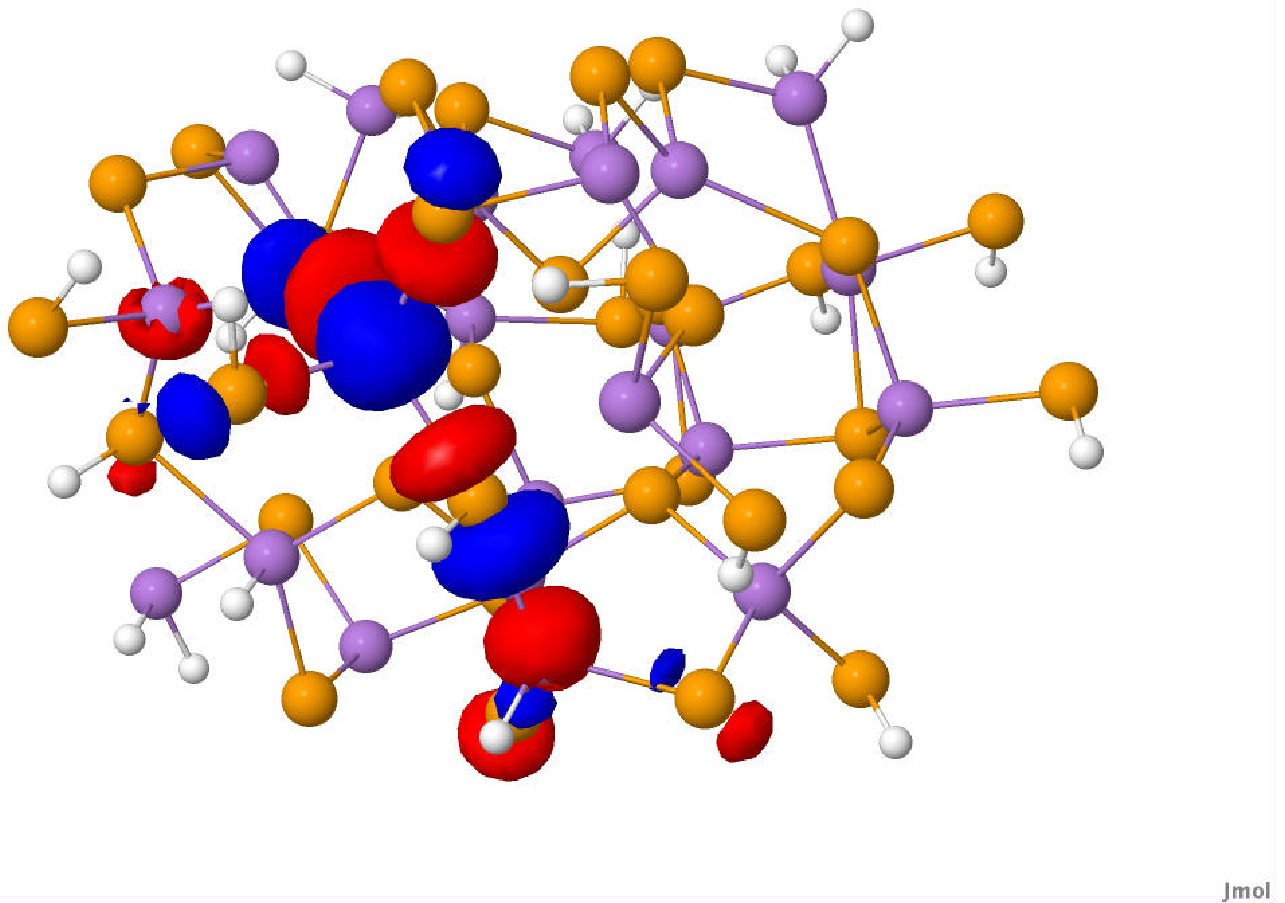}
  \caption{\label{relaxed} Same as Fig.~\ref{unrelaxed}, but following
    geometric optimization. The midgap state is seen to move closer to
    the gap center while the midgap state wave-function becomes more
    extended and less compact.}
\end{figure}

\section{Localization and shape anisotropy of electronic states for
  samples containing an odd number of electrons}

\begin{figure}[H]
  \centering
  \includegraphics[width=.9\figurewidth]{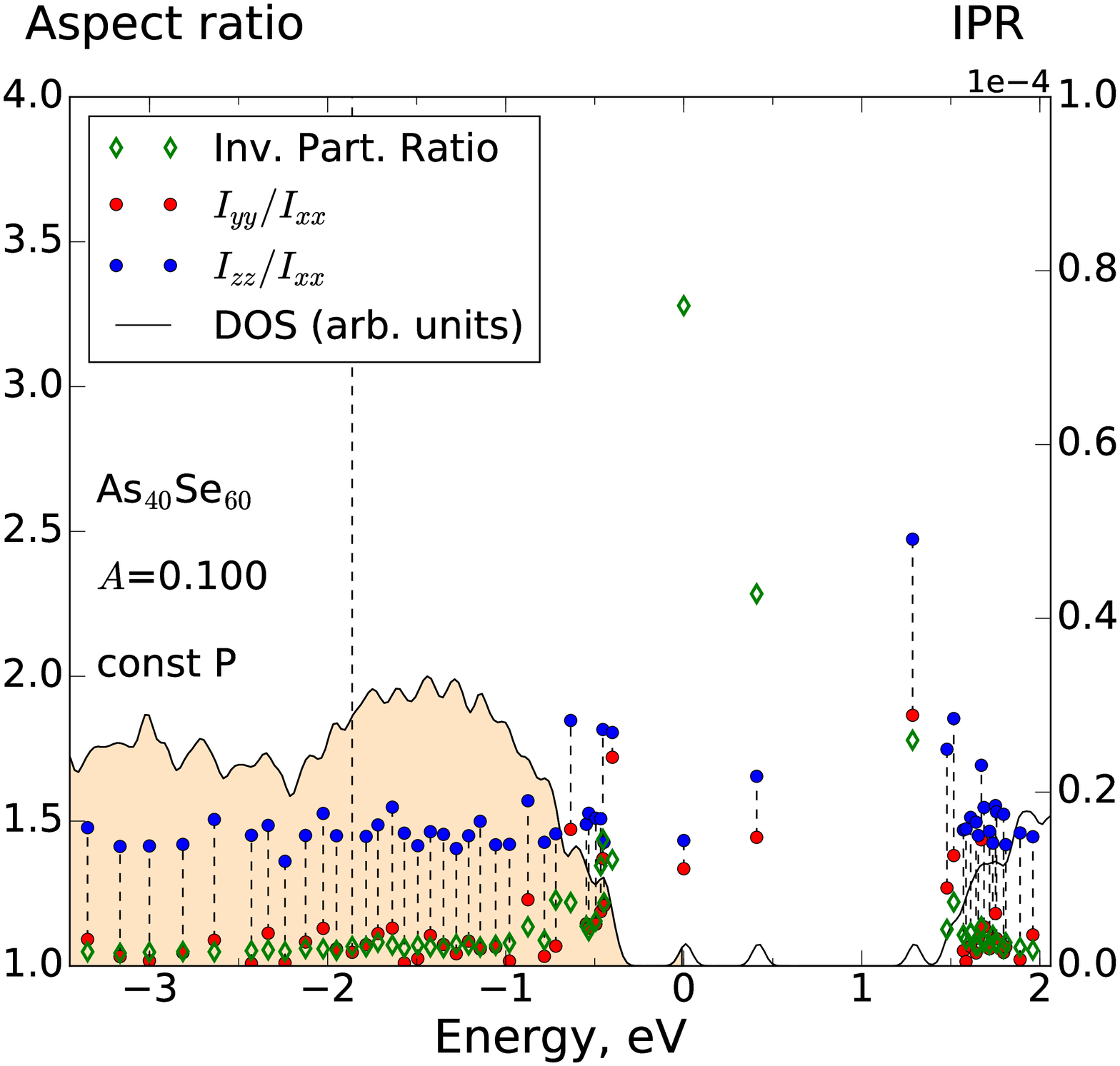}
  \caption{\label{inertiaratio1} Odd number of electrons, protocol I:
    Aspect ratios $I_{yy}/I_{xx}$ and $I_{zz}/I_{xx}$ and
    corresponding inverse participation ratio (IRP) for select states
    in the band gap and its vicinity for a As$_{40}$Se$_{60}$ sample
    that hosts at least one midgap state. The thin solid line shown
    the electronic density of states, shading indicates filled states.
  }
\end{figure}

\begin{figure}[H]
  \centering
  \includegraphics[width=.9\figurewidth]{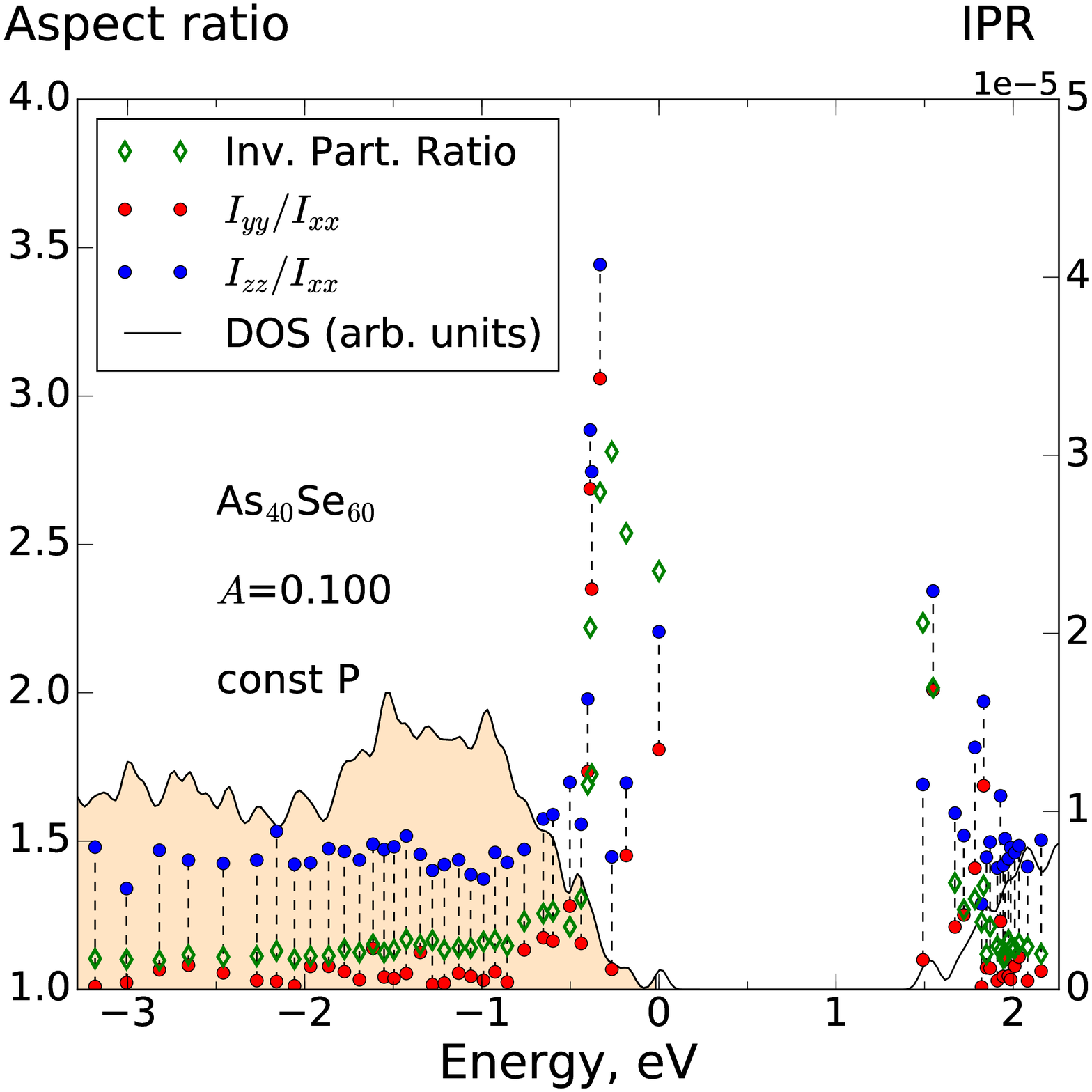}
  \caption{\label{inertia-ratio2} Same as Fig.~\ref{inertiaratio1} but
for protocol II. }
\end{figure}

\begin{figure}[H]
  \centering
  \includegraphics[width=.9\figurewidth]{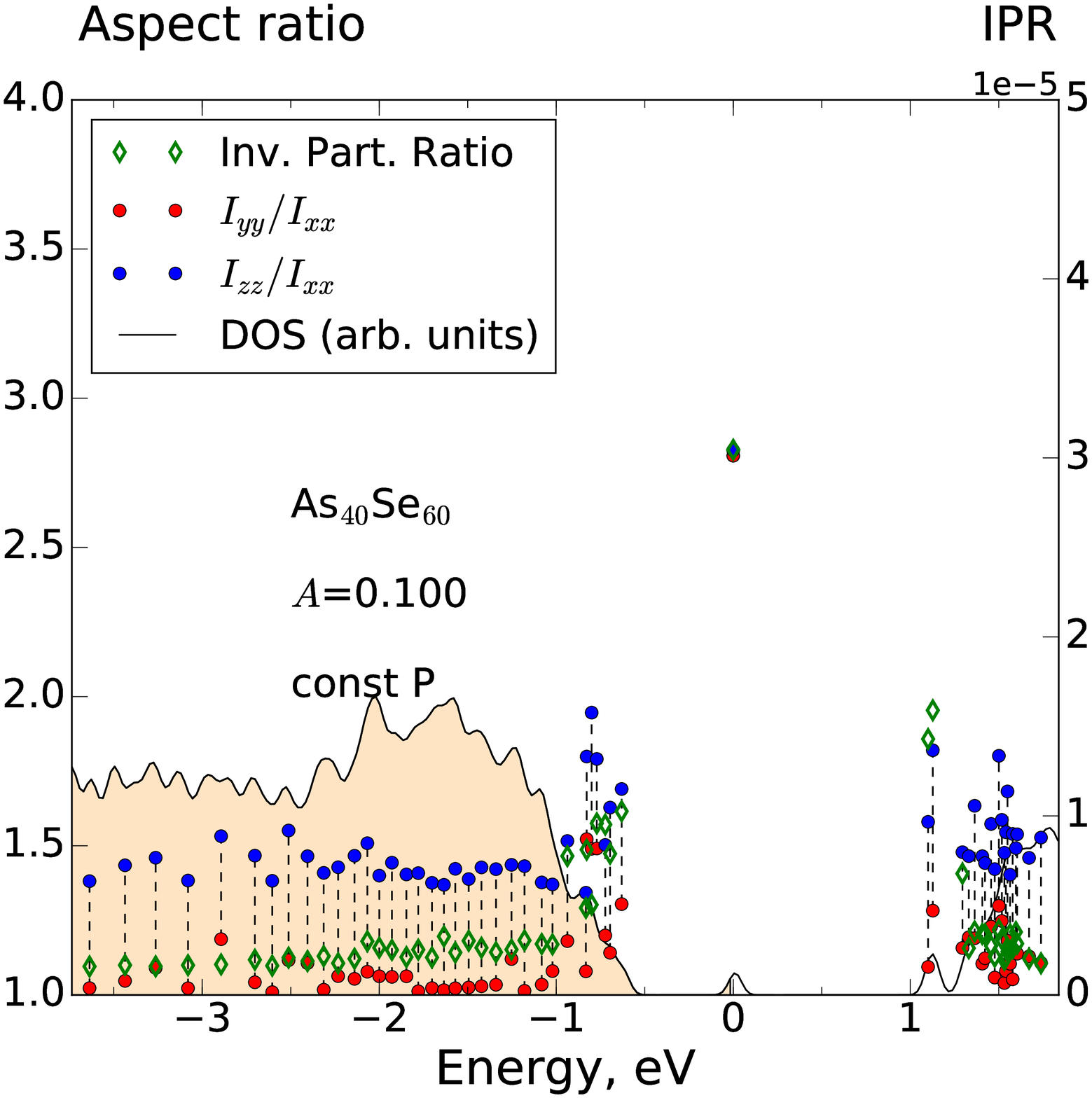}
  \caption{\label{inertia-ratio3} Same as Fig.~\ref{inertiaratio1} but
    for protocol III.  }
\end{figure}

%


\end{document}